\documentclass[a4paper,fleqn,usenatbib]{mnras}
\usepackage[T1]{fontenc}
\usepackage{times}

\usepackage{color,soul}
\usepackage{graphicx}	% Including figure files
\usepackage{amsmath}	% Advanced maths commands
\usepackage{amssymb}	% Extra maths symbols
\usepackage{etoolbox}
\makeatletter
\patchcmd\@combinedblfloats{\box\@outputbox}{\unvbox\@outputbox}{}{%
   \errmessage{\noexpand\@combinedblfloats could not be patched}%
}%
 \makeatother

\def\ba{\begin{eqnarray}}
\def\ea{\end{eqnarray}}

\newcommand{\md}{\mathrm{d}}

\newcommand{\Rg}{\mathbf{R}_\mathrm{g}}
\newcommand{\Rb}{\mathbf{R}_\mathrm{b}}
\newcommand{\rp}{r_\mathrm{p}}
\newcommand{\ra}{r_\mathrm{a}}

\newcommand{\nn}{\nonumber}

 \title[Secular dynamics of binaries I: general formulation]{Secular dynamics of binaries in stellar clusters I: general formulation and dependence on cluster potential}

\author[C. Hamilton \& R. R. Rafikov]{
  Chris Hamilton$^{1}$\thanks{E-mail: ch783@cam.ac.uk} and Roman R. Rafikov$^{1,2}$\\
$^1$Department of Applied Mathematics and Theoretical Physics, University of
Cambridge, Wilberforce Road, Cambridge CB3 0WA, UK\\
$^2$Institute for Advanced Study, Einstein Drive, Princeton, NJ 08540, USA}

\date{Accepted XXX. Received YYY; in original form ZZZ}

\pubyear{2019}

\begin{document}
\label{firstpage}
\pagerange{\pageref{firstpage}--\pageref{lastpage}}
\maketitle

% Abstract of the paper
\begin{abstract}
Orbital evolution of binary systems in dense stellar clusters is important in a variety of contexts: origin of blue stragglers, progenitors of compact object mergers, millisecond pulsars, and so on. Here we consider the general problem of secular evolution of the orbital elements of a binary system driven by the smooth tidal field of an axisymmetric stellar cluster (globular, nuclear, etc.) in which the binary orbits. We derive a secular Hamiltonian (averaged over both the inner Keplerian orbit of the binary and its outer orbit within the cluster) valid to quadrupole order for an arbitrary cluster potential and explore its characteristics. This doubly-averaged `tidal' Hamiltonian depends on just two parameters, which fully absorb the information about the background cluster potential and the binary's orbit within it: a dimensional parameter $A$ setting the secular timescale, and a dimensionless parameter $\Gamma$ which determines the phase portrait of the binary's inner orbital evolution. We examine the dependence of $A$ and $\Gamma$ on cluster potential (both spherical and axisymmetric) and on the binary orbit within the cluster. Our theory reproduces known secular results --- such as Lidov-Kozai evolution and the effect of the Galactic tide on Oort Cloud comets --- in appropriate limits, but is more general. It provides a universal framework for understanding dynamical evolution of various types of binaries driven by the smooth tidal field of any axisymmetric potential. In a companion paper (Hamilton \& Rafikov 2019b) we provide a detailed exploration of the resulting orbital dynamics.
\end{abstract}

% Select between one and six entries from the list of approved keywords.
% Don't make up new ones.
\begin{keywords}
gravitation -- celestial mechanics -- stars: kinematics and
dynamics -- galaxies: star clusters: general -- binaries: general
\end{keywords}

%%%%%%%%%%%%%%%%%%%%%%%%%%%%%%%%%%%%%%%%%%%%%%%%%%

%%%%%%%%%%%%%%%%% BODY OF PAPER %%%%%%%%%%%%%%%%%%
	
\section{Introduction}

Orbital motion of two bound point masses perturbed by external forces represents one of the oldest problems in celestial mechanics \citep{Murray1999}. It naturally emerges in a variety of astrophysically relevant situations: classical secular Laplace-Largange evolution of planetary systems \citep{Murray1999}, Lidov-Kozai oscillations forced by a distant companion in a triple system \citep{Lidov1962,Kozai1962}, evolution of Oort Cloud comets driven by the Galactic tide \citep{Heisler1986}, and so on. Due to the weakness of the external perturbation, the evolution is usually {\it secular} in nature and operates on long time scales.

Dense stellar systems --- globular and open clusters, nuclear star clusters, galaxies themselves (hereafter collectively called `clusters') --- represent ideal environments for perturbing binaries. Historically, these perturbations were considered predominantly in the context of encounters with individual passing stars \citep{Heggie1975,Hut1983a,Hut1983b}, which should occur rather frequently in dense clusters (although not always in the perturbative limit). In particular, secular changes of binary orbital elements caused by a passage of a single star were calculated by \citet{Rasio1995}, \citet{Heggie1996}, and \citet{Hamers2018}. 

At the same time, the gravitational effect of the smooth mass distribution of the full cluster on binary orbital evolution has been explored mainly in the context of Oort Cloud formation and evolution driven by the Galatic tide (\citealt{Heisler1986}, hereafter HT86). In this case the `binary' is the Sun-comet system, and Galactic tides can be at least as important as stellar encounters for the orbital evolution of Oort Cloud comets. Effects of the spatially smooth Galactic tide have also been studied in the context of wide stellar binaries \citep{Jiang2010} and long-period planetary systems in the Galactic bulge \citep{Veras2013a}. 

Interest in binaries in stellar clusters has been renewed recently by the discoveries of mergers of compact-object binaries by the {\it LIGO} collaboration (e.g. \citealt{Abbott2016}, \citealt{Abbott2017}). A number of channels for the production and orbital evolution of merging binaries in dense stellar systems have been proposed following these discoveries \citep{Wen2003,OLeary2006,Miller2009,Bae2014,Rodriguez2015,Rodriguez2016a,Rodriguez2016b,Antonini2016,Askar2017,Chatterjee2017,Petrovich2017,Hamers2018a,Hoang2018,Rodriguez2018,Samsing2018,Fragione2018}. Some of them involve Lidov-Kozai (\citealt{Lidov1962,Kozai1962}; hereafter LK) coupling with a third distant companion\footnote{\citet{Silsbee2017} and \citet{Antonini2017} also considered Lidov-Kozai cycles in isolated triples to explain {\it LIGO} events.}, which is either captured from the cluster environment (e.g. \citealt{Antonini2016}) or is effectively represented by a central super-massive black hole \citep{Antonini2012,Petrovich2017,Hoang2018}. Under certain conditions, LK oscillations can naturally drive the binary orbit to become highly eccentric, boosting gravitational wave emission and substantially speeding up binary coalescence. Similar ideas (with different sources of dissipation at periapsis) have been previously considered for explaining the origin of other exotic objects typically found in stellar clusters, such as blue stragglers (e.g. \citealt{Knigge2009,Perets2009}). 

In this work we consider the general problem of secular evolution of binary orbital elements driven by the tidal field that arises due to the smooth mass distribution of an axisymmetric cluster in which the binary moves. We do this for an arbitrary axisymmetric cluster potential and set no constraints on the type of orbit that the binary can have in the cluster. This allows us to formulate the most general framework for treating tide-driven secular evolution of binaries, applicable to a variety of astrophysical systems.  

In doing this we neglect the stochastic effect of individual stellar passages on the orbital elements of the binary. Separating the cumulative effect of multiple stellar encounters from the smooth Galactic tide is a non-trivial exercise, as demonstrated previously by \cite{Collins2010} and \cite{Jiang2010}. Nevertheless, for the purposes of clarity, we prefer to focus here on the effects of the mean tidal field due to the smooth mass distribution inside the cluster --- effects of encounters with individual stars will be incorporated later.

This paper (the first in a series) is devoted to the derivation of the general Hamiltonian governing secular orbital evolution of a binary in an arbitrary axisymmetric cluster potential, as well as to exploring the dependence of some characteristics of this Hamiltonian on the properties of the cluster potential and the binary's orbit within it. It is structured as follows. In \S\ref{setting} we derive the tidal Hamiltonian for the dynamical evolution of binary orbital elements due to any axisymmetric perturbation when expanded to quadrupole order.  In \S\S\ref{sect:avgH}-\ref{secham} we average the tidal potential over both the binary's inner orbit and then over many orbits of the binary around the cluster, arriving at a simple doubly-averaged (secular) Hamiltonian which describes long-term evolution of the binary's orbital elements. The coefficients entering this secular Hamiltonian depend on the potential of the host system and the binary's barycentric orbit within this potential, and we explore this dependence in detail in \S \ref{abgamma}. We verify the time-averaging procedure numerically in \S\ref{numveravg}.  In \S\ref{sect:disc}   we discuss the limitations of our theory, and show how our general results are connected with various special cases already explored by others (also in Appendices \ref{RecoverLK} \& \ref{RecoverHT}).

In a companion paper (\citealt{Hamilton2019b}, hereafter `Paper II') we provide a complete study of the dynamics that result from the secular tidal Hamiltonian derived in this work, calculate the timescale and amplitude of eccentricity oscillations, and verify our theory with direct N-body integration.

%%%%%%%%%%%%%%%%%%%%%%%%%%%%%%%%%%%%%%%%%%
%%%%%%%%%%%%%%%%%%%%%%%%%%%%%%%%%%%%%%%%%%

	\section{Hamiltonian with cluster tides} 
	\label{setting}
	
%%%%%%%%%%%%%%%%%%%%%%%%%%%%%%%%%%%%%%%%%%

Let us consider a binary system with semi-major axis $a$ and eccentricity $e$, consisting of point masses $m_1$ and $m_2$. Binary components gravitationally interact with each other and a fixed smooth background potential $\Phi$ of a much more massive system, which we will later take to be axisymmetric.  The application we have most readily in mind is that of binary stars in the mean field potential of a globular or nuclear star cluster, and for this reason we will frequently refer to $m_1$ and $m_2$ as `stars' and to the background system as `the cluster'.  However it should be borne in mind that our analysis works for any system of two gravitationally bound objects (binary black holes, comet-Sun system, etc.) moving in any axisymmetric potential (galaxy, open cluster, young stellar cluster, etc).  
	
\begin{figure}
\includegraphics[width=0.99\linewidth]{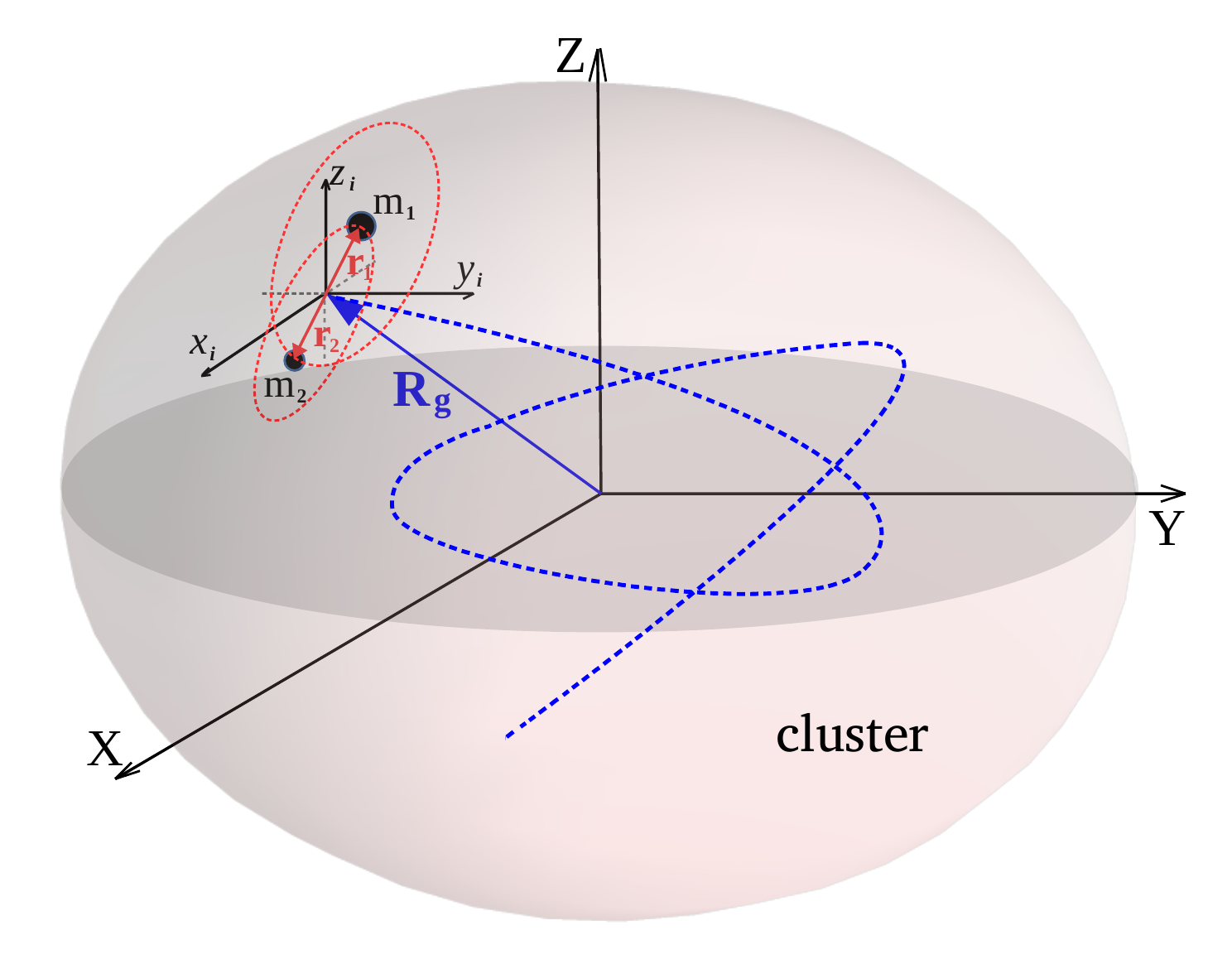}
\caption{Illustration of the binary within the axisymmetric cluster.  The binary's barycentre $\Rb$ coincides, to sufficient accuracy, with a `guide' radius vector $\Rg$ moving as a test particle in the cluster-centric coordinate system $(X,Y,Z)$.  The symmetry axis of the cluster is $Z$. Binary inclination $i$ is measured relative to the $(X,Y)$ plane and the longitude of the ascending node $\Omega$ of the binary is measured with respect to the $X$ axis (\S\ref{orbels}).  Note that the trajectory of $\Rg$ (illustrated with a blue dashed line) is not closed in a general axisymmetric cluster potential.}
\label{picture}
\end{figure}

Throughout this work we will refer to the binary's orbit around the cluster as the `outer orbit', while the orbit of the binary components about their common barycentre will be called the `inner orbit', to coincide with the standard terminology in LK studies (e.g. \citealt{Naoz2016}). To describe the outer and inner orbits we set up two coordinate systems --- see Figure \ref{picture} for illustration. 

The first, given by $\mathbf{R}=(X,Y,Z)$, has its origin at the centre of the cluster. In this coordinate system, the radius vector of the outer orbit, i.e. from the cluster centre to the barycentre of the binary is given by $\Rb = (X_\mathrm{b},Y_\mathrm{b},Z_\mathrm{b})$.  The second (non-inertial) coordinate system has its origin at $\Rb$, and its axes are fixed to be aligned with those of the first system, so only its origin moves. The position of star $i=1,2$ in the non-inertial system is then given by $\mathbf{r}_i=(x_i,y_i,z_i)$. The position of star $i$ relative to the centre of the cluster is $\mathbf{R}_i = \Rb + \mathbf{r}_i = (X_\mathrm{b}+x_i, Y_\mathrm{b}+y_i,Z_\mathrm{b}+z_i)$ and the barycentre is at $\Rb=(m_1\mathbf{R}_1+m_2\mathbf{R}_2)/(m_1+m_2)$.  

The equation of motion of star $i=1,2$ is then
\begin{equation}
	  \frac{\md^2(\mathbf{r}_i+\Rb)}{\md t^2} = - (\nabla \Phi)_{\Rb+\mathbf{r}_i} - \frac{Gm_j}{\vert \mathbf{r}_i-\mathbf{r}_j\vert^3}( \mathbf{r}_i-\mathbf{r}_j), 
	 \label{one}
\end{equation} 
for $i\neq j$, where the subscript on derivatives means that we evaluate the derivative at $\Rb + \mathbf{r}_i$.  

Defining the relative position $\mathbf{r} = (x,y,z) \equiv \mathbf{r}_1-\mathbf{r}_2$, and $\mu\equiv G(m_1+m_2)$, one obtains from (\ref{one})
\begin{align} 
\frac{\md^2 \mathbf{r}}{\md t^2} = - \left[(\nabla \Phi)_{\Rb+\mathbf{r}_1}- (\nabla \Phi)_{\Rb+\mathbf{r}_2}\right]  - \frac{\mu \mathbf{r}}{r^3}, 
\label{d2rdt2_0} 
\end{align} 
which is the general equation of relative motion of the binary components.

%%%%%%%%%%%%%%%%%%%%%%%%%%%%%%%%%%%%%%%%%%

\subsection{Tidal approximation}
\label{sect:tidalH}

%%%%%%%%%%%%%%%%%%%%%%%%%%%%%%%%%%%%%%%%%%

We now employ the tidal approximation, which means that in equation (\ref{d2rdt2_0}) we expand the potential $\Phi(\Rb+\mathbf{r}_i)$ around $\Rb$.  The Cartesian components of the vector $\nabla \Phi$ at position $\Rb+\mathbf{r}_i$ are 
\begin{align} 
[(\nabla \Phi)_{\Rb+\mathbf{r}_i}]_\alpha &= \left( \frac{\partial \Phi}{\partial R_\alpha}\right)_{\Rb} \nn \\ &+ \sum_{\beta} \left( \frac{\partial^2\Phi}{\partial R_\alpha \partial R_\beta}\right)_{\Rb}r_{i,\beta} + O(r_i^2),
\label{eq:expansion}
\end{align} 
where $\alpha$, $\beta$ are the coordinate indices, so that e.g. $R_\alpha$ and $r_{i,\alpha}$ represent the components of $\mathbf{R}= (X,Y,Z)$ and $\mathbf{r}_i = (x_i,y_i,z_i)$ respectively. We expect that terms $O(r_i^2)$ will be subdominant because the distance to the centre of the cluster ($\sim \vert \Rb \vert$) is much greater than the binary separation $r$. In this approximation we find 
\begin{align}
\frac{\md^2 \mathbf{r}}{\md t^2} = - \sum_{\beta}  r_\beta
\frac{\partial}{\partial R_\beta}\left(\nabla\Phi\right)_{\Rb} - \frac{\mu \mathbf{r}}{r^3},
\label{d2ridt2} 
\end{align} 
with $r_\alpha$ the components of $\mathbf{r}=(x,y,z)$.  Keeping in mind that, since the axes of the two coordinate systems are aligned, we may interchange $\partial/\partial R_\alpha$ with $\partial/\partial r_\alpha$, we can also write this as
\begin{align} 
\frac{\md^2 \mathbf{r}}{\md t^2} = - (\mathbf{r} \cdot\nabla)
\left(\nabla\Phi\right)_{\Rb} - \frac{\mu \mathbf{r}}{r^3}. 
\label{d2rdt2} 
\end{align} 
	
With $x,y,z$ as our canonical coordinates and $p_x=\dot{x}, p_y=\dot{y}, p_z=\dot{z}$ as the corresponding momenta, these equations of motion may be derived from the time-dependent Hamiltonian  
\begin{equation} 
H = H_0 + H_1, 
\end{equation}
where 
\begin{align} 
\label{H0} 
H_0 =& \frac{1}{2}\mathbf{p}^2 - \frac{\mu}{r}, \\  
H_1 =& \left( \frac{\partial^2 \Phi}{\partial x^2}\right)_{\Rb}\frac{x^2}{2} + \left( \frac{\partial^2 \Phi}{\partial y^2}\right)_{\Rb}\frac{y^2}{2} + \left( \frac{\partial^2 \Phi}{\partial z^2}\right)_{\Rb}\frac{z^2}{2} \nn \\ &+ \left(\frac{\partial^2\Phi}{\partial x \partial y}\right)_{\Rb}xy  + \left(\frac{\partial^2\Phi}{\partial x \partial z}\right)_{\Rb}xz  +   \left(\frac{\partial^2 \Phi}{\partial y \partial z}\right)_{\Rb}yz  ,    
\label{H1non} 
\end{align} 
and $H_1\ll H_0$. To compress the notation we write $(\partial^2\Phi/\partial r_\alpha \partial r_\beta)_{\Rb}\equiv \Phi_{\alpha\beta}$ so that the perturbing  (`tidal') Hamiltonian reads 
\begin{align} 
H_1 =& \frac{1}{2}\sum_{\alpha \beta}\Phi_{\alpha \beta}(\Rb)\, r_\alpha r_\beta, 
\label{H1comp} 
\end{align} 
where we sum over $\alpha, \beta = x,y,z$.  

The dominant part of the Hamiltonian $H_0$ corresponds to the motion of an isolated binary star about its own barycentre, and has no explicit time dependence.  The perturbing term $H_1$ takes into account the tidal effects of the external potential, which will drive the secular evolution of the binary orbital elements. It implicitly depends on time through $\Rb(t)$, which we look at next.

According to equation (\ref{one}) the evolution of $\Rb$ is governed by
\begin{align} 
\frac{\md^2 \Rb}{\md t^2} &= - \frac{m_1 (\nabla \Phi)_{\Rb+\mathbf{r}_1}+ m_2(\nabla \Phi)_{\Rb+\mathbf{r}_2}}{m_1+m_2}\nonumber\\
&= - (\nabla \Phi)_{\Rb}\left[1+O\left(r^2/\vert \mathbf{R}_\mathrm{b}\vert^2\right)\right]. 
\end{align}
The small correction terms on the right hand side of this equation mean that, in general, the motion of $\Rb$ in the cluster does not coincide exactly with that of a test particle. However, at the level of accuracy needed in this work we can neglect this difference and assume that $\Rb$ coincides with the `guide' radius vector $\Rg$, which evolves according to the equation of motion of a test particle in the cluster potential,
\begin{align}  
\md^2 \Rg/\md t^2 = -(\nabla \Phi)_{\Rg}.
\label{eq:R_g}
\end{align}
In other words, in the following we set $\Rb=\Rg$ and calculate $\Rg(t)$ using equation (\ref{eq:R_g}).

Our neglect of the terms quadratic and higher order in $r_i$ in equation (\ref{eq:expansion}) is equivalent to the so-called `quadrupole approximation' in the hierarchical three-body problem.
Keeping the next (quadratic) term in the expansion would correspond to the `octupole approximation', and so on. In Appendix \ref{Higher_Order} we describe the extension of our tidal Hamiltonian to octupole order and provide a connection to the LK problem in the octupole approximation.

%%%%%%%%%%%%%%%%%%%%%%%%%%%%%%%%%%%%%%%%%%

\subsection{Orbital elements and Delaunay variables}	\label{orbels}

%%%%%%%%%%%%%%%%%%%%%%%%%%%%%%%%%%%%%%%%%%
	
We now introduce standard orbital elements in the frame of the binary. The reference direction is taken to be the $X$ direction and the reference plane the $(X,Y)$ plane (see Figure \ref{picture}; we will later take the $Z$ axis to be the symmetry axis of the potential but the assumption of axisymmetry is not needed at the moment). We define binary argument of pericentre $\omega$, inclination $i$, longitude of ascending node $\Omega$ and mean anomaly $M$ relative to this reference plane and direction.  When written in orbital elements the relative coordinates $\mathbf{r}=(x,y,z)$ become \citep{Murray1999}: \begin{align} 
\nn x =& a\left(\cos \Omega \left[(\cos E - e)\cos\omega - \sqrt{1-e^2}\sin E\sin\omega\right]  \right. \\ &\left. - \cos i \sin \Omega \left[ (\cos E -e)\sin\omega + \sqrt{1-e^2}\sin E \cos \omega \right] \right), 
\label{xoe} 
\end{align}
%%%		
\begin{align} 
\nn y =& a\left(\sin \Omega \left[(\cos E - e)\cos\omega - \sqrt{1-e^2}\sin E\sin\omega\right]  \right. \\ &\left. + \cos i \cos \Omega \left[ (\cos E -e)\sin\omega + \sqrt{1-e^2}\sin E \cos \omega \right] \right),  
\label{yoe}  
\end{align}
%%%
\begin{align} z =& a \sin i \left[ (\cos E -e)\sin\omega + \sqrt{1-e^2}\sin E \cos \omega \right],  
\label{zoe}  
\end{align} 
and $M=E-e\sin E$ where $E$ is the eccentric anomaly. It is important that the orbital elements are defined with respect to a reference frame with axis directions fixed in time (cf. \citealt{Brasser2001, Veras2013a, Correa-Otto2017}). In the limit of the cluster tide going to zero these orbital elements stay constant.  

For dynamical studies it is often more convenient to use Delaunay variables, in which the actions 
\begin{align}
L=\sqrt{\mu a}; \,\,\,\,\,\,\,\,\,\,\,\,
J=L\sqrt{1-e^2}; \,\,\,\,\,\,\,\,\,\,\,
J_z&=J\cos i, 
\end{align}  
are complemented by their conjugate angles $M$, $\omega$, $\Omega$. We will use them extensively in Paper II. 	

Since Delaunay variables are angle-action variables, we can easily identify the conserved quantities in the Hamiltonian. The dominant part of the Hamiltonian \eqref{H0} reads
\begin{align} 
H_0 = -\frac{\mu}{2a} = -\frac{\mu^2}{2L^2}, 
\end{align}
while the perturbing Hamiltonian is given by equation \eqref{H1comp} with $x,y$ and $z$ given by equations \eqref{xoe}, \eqref{yoe} and \eqref{zoe}   respectively (or their Delaunay equivalents).

%%%%%%%%%%%%%%%%%%%%%%%%%%%%%%%%%%%%%%%%%%
%%%%%%%%%%%%%%%%%%%%%%%%%%%%%%%%%%%%%%%%%%
	
\section{Averaging the tidal Hamiltonian} 
\label{sect:avgH}
	
%%%%%%%%%%%%%%%%%%%%%%%%%%%%%%%%%%%%%%%%%%

Dynamics of binaries in stellar clusters benefits from a natural separation of scales. For example, a Solar mass binary with $a=20\,\mathrm{AU}$ has an inner orbital period of $\sim 100$ years, while its outer orbit around a globular cluster might have a period of $\sim 10^5$ years. As we show in Paper II, the resulting secular evolution of the binary's orbital elements due to the tidal potential of the cluster may take $\sim 10^8$ years. 

This naturally allows us to simplify our Hamiltonian (\ref{H1comp}), first by integrating out the fast evolution of the mean anomaly $M$ of the inner orbit (`single-averaging', see \S \ref{avgmeananom}),  and then by also integrating over many (outer) orbits of the binary around the cluster (`double averaging', see \S \ref{davg}).  

%%%%%%%%%%%%%%%%%%%%%%%%%%%%%%%%%%%%%%%%%%
	
\subsection{Singly-averaged Hamiltonian: averaging over the mean anomaly $M$} 
\label{avgmeananom}
	
%%%%%%%%%%%%%%%%%%%%%%%%%%%%%%%%%%%%%%%%%%
	
We begin by averaging over the shortest timescale in the problem, namely over the inner orbital motion of the binary components around their common barycentre. Our singly-averaged Hamiltonian is 
\begin{align} 
\langle H \rangle_M = H_0 + \langle H_1 \rangle_M, 
\end{align} 
where the average of a quantity $\mathcal{F}$ over the mean anomaly $M$ is defined as 
\begin{align} 
\langle \mathcal{F} \rangle_M &\equiv \frac{1}{2\pi}\int_0^{2\pi} \mathcal{F} \,\md M  = \frac{1}{2\pi}\int_0^{2\pi}  (1-e\cos E) \, \mathcal{F} \,\md E.
\end{align} 
The coefficients $\Phi_{\alpha \beta}$ depend on time only through $\Rg(t)$, which is a `slow' variable, so 
\begin{align} 
\label{H1M}
\langle H_1 \rangle_M = \frac{1}{2}\sum_{\alpha \beta}\Phi_{\alpha \beta} \langle r_\alpha r_\beta \rangle_M .
\end{align}
For reference, the full algebraic expressions for $\langle r_\alpha r_\beta \rangle_M$ in terms of orbital elements are given in Appendix \ref{algebraic_expressions}.  The singly-averaged Hamiltonian \eqref{H1M} incorporating these expressions is completely general and can be used to describe orbital evolution of binaries moving in an arbitrary external potential.

Obviously we have eliminated the angle $M$, therefore the conjugate action $L=\sqrt{\mu a}$ is conserved, and so the binary's semi-major axis $a$ is constant. When written in Delaunay variables, the singly-averaged Hamiltonian $\langle H_1\rangle_M$ is a function of $J, J_z, \omega, \Omega$ and the time $t$ through the time-dependent coefficients $\Phi_{\alpha \beta}(\Rg(t))$.

%%%%%%%%%%%%%%%%%%%%%%%%%%%%%%%%%%%%%%%%%%
\subsubsection{Example: orbits in a harmonic potential} \label{harmonicpot}
 
For illustration, as well as to connect to subsequent results, we consider a binary orbiting in a globular cluster with a triaxial constant-density core.  For orbits in this core the potential is that of a three-dimensional harmonic oscillator with frequencies $\kappa_\alpha$, namely $\Phi=\sum_{\alpha}\frac{1}{2}\kappa_\alpha^2 R_\alpha^2$. Then $\Phi_{\alpha\beta} = \kappa_\alpha^2 \delta_{\alpha \beta}$ so the singly-averaged Hamiltonian \eqref{H1M} becomes 
\begin{align} 
\langle H_1 \rangle_M = \frac{1}{2}\left[\kappa_x^2 \langle x^2 \rangle_M + \kappa_y^2 \langle y^2 \rangle_M + \kappa_z^2 \langle z^2 \rangle_M \right]. 
\end{align} 

Let us now consider an axisymmetric core where the symmetry axis is the $Z$ axis. Then $\kappa_x = \kappa_y$ and the binary's outer orbit fills a section of a cylindrical surface with elliptical cross-section (aligned with the $Z$ axis). Using equations (\ref{eq:x2})-(\ref{eq:z2}) we end up with 
\begin{align} 
\nn \langle H_1 \rangle_{M} &=\frac{\kappa_+^2 a^2}{8}\\ & \times\left[ (2+3e^2)\left(1+\frac{\kappa^2_-}{\kappa^2_+}\cos^2i\right) +5\frac{\kappa^2_-}{\kappa^2_+} e^2\sin^2 i \cos 2\omega \right], \label{axiharmonic}
\end{align}	
where $\kappa^2_\pm \equiv \kappa^2_x \pm \kappa^2_z$. 

Note that the dependence on the longitude of the ascending node $\Omega$ has dropped out of this Hamiltonian, so the $z$-component of binary angular momentum $J_z$ is conserved.  That is, in the reference frame of a binary orbiting an axisymmetric harmonic potential, the perturbation due to the tidal field of the cluster effectively becomes axisymmetric after averaging only over the inner orbit of the binary (single-averaging), not its outer orbit around the cluster (double-averaging). This is despite the fact that the outer orbit itself is not axisymmetric in general, even after averaging over a long time interval (its projection onto the $(X,Y)$ plane is an ellipse centred at the origin). This property does not hold for arbitrary potentials. 
		
Things simplify further if we assume the core to be spherically symmetric.  Without loss of generality we can then assume the outer orbit of the binary to be in the $Z=0$ plane, and put all frequencies equal to $\kappa$ (so that $\kappa_-=0$). We find that %\begin{align} \nn \langle H_1 \rangle_{M} = &\frac{L^2\kappa^2}{4\mu^2}( 5L^2-3J^2). \end{align} 
\begin{align} 
\langle H_1 \rangle_{M} = &\frac{\kappa^2a^2}{4}(2+3e^2)=
\frac{\kappa^2L^4}{4\mu^2}\left(5-3\frac{J^2}{L^2}\right).
\end{align} 
This singly-averaged Hamiltonian is now also independent of the argument of pericentre $\omega$.  As a result, in this case there is no evolution of eccentricity or inclination of the binary. The only variation of its orbital elements is apsidal precession at the rate
\begin{align}   
\dot\omega=\frac{\partial \langle H_1 \rangle_{M}}{\partial J}=-\frac{3}{2}\frac{\kappa^2}{n_{\rm K}}\sqrt{1-e^2},
\end{align}  
independent of the orientation of the binary orbit (i.e. $\omega$, $\Omega$, $i$). Here $n_{\rm K}=\sqrt{\mu/a^3}$ is the Keplerian mean motion of the binary.

%%%%%%%%%%%%%%%%%%%%%%%%%%%%%%%%%%%%%%%%%%

\subsection{Doubly-averaged Hamiltonian: averaging over time} 
\label{davg}

%%%%%%%%%%%%%%%%%%%%%%%%%%%%%%%%%%%%%%%%%%
		
As we already mentioned, binary orbital elements change significantly on timescales that are much longer than the outer orbital period of the binary around its host system. For that reason, it makes sense to average $\langle H\rangle_M$ over many outer orbits.  Indicating such time-averages with an over bar, we write: 
\begin{equation} 
\overline{\langle H \rangle}_{M} = -\frac{\mu^2}{2L^2} + \overline{\langle H_1 \rangle}_{M}, 
\end{equation} 
where the doubly-averaged perturbing Hamiltonian $\overline{\langle H_1 \rangle}_{M}$ differs from its singly-averaged predecessor $\langle H_1 \rangle_M$ (equation \eqref{H1M}) only in that each $\Phi_{\alpha \beta}$ is now replaced by its time-averaged value $\overline{\Phi}_{\alpha \beta}$: 
\begin{align} 
\label{H1MtSum} 
\overline{\langle H_1 \rangle}_{M} = \frac{1}{2}\sum_{\alpha\beta}\overline{\Phi}_{\alpha \beta} \langle r_\alpha r_\beta \rangle_M. 
\end{align} 
This works because the outer orbit $\Rg(t)$ only enters $\Phi_{\alpha\beta}$ and not $r_\alpha r_\beta$.

Equation \eqref{H1MtSum} is the doubly-averaged perturbing Hamiltonian and is the main result of this section.  It describes the secular evolution of the orbital elements of any binary perturbed by a smooth external potential $\Phi$. However in its current abstract form it is not of much use.  In the following section we show how the time-averages $\overline{\Phi}_{\alpha\beta}$ may be calculated for cluster potentials possessing certain symmetries, culminating in the expressions \eqref{H1Mt}, \eqref{H1Star}.

%%%%%%%%%%%%%%%%%%%%%%%%%%%%%%%%%%%%%%%%%%
%%%%%%%%%%%%%%%%%%%%%%%%%%%%%%%%%%%%%%%%%%

\section{Time-averaging in axisymmetric potentials}	
\label{sec_Time_Averaging_Procedure}

%%%%%%%%%%%%%%%%%%%%%%%%%%%%%%%%%%%%%%%%%%

So far we did not need to specify anything about the potential $\Phi$.  However, we will now focus on binaries orbiting in fixed axisymmetric potentials (\S\S \ref{sect:cyl}-\ref{Non-Com}). We describe the time-averaging procedure in spherical clusters (\S \ref{tapspherical}) and then extend it to general axisymmetric potentials (\S \ref{tapaxi}).

%%%%%%%%%%%%%%%%%%%%%%%%%%%%%%%%%%%%%%%%%%

\subsection{\texorpdfstring{$\Phi_{\alpha\beta}$}{} in cylindrical coordinates}
\label{sect:cyl}

%%%%%%%%%%%%%%%%%%%%%%%%%%%%%%%%%%%%%%%%%%
		
In an axisymmetric cluster we can choose the symmetry axis of the potential to be the $Z$ axis (like in Figure \ref{picture}).  Then it makes sense to write down the derivatives $\Phi_{\alpha \beta}$ in the cylindrical $(R,\phi,Z)$ coordinate system with origin at the centre of the cluster, where $R=\sqrt{X^2+Y^2}$ and $\phi = \tan^{-1}(Y/X)$.  The axisymmetric potential may then be expressed as $\Phi(R,Z)$, and we find
\begin{align} 
\label{phixxnon} 
\nn \Phi_{xx} &= \frac{1}{2} \left[\left( \frac{\partial^2\Phi}{\partial R^2}\right)_{\Rg}+\left( \frac{1}{R}\frac{\partial\Phi}{\partial R}\right)_{\Rg}\right] \\ &+\frac{1}{2}\cos 2\phi_\mathrm{g} \left[\left( \frac{\partial^2\Phi}{\partial R^2}\right)_{\Rg}-\left( \frac{1}{R}\frac{\partial\Phi}{\partial R}\right)_{\Rg}\right], \\
\label{phiyynon} 
\nn \Phi_{yy} &= \frac{1}{2} \left[\left( \frac{\partial^2\Phi}{\partial R^2}\right)_{\Rg}+\left( \frac{1}{R}\frac{\partial\Phi}{\partial R}\right)_{\Rg}\right] \\ &-\frac{1}{2}\cos 2\phi_\mathrm{g} \left[\left( \frac{\partial^2\Phi}{\partial R^2}\right)_{\Rg}-\left( \frac{1}{R}\frac{\partial\Phi}{\partial R}\right)_{\Rg}\right], \\
\label{phizznon}
\Phi_{zz} &=\left( \frac{\partial^2\Phi}{\partial Z^2}\right)_{\Rg},\\  
\label{phixynon} 
\Phi_{xy} &=  \frac{1}{2}\sin 2\phi_\mathrm{g} \left[ \left( \frac{\partial^2\Phi}{\partial R^2}\right)_{\Rg}-\left( \frac{1}{R}\frac{\partial\Phi}{\partial R}\right)_{\Rg}\right],  \\
\Phi_{xz}&= \cos \phi_\mathrm{g} \left( \frac{\partial^2\Phi}{\partial R\partial Z}\right)_{\Rg},
\label{phixznon} \\ 
\Phi_{yz}&=\sin \phi_\mathrm{g} \left( \frac{\partial^2\Phi}{\partial R\partial Z}\right)_{\Rg} 
\label{phiyznon}. 
\end{align}
Here $\phi_\mathrm{g}$ is the azimuthal coordinate of $\Rg$, namely $\tan^{-1}({Y_\mathrm{g}}/{X_\mathrm{g}})$, and again the subscripts on derivatives mean `evaluated at position $\Rg(t)$'.  The coefficients $\Phi_{\alpha \beta}$ have certain symmetry properties which will become important when we consider their time-averaged values in \S\ref{Non-Com}.  
	
%%%%%%%%%%%%%%%%%%%%%%%%%%%%%%%%%%%%%%%%%%

\subsection{Orbit families and non-commensurable frequencies} 
\label{Non-Com}

%%%%%%%%%%%%%%%%%%%%%%%%%%%%%%%%%%%%%%%%%%
	
We now consider which orbit families are possible in general axisymmetric potentials.  Numerical orbit integration confirms that most orbits in axisymmetric potentials are regular and appear to respect a third integral of motion $I_3$ alongside energy $\mathcal{E}$ and the $Z$-component of angular momentum $\mathcal{L}_Z$  \citep{Binney2008,Merritt2013}. The vast majority of these regular orbits form a tube, or `torus', around the symmetry axis:  in an oblate potential they are short-axis tube orbits, while in a prolate potential they are (inner- or outer-) long axis tube orbits. Other possibilities include near-resonant regular orbits and irregular (chaotic) orbits, but both of these are typically scarce compared to the tubes.
	
We will ignore chaotic orbits in what follows since they are very rare in axisymmetric potentials \citep{Regev2006}.   We are left with tube orbits and (near-)resonant non-tube orbits.  The resonant family corresponds to $\Rg(t)$ having commensurable frequencies.  Mathematically, if we denote the frequencies of motion of $\Rg(t)$ in each direction by the vector $\mathbf{\Omega} = (\Omega_R, \Omega_\phi, \Omega_Z)$, we must consider whether there exists any triple of integers $\mathbf{n}=(n_1,n_2,n_3)$ such that 
\begin{align} 
\label{commens} 
\mathbf{n} \cdot \mathbf{\Omega} = 0. 
\end{align} 
The role of commensurabilities and near-commensurabilities will be discussed in \S\S \ref{numveravg}-\ref{sect:disc}.

If the frequencies are non-commensurable (i.e. relation \eqref{commens} does not hold), then the outer orbit of the binary will trace out a non-repeating path around the cluster.  Over time this path will densely fill a 3D axisymmetric torus whose symmetry axis is $Z$.  We may therefore replace the time-average of a function following an orbit $\Rg(t)$ with a weighted (by the time the orbit spends at a given point) volume-average over the torus. 

Since the torus is axisymmetric, time-averaging over non-commensurable orbits inevitably involves integrating the expressions (\ref{phixxnon})-(\ref{phiyznon}) over azimuthal angle $\phi_\mathrm{g}$. As a result, time-averages of $\Phi_{\alpha\beta}$ become 
\begin{align} 
\overline{\Phi}_{xx} &= \overline{\Phi}_{yy} = \frac{1}{2} \overline{\left[\left( \frac{\partial^2\Phi}{\partial R^2}\right)_{\Rg}+\left( \frac{1}{R}\frac{\partial\Phi}{\partial R}\right)_{\Rg}\right]}, 
\label{phixxyy} \\
%%%%%%
\overline{\Phi}_{zz} &=\overline{\left( \frac{\partial^2\Phi}{\partial Z^2}\right)}_{\Rg},
\label{phizz}\\  
%%%%%%
\overline{\Phi}_{xy} &=  0, 
\label{phixy}  \\
%%%%%%
\overline{\Phi}_{xz} &= \overline{\Phi}_{yz}=0,
\label{phiyzxz}
\end{align}  
see equations (\ref{phixxnon})-(\ref{phiyznon}). In practice, vanishing of $\overline{\Phi}_{xy}$, $\overline{\Phi}_{xz}$, $\overline{\Phi}_{yz}$ typically requires averaging over many outer orbits --- see \S \ref{numveravg}.

%%%%%%%%%%%%%%%%%%%%%%%%%%%%%%%%%%%%%%%%%%
	
\subsection{Time-averages in spherical potentials}
\label{tapspherical}

%%%%%%%%%%%%%%%%%%%%%%%%%%%%%%%%%%%%%%%%%%

In spherical potentials the outer orbit of the binary remains in the same plane, which can be chosen to coincide with the $(X,Y)$ plane. In this plane the coefficients $\Phi_{xz}$ and $\Phi_{yz}$ vanish identically. In other words, equation (\ref{phiyzxz}) holds true even without averaging over the outer orbit. At the same time, $\overline{\Phi}_{xy}$ asymptotically tends to zero only upon averaging over many orbits, as we will see later in \S \ref{sect:sph_validity}.

In the $(X,Y)$ plane the path of $\Rg$ is a rosette, assuming it has non-commensurable radial and azimuthal frequencies; see Figure \ref{convplots2} for illustration. Over time the rosette densely fills an axisymmetric annulus with inner and outer radii corresponding to the pericentre $r_\mathrm{p}$ and apocentre $r_\mathrm{a}$ of the outer orbit $\Rg(t)$.  When discussing spherical potentials we will sometimes refer to this as the `axisymmetric annulus approximation'. 
	
In this case it is easy to write down an analytical formula for the averages $\overline{\Phi}_{\alpha \beta}$ in terms of $r_\mathrm{p}$ and $r_\mathrm{a}$, as averaging over $\md t$ can be replaced with averaging over $\md R$ via $\md t=v_R^{-1}\md R$, where $v_R=[2(\mathcal{E}-\Phi(R))-\mathcal{L}^2/R^2]^{1/2}$ is the radial velocity. Specific energy $\mathcal{E}$ and angular momentum $\mathcal{L}$ of the outer orbit $\Rg(t)$ in a spherical potential $\Phi$ can be explicitly expressed as function of $\rp$ and $\ra$ as follows:
\begin{align} 
\mathcal{E}(r_\mathrm{p},r_\mathrm{a}) &= \frac{\ra^2\Phi(\ra)-\rp^2\Phi(\rp)}{\ra^2-\rp^2},\\
\mathcal{L}(r_\mathrm{p},r_\mathrm{a}) &= \sqrt{\frac{2[\Phi(\ra)-\Phi(\rp)]}{\rp^{-2}-\ra^{-2}}}. 
\end{align}
With this in mind, we can write the time-average of any radially-dependent function $\mathcal{F}(R)$ as
\begin{align}
\overline{\mathcal{F}}= \frac{\int_{r_\mathrm{p}}^{r_\mathrm{a}} \md R \, \mathcal{F}(R) \left[2(\mathcal{E}-\Phi(R))-\mathcal{L}^2/R^2\right]^{-1/2}}{\int_{r_\mathrm{p}}^{r_\mathrm{a}} \md R  \left[2(\mathcal{E}-\Phi(R))-\mathcal{L}^2/R^2\right]^{-1/2}}. 
\label{2Davgeqn}
\end{align} 
Remembering that only the azimuthally-averaged versions of $\Phi_{\alpha \beta}$ provide non-zero contributions (see \S \ref{Non-Com} and equations (\ref{phixxyy})-(\ref{phiyzxz})), we see that in spherical potentials the time-averages $\overline{\Phi}_{\alpha \beta}$ can be calculated in a straightforward fashion via integration over radius $R$.

%%%%%%%%%%%%%%%%%%%%%%%%%%%%%%%%%%%%%%%%%%

\subsection{Time-averages in axisymmetric potentials}
\label{tapaxi}

%%%%%%%%%%%%%%%%%%%%%%%%%%%%%%%%%%%%%%%%%%

We would like to generalise the approach of \S\ref{tapspherical} to axisymmetric potentials $\Phi(R,Z)$.  This would involve averaging  $\Phi_{\alpha \beta}$ over the $(R,Z)$ cross-section of an axisymmetric torus filled by the outer orbit of the binary --- see the central columns of Figures \ref{OPlotsF} \& \ref{OPlotsMN} for examples of such cross-sections. However, there are several difficulties with this approach.  

First, each $\mathrm{d}R\,\mathrm{d}Z$ element of the cross-section enters the averaging procedure with a certain weight proportional to the time the orbit spends in it. Calculating this weight is not trivial and involves the knowledge of the meridional velocity ($v_R$, $v_Z$) structure. For a general axisymmetric potential this calculation cannot be done analytically. 

Second, even the shape of the outer boundary of the cross-section cannot be determined analytically for a general axisymmetric potential. The difficulty is that the knowledge of $\mathcal{E}$ and $Z$-component of angular momentum $\mathcal{L}_Z$ (which are integrals of motion in a general axisymmetric potential) is not sufficient to determine the shape of the meridional cross section of the torus: one also needs to know a third integral of motion $I_3$.  The exact shape of the torus cross-section is known only for orbits in Staeckel potentials \citep{Binney2008}, since only for those do we have analytic expression for the third integral $I_3$.  Even then, writing down a formula for the time-averaged coefficients $\overline{\Phi}_{\alpha \beta}$ is tiresome because of the awkward confocal-ellipsoidal coordinate system involved \citep{Binney2008} and the complicated functional form of the third integral.  

For these reasons, in this work we usually\footnote{There are special cases in certain axisymmetric potentials where we can compute $A,\Gamma$ (semi-)analytically, see \S\ref{sect:gen} \& \S\ref{PlaneAxi}.} compute $A$ and $\Gamma$ in axisymmetric potentials by directly integrating the orbit of a guide $\Rg(t)$ numerically using equation (\ref{eq:R_g}) for a given set of initial conditions $(R,v_R,\phi,v_\phi,Z,v_Z)$, where $v_R$ is the velocity in the direction of increasing $R$, etc. This orbit is then used to carry out the time-average of $\Phi_{\alpha \beta}$ using a method outlined in Appendix \ref{Numerical_Time_Averages}.

Note that, unlike in the spherically symmetric case, $\overline{\Phi}_{xz}$ and $\overline{\Phi}_{yz}$ no longer vanish identically due to a symmetry of the potential. Nevertheless, equations (\ref{phixy})-(\ref{phiyzxz}) are still fulfilled upon averaging over many outer orbits.

%%%%%%%%%%%%%%%%%%%%%%%%%%%%%%%%%%%%%%%%%%
%%%%%%%%%%%%%%%%%%%%%%%%%%%%%%%%%%%%%%%%%%

\section{The secular Hamiltonian} 
\label{secham}

%%%%%%%%%%%%%%%%%%%%%%%%%%%%%%%%%%%%%%%%%%

Despite the fact that in general axisymmetric potentials we cannot write down a useful analytic expression for time-averages, we can still continue our derivation of the secular Hamiltonian owing to the symmetries of the $\overline{\Phi}_{\alpha\beta}$ coefficients (equations \eqref{phixxyy}-\eqref{phiyzxz}). These symmetry properties allow us to greatly simplify the doubly-averaged perturbing Hamiltonian \eqref{H1MtSum} so that it reads: 
\begin{align} 
\label{H1MtSimp} 
\overline{\langle H_1 \rangle}_{M}=  \frac{1}{2}\overline{\Phi}_{xx} \langle x^2 + y^2 \rangle_M + \frac{1}{2}\overline{\Phi}_{zz} \langle z^2 \rangle_M. 
\end{align}   
Let us define the quanitities 
\begin{align} 
A \equiv \overline{\Phi}_{zz} + \overline{\Phi}_{xx}, \,\,\,\,\,\,\, B \equiv \overline{\Phi}_{zz} - \overline{\Phi}_{xx}, \,\,\,\,\,\,\, \Gamma \equiv B/3A, 
\label{ABGamDef} 
\end{align} 
(note that the constants $A$ and $B$ are \textit{not} related to the Oort constants!)
and write $x,y,z$ in terms of orbital elements using equations (\ref{eq:x2})-(\ref{eq:z2}). Then the Hamiltonian \eqref{H1MtSimp} becomes 
\begin{align} 
\label{H1Mt} 
\overline{\langle H_1 \rangle}_{M} = CH_1^* \,\,\,\,\,\,\, \mathrm{where} \,\,\,\,\,\,\, C= Aa^2/8,
\end{align}
and $H_1^*$ is the `dimensionless Hamiltonian'
\begin{align} 
H_1^* = (2+3e^2)(1-3\Gamma \cos^2 i)-15\Gamma e^2\sin^2 i \cos 2\omega. 
\label{H1Star} 
\end{align}	
Note that $H_1^*$ involves only a single dimensionless parameter $\Gamma$, while $C$ depends on $A$ (which has units of (frequency)$^2$).  In Paper II we will see that $\Gamma$ determines the phase space structure of the Hamiltonian while $A$ sets the timescale for secular evolution. All the information about the cluster properties and the characteristics of the (outer) orbit of the binary enter the Hamiltonian through these two parameters only. In \S\ref{abgamma} we investigate in detail how these key quantities depend on the form of the background potential and the outer orbit of the binary within the potential.

The dependence of the Hamiltonian upon the longitude of ascending node $\Omega$ has dropped out under time-averaging and so the $z$-component of angular momentum $J_z$ is conserved, as we would expect for an axisymmetric perturbation (which the cluster potential looks like from the binary frame upon averaging over many outer orbits).  The dimensionless quantity $J_z/L = \sqrt{1-e^2}\cos i$ is an integral of motion, which implies that variations of eccentricity must come at the expense of changes in inclination and vice versa, just as in the LK mechanism \citep{Lidov1962,Kozai1962,Naoz2013}.  

Finally, we note that the doubly-averaged perturbing Hamiltonian \eqref{H1Mt} appears very similar to the singly-averaged one derived for the example of an axisymmetric harmonic potential in \S \ref{harmonicpot} (equation \eqref{axiharmonic}).  Indeed, comparing equations \eqref{axiharmonic} and \eqref{H1Mt} one might be tempted to say that axisymmetric harmonic potentials have $\Gamma=-\kappa_-^2/3\kappa_+^2$.  However, this correspondence is a mathematical coincidence:  the assumption of non-commensurability (\S\ref{Non-Com}) does not apply to harmonic potentials, for which all orbits are closed non-precessing ellipses.  Despite their similarities the Hamiltonians \eqref{axiharmonic} and \eqref{H1Mt} are different objects derived under different approximations.

%%%%%%%%%%%%%%%%%%%%%%%%%%%%%%%%%%%%%%%%%%
\subsection{Orbits in a Kepler potential: link to the Lidov-Kozai mechanism} 
\label{linkLK}
	
Another example of such a mathematical coincidence is represented by the well known test particle quadrupole Lidov-Kozai	(LK) problem \citep{Lidov1962,Kozai1962}. The Hamiltonian for this problem takes the form \eqref{H1Star} if we were to set $\Gamma=1$. However, we have derived this Hamiltonian under the assumption that $\Rg$ fills an axisymmetric torus, while in the LK case $\Rg$ moves in a Keplerian ellipse, which in the test particle limit does not precess. Nevertheless, it is known that for elliptical orbits the time-averaged tidal Keplerian potential is exactly axisymmetric to quadrupole order (e.g. \citealt{Katz2011,Naoz2011}), and so \eqref{H1Mt} \textit{does} in fact hold.  

In Appendix \ref{RecoverLK} we show explicitly how the LK Hamiltonian is recovered in the `test particle quadrupole' approximation from the singly-averaged equation \eqref{H1M} in the limit that the background potential $\Phi$ in which the binary orbits arises from a point mass at the origin. We recover the LK Hamiltonian exactly if we set $\Gamma=1$ in \eqref{H1Star}.

%%%%%%%%%%%%%%%%%%%%%%%%%%%%%%%%%%%%%%%%%%
\subsection{Epicyclic orbits in a disk: link to \citet{Heisler1986}}
\label{sect:epi}
	
For wide binaries in the solar neighbourhood, the tidal potential of the Galactic disk can provide the dominant torque, as shown by \citet{Heisler1986} for the Oort Cloud comets. Averaged over many orbits of the Sun around the Galaxy, the Galactic disk provides an axisymmetric tide onto the binary. In Appendix \ref{RecoverHT} we show how to calculate $\overline{\Phi}_{\alpha\beta}$ in the case where $\Rg$ performs epicyclic motion in a disk. It is then easy to recover the tidal Hamiltonian of HT86 from \eqref{H1Mt}. We reproduce the dimensionless version of HT86's Hamiltonian by setting $\Gamma=1/3$ in \eqref{H1Star}.

%%%%%%%%%%%%%%%%%%%%%%%%%%%%%%%%%%%%%%%%%%
%%%%%%%%%%%%%%%%%%%%%%%%%%%%%%%%%%%%%%%%%%

\section{Dependence of Hamiltonian coefficients \texorpdfstring{$A$}{} and \texorpdfstring{$\Gamma$}{} on the cluster potential and binary orbit}
\label{abgamma}

%%%%%%%%%%%%%%%%%%%%%%%%%%%%%%%%%%%%%%%%%%

All of the information about the effect of the tidal potential on secular dynamics of the binary is contained in the two crucial quantities $A$ and $\Gamma$, which are constructed from the time-averages $\overline{\Phi}_{zz}$ and $\overline{\Phi}_{xx}$, see equation \eqref{ABGamDef}. 

The quantity $A$ is a direct measure of the strength of the potential. Its influence on the dynamics is trivial: it enters the problem only as a proportionality constant of the Hamiltonian (equation \eqref{H1Mt}), and therefore merely sets the (inverse of the) secular timescale. In addition, $A$ is also a fairly intuitive quantity: if a binary is in a strong tidal potential we expect it will have a large $A$.  

The meaning of $\Gamma$ is less intuitive than $A$ although its influence upon the system is quite profound. In Paper II we will see that the phase portrait of the Hamiltonian $H_1^*$ undergoes several bifurcations as we change the value of $\Gamma$, altering the dynamics completely.  In particular, we show that there are four qualitatively different regimes --- (i) $\Gamma>1/5$, (ii) $0 < \Gamma \leq 1/5$, (iii) $-1/5 < \Gamma \leq 0$, and (iv) $\Gamma \leq -1/5$.  The value of $\Gamma$ is so important because, for instance, high-eccentricity excitation is ubiquitous for binaries in regime (i), whereas it is much less readily achieved by binaries in regime (ii).  Hence, most effort in this section will be directed towards understanding which cluster potentials and outer binary orbits give rise to which values of $\Gamma$.  So far we have seen that $\Gamma = 1$ for any orbit in a Keplerian potential, and that $\Gamma = 1/3$ for epicyclic orbits in a thin disk. In this section we explore in more detail how the values of $\Gamma$ (and $A$) depend on the form of the background potential $\Phi$ and the binary's outer orbit $\Rg$ within it. 

We start by stating some general properties of $A$ and $\Gamma$ in \S \ref{sect:gen}. We then discuss the behavior of these parameters in certain spherical (\S \ref{AGamSph}) as well as axisymmetric (\S \ref{AGammaAxi}) potentials.

%%%%%%%%%%%%%%%%%%%%%%%%%%%%%%%%%%%%%%%%%%

\subsection{General properties of \texorpdfstring{$A$}{} and \texorpdfstring{$\Gamma$}{}}	
\label{sect:gen}

%%%%%%%%%%%%%%%%%%%%%%%%%%%%%%%%%%%%%%%%%%

%%%%%%%%%%%%%%%%%%%%%%%%%%%%%%%%%%%%%%%%%%
\begin{figure}
\centering
\includegraphics[width=\linewidth]{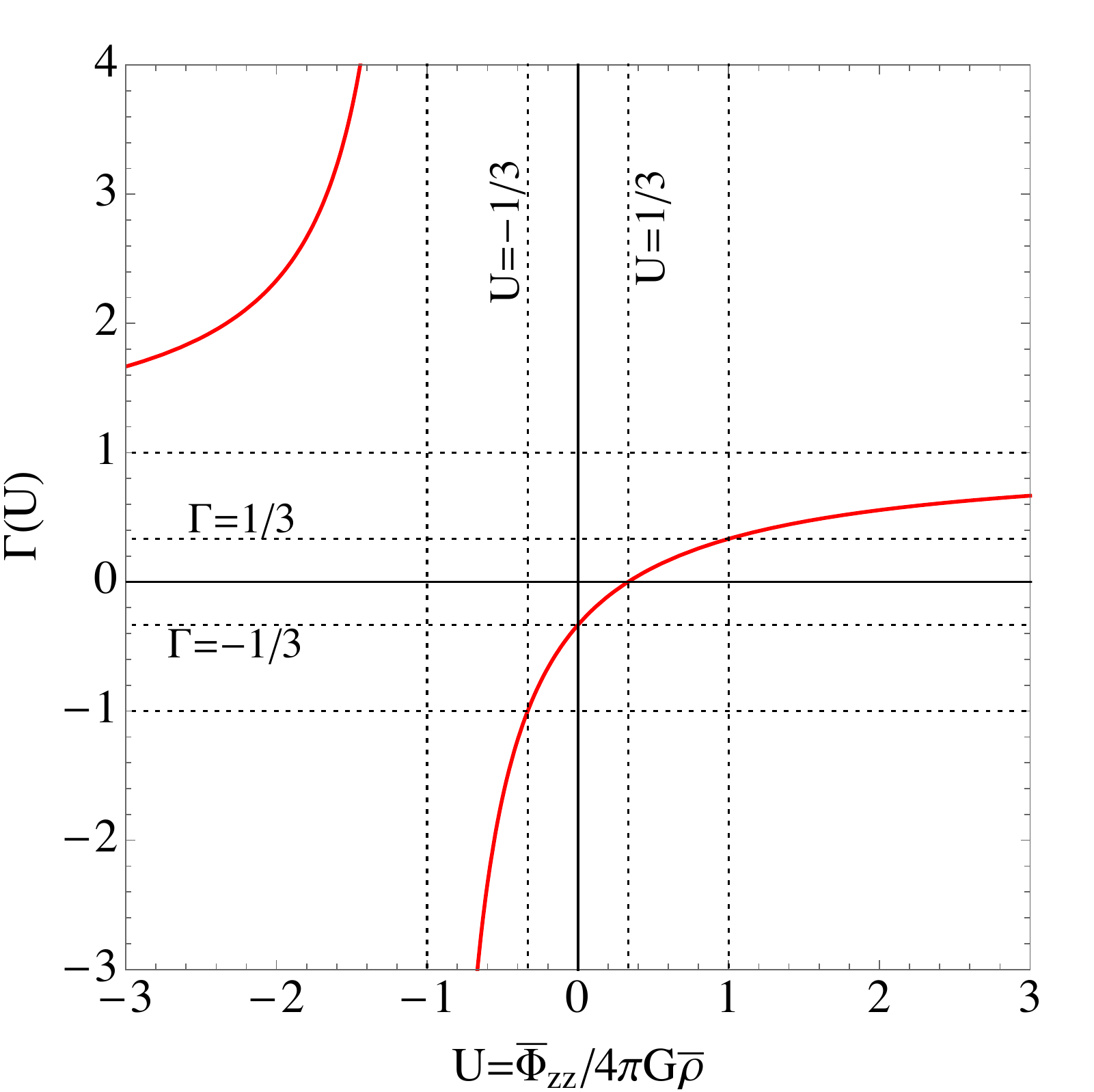}
\caption{Plot of the function $\Gamma(U)$ defined by equation \eqref{GammaU} with $U=\overline{\Phi}_{zz}/(4\pi G \overline{\rho})$ given by equation \eqref{Udef}. See text for details.}
\label{GammaOfU}
\end{figure}
%%%%%%%%%%%%%%%%%%%%%%%%%%%%%%%%%%%%%%%%%%

Writing down Poisson's equation in cylindrical coordinates 
\begin{align}  
\frac{\partial^2\Phi}{\partial R^2}+\frac{1}{R}\frac{\partial\Phi}{\partial R}+\frac{\partial ^2\Phi}{\partial Z^2}=4\pi G \rho,
\label{eq:Poisson}
\end{align} 
and using equations (\ref{phixxyy})-(\ref{phizz}) \& (\ref{ABGamDef}),
one can easily show that in a general axisymmetric potential 
\begin{align} 
A = \frac{1}{2}(\overline{\Phi}_{zz} + 4\pi G\overline{\rho}), \,\,\,\,\,  B = \frac{1}{2}(3\overline{\Phi}_{zz} - 4\pi G\overline{\rho}),
\end{align}  
where $\overline{\rho}$ is the cluster density in the vicinity of the binary time-averaged over many outer orbital periods. Then, defining the dimensionless ratio 
\begin{align} 
U \equiv \frac{\overline{\Phi}_{zz}}{4\pi G \overline{\rho}}
=-1+\frac{A}{2\pi G \overline{\rho}}, 
\label{Udef} 
\end{align} 
we can write $\Gamma$ quite generally as 
\begin{align} 
\Gamma(U) = \frac{3U-1}{3(U+1)}. 
\label{GammaU} 
\end{align} 
The function $\Gamma(U)$ is plotted in Figure \ref{GammaOfU}. 

In principle there are no limits on the values $U$ can take, although in practice, achieving values of $U$ less than $-1$ (and hence $\Gamma > 1$) may require rather contrived orbits.  An example of such an orbit with $U < -1$ and $\Gamma>1$ is given in Appendix \ref{sect:AGamma} (see Figure \ref{fig:App_D_Ex2}). Note that $U<-1$ necessarily implies that $A<0$, see equation (\ref{Udef}).

Somewhat stronger statements can be formulated for realistic spherically symmetric cluster potentials, as we show in Appendix \ref{sect:AGamma}. In particular, one can demonstrate that in such potentials $A>0$, $B\geq 0$, and $0\leq \Gamma \leq 1$. In non-spherical potentials negative values of $\Gamma$ become possible for certain binary orbits as we will show in \S \ref{AGammaAxi}.

It is instructive to consider the values of $\Gamma$ for some specific potentials $\Phi$. 
\begin{itemize}

\item In the case of a Keplerian cluster potential, i.e. a spherical point mass potential with vanishingly small density $\rho$ outside the centre, one has $\Phi_{zz}>0$, $\overline{\rho}\to 0$, $U \to +\infty$ and $\Gamma \to 1$ (see \S \ref{linkLK}). 

\item In a spherical harmonic potential, symmetry dictates that $\Phi_{zz}=(1/3)\nabla^2\Phi=(4\pi/3)G\rho$ so that $U=1/3$ and $\Gamma=0$ (see \S \ref{harmonicpot}). 

\item In a spherical cluster with a cusped density distribution $\rho \propto r^{-\beta}$ with $0<\beta<3$ (having finite mass at the centre) we have $\Gamma=\beta/[3(4-\beta)]$, see Appendix \ref{SphericalABGamma}. 

\item In a thin galactic disk, assuming that $\Phi_{zz}$ dominates over other spatial derivatives in Poisson's equation, one has $\Phi_{zz} \approx 4\pi G\rho$; hence we find $U=1$ and $\Gamma =1/3$ (see \S \ref{sect:epi}). 

\item In the opposite limit of a `cylindrical' (or highly prolate) potential $\Phi=\Phi(R)$ with no $Z$-dependence, one has $\Phi_{zz}=0$, $U=0$ and $\Gamma=-1/3$. 
\end{itemize}
The values (or ranges) of $U$ and $\Gamma(U)$ for these and some other types of cluster potential are summarized in Table \ref{GammaUTable}. We stress again that even though applying the `axisymmetric annulus' approximation gives the correct results for Keplerian and spherical harmonic potentials, this is a mathematical coincidence unless the outer orbit of the binary in these potentials is circular (see \S \ref{harmonicpot} \& \ref{linkLK}). 

%%%%%%%%%%%%%%%%%%%%%%%%%%%%%%%%%%%%%%%%%%
\begin{figure*}
\includegraphics[width=0.98\linewidth, height=0.8\linewidth]{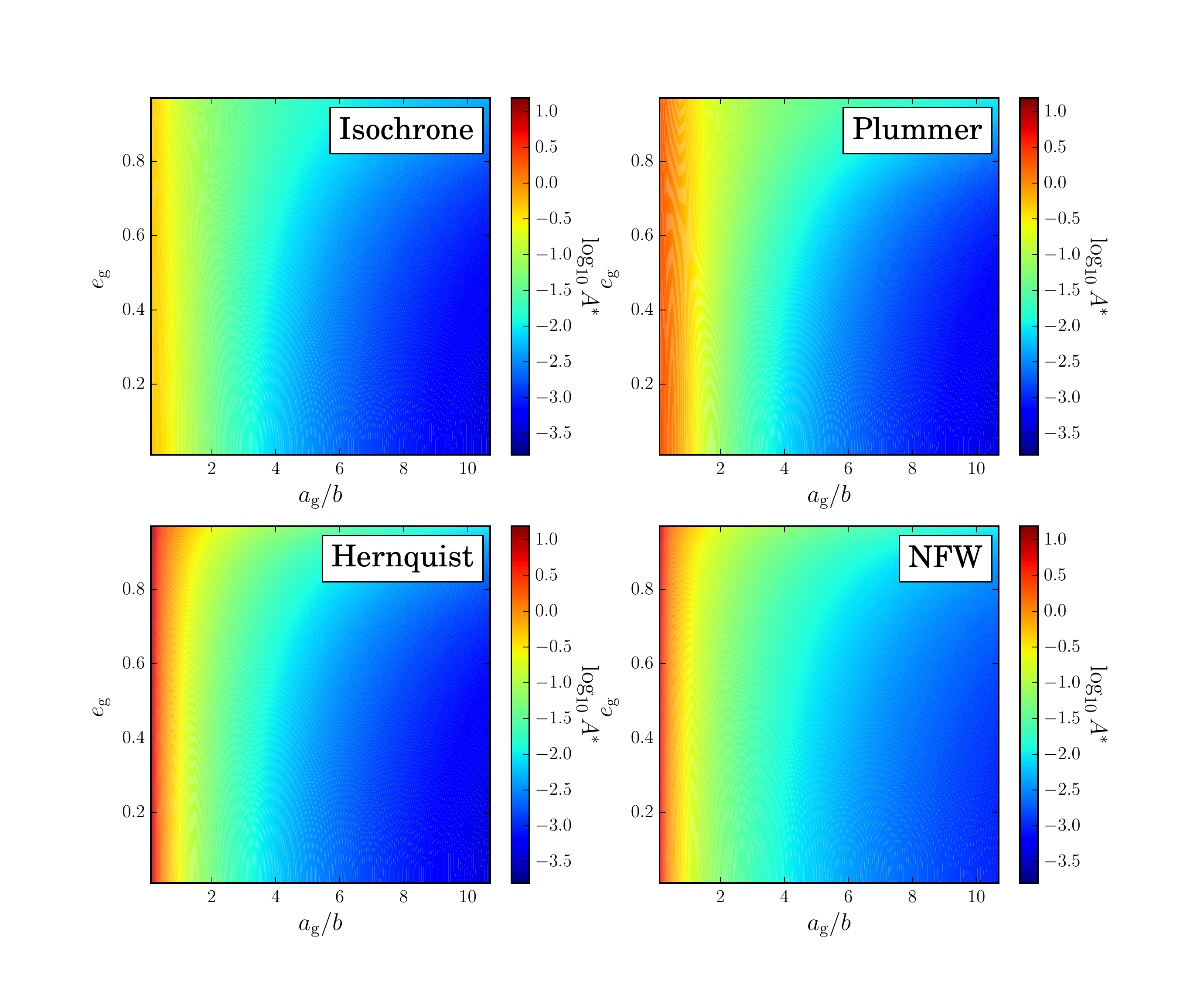}
\caption{Contour plots of $\log_{10}A^*$, where $A^* \equiv A/(GM/b^3)$, as a function of generalized semi-major axis $a_\mathrm{g}$ and eccentricity $e_\mathrm{g}$ for binary orbits in four spherical potentials (equations \eqref{IsoPot}-\eqref{NFWPot}), each with scale radius $b$. The value of $A^*$ in general depends both on $a_\mathrm{g}$ and $e_\mathrm{g}$.}
\label{ABPlots}
\end{figure*} 
%%%%%%%%%%%%%%%%%%%%%%%%%%%%%%%%%%%%%%%%%%

%%%%%%%%%%%%%%%%%%%%%%%%%%%%%%%%%%%%%%%%%%

\subsection{Behavior of Hamiltonian characteristics in some spherical potentials}	
\label{AGamSph}

%%%%%%%%%%%%%%%%%%%%%%%%%%%%%%%%%%%%%%%%%%

In spherical potentials the values of $\overline{\Phi}_{\alpha\beta}$ that enter $A$ and $\Gamma$ are computed using the analytic expression \eqref{2Davgeqn}, which for a fixed potential depends only on the peri/apocentre $(\rp,\ra)$ of the binary's outer orbit $\Rg(t)$.  We can define the outer orbit's generalised semi-major axis $a_\mathrm{g}$ and generalised eccentricity $e_\mathrm{g}$ in terms of the peri/apocentre as
\begin{align} 
a_\mathrm{g} \equiv \frac{1}{2}(\ra+\rp); \,\,\,\, \,\,\,\,\, e_\mathrm{g} \equiv \frac{\ra-\rp}{\ra+\rp}. 
\end{align} 
These reduce to the usual orbital elements in the case of a Keplerian potential. These variables fully characterize the outer orbit of the binary in a given spherical potential. 

In any spherical potential with scale radius $b$ and total mass $M$ we can also construct the dimensionless parameter $A^*\equiv A/(GM/b^3)$; this normalization arises because $A$ is constructed from the second derivatives of the potential, which are of order\footnote{Note that $2\pi/\sqrt{GM/b^3}$ is roughly $\sim T_{\phi}$, the characteristic azimuthal orbital period of the binary around the cluster, so that $A \sim 4\pi^2A^*/T_\phi^2$.} $GM/b^3$, see equation (\ref{ABGamDef}). This allows us to estimate
\begin{align} 
A = 226\,\, \mathrm{Myr}^{-2} \times \left(\frac{A^*}{0.5}\right) \left(\frac{M}{10^5 M_\odot}\right)\left(\frac{b}{\mathrm{pc}}\right)^{-3}. 
\end{align}   
Both $A^*$ and $\Gamma$ are dimensionless numbers which, for a given potential, depend only on $a_\mathrm{g}$ and $e_\mathrm{g}$. In the following we will explore the dependence of $A^*$ (rather than $A$, which also depends on the cluster mass and size) and $\Gamma$ on the shape of the potential and the binary orbit in it. 

%%%%%%%%%%%%%%%%%%%%%%%%%%%%%%%%%%%%%%%%%%
\begin{figure}
\includegraphics[width=0.99\linewidth]{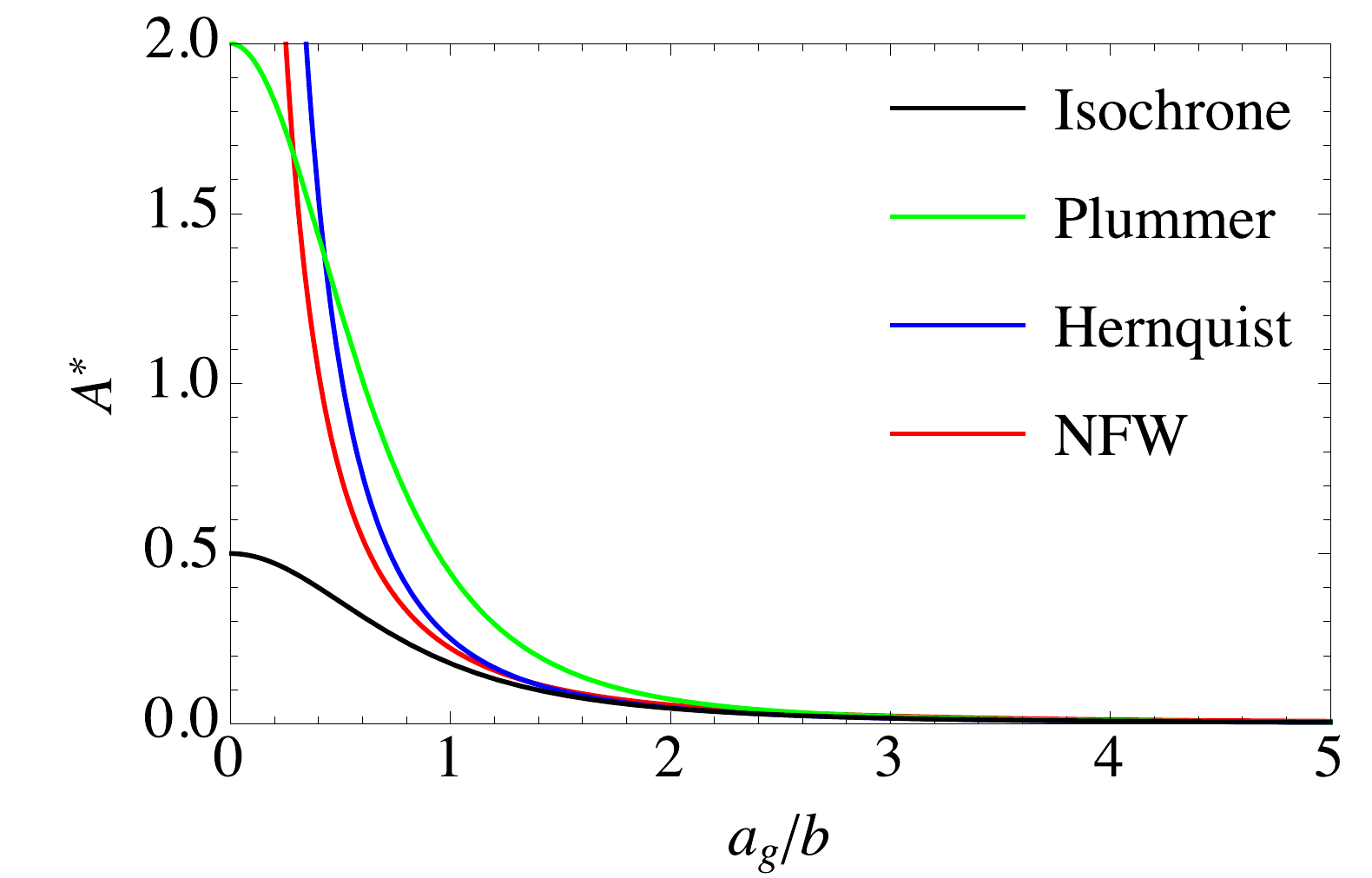}
\caption{Plots of $A^*$ for circular outer orbits ($e_\mathrm{g}=0$) as a function of $a_\mathrm{g}/b$ (where $b$ is a scale radius) in the same four potentials as in Figure \ref{ABPlots}.}
\label{ACirc}
\end{figure} 
%%%%%%%%%%%%%%%%%%%%%%%%%%%%%%%%%%%%%%%%%%

%%%%%%%%%%%%%%%%%%%%%%%%%%%%%%%%%%%%%%%%%%
\begin{figure*}
\includegraphics[width=0.98\linewidth, height=0.8\linewidth]{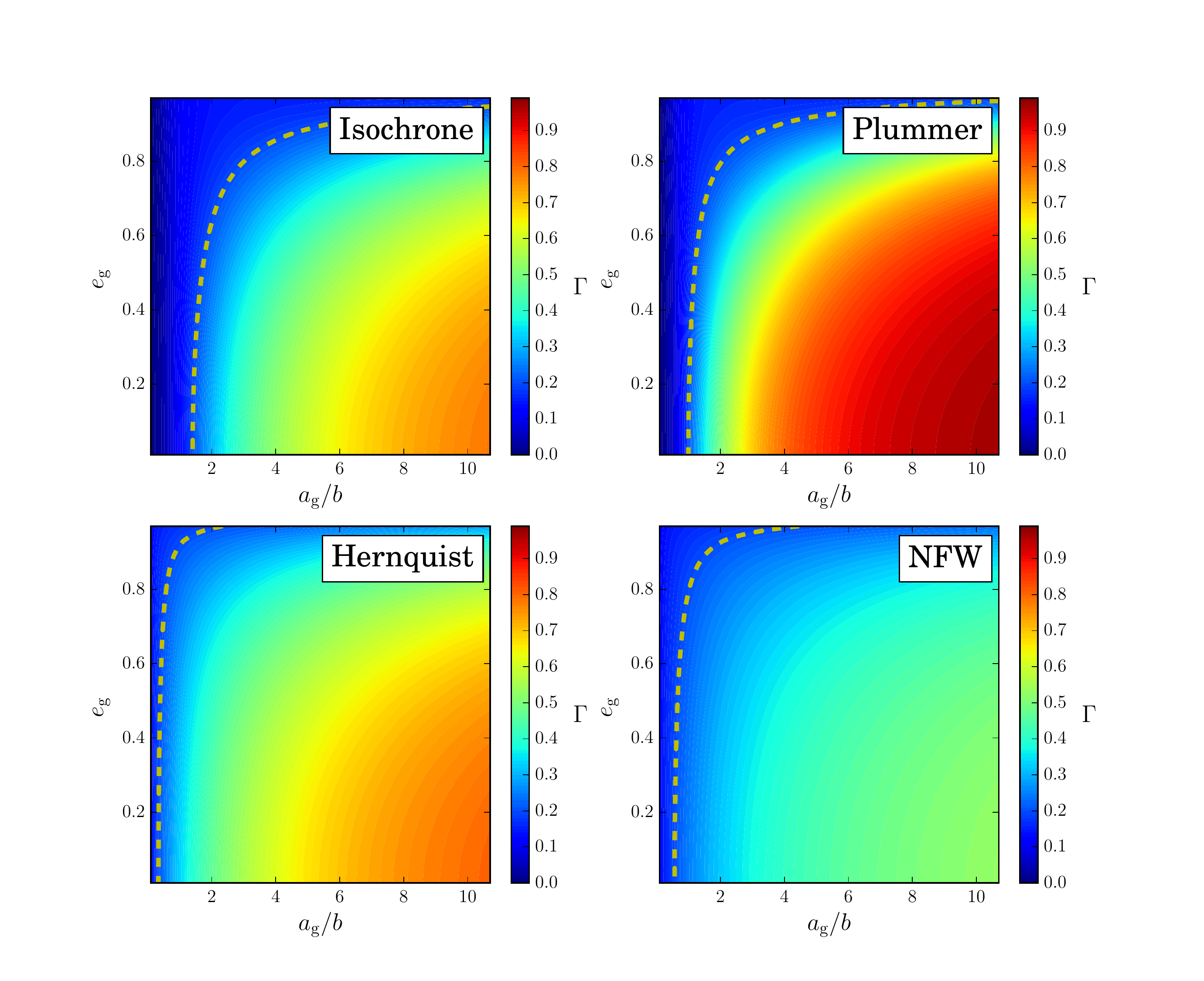}
\caption{Same as Figure \ref{ABPlots} but showing contour plots of the parameter $\Gamma$.  Note that $\Gamma \to 0$ at the centre of the cored potentials (isochrone and Plummer).  The dashed yellow in each plot corresponds to $\Gamma=1/5$, which is the location of an important bifurcation in the dynamical phase portrait, as we show in  Paper II.}
\label{GammaPlots}
\end{figure*} 
%%%%%%%%%%%%%%%%%%%%%%%%%%%%%%%%%%%%%%%%%%

%%%%%%%%%%%%%%%%%%%%%%%%%%%%%%%%%%%%%%%%%%
\begin{figure}
\centering
\includegraphics[trim={0.2cm 0 0.8cm 0},clip,width=0.995\linewidth]{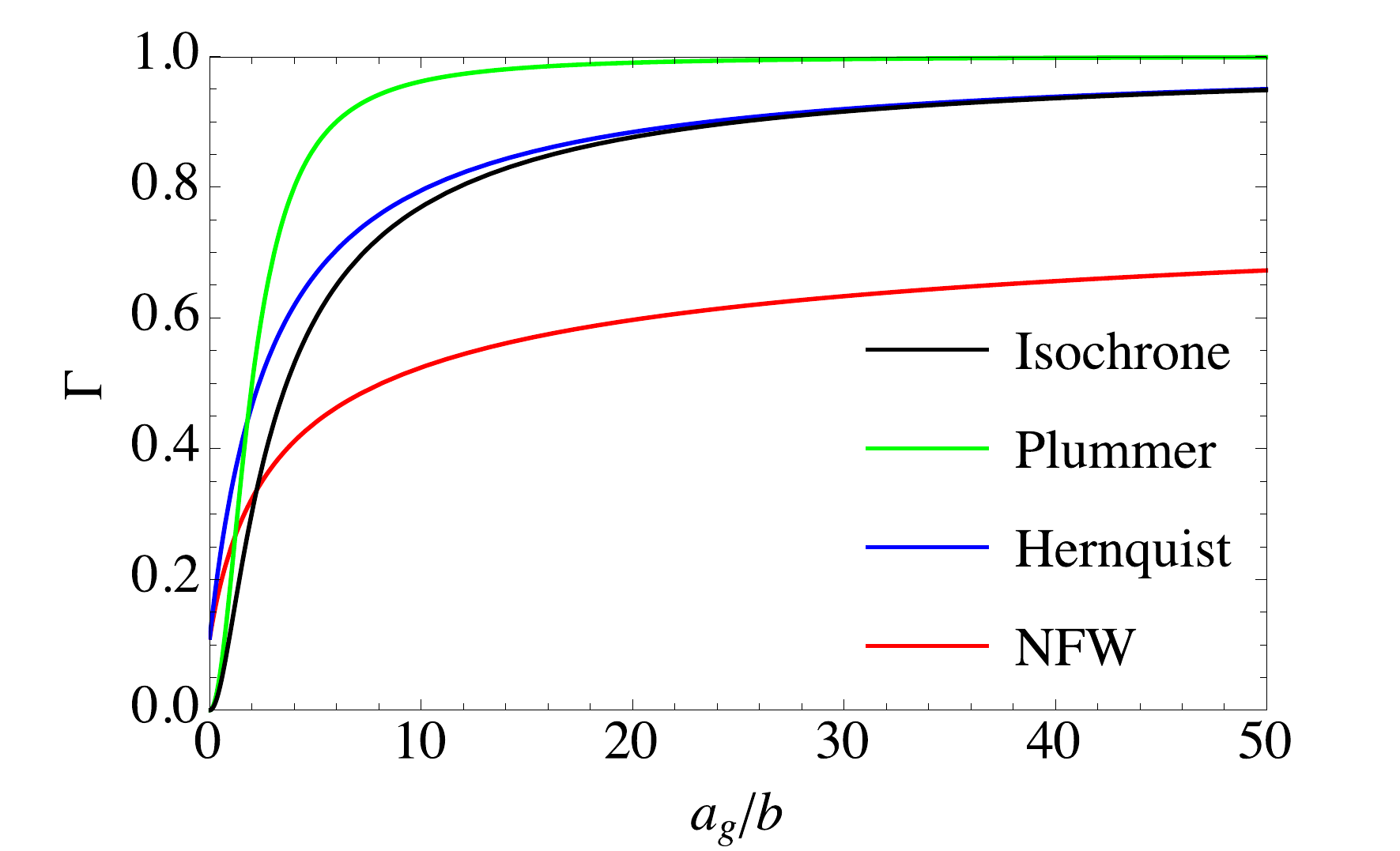}
\caption{Behavior of the parameter $\Gamma$ for circular orbits ($e_\mathrm{g}=0$) as a function of $a_\mathrm{g}/b$ (where $b$ is a scale radius) in the same four potentials as in Figure \ref{GammaPlots}.  This plot demonstrates that $\Gamma$ converges to $1$ for potentials that are Keplerian at large radii. In the opposite limit $a_\mathrm{g}/b \to 0$, we see that $\Gamma \to 0$ for the cored potentials and $\Gamma \to 1/9$ for NFW, as expected (\S\ref{GamSph}).}
\label{GammaConvergence}
\end{figure}
%%%%%%%%%%%%%%%%%%%%%%%%%%%%%%%%%%%%%%%%%%

We use the following spherically symmetric potentials: \\
(i) the isochrone potential (which has a constant-density core and half mass radius $r_\mathrm{h}=3.06b$)
\begin{align} 
\label{IsoPot} 
\Phi_\mathrm{iso}(r) = -GM/(b+\sqrt{b^2+r^2}),
\end{align} 
(ii)  the Plummer potential (also has a core and  $r_\mathrm{h}=1.31b$)
\begin{align}
\label{PlumPot}
\Phi_\mathrm{Plum}(r) = -GM/\sqrt{b^2+r^2},
\end{align} 
(iii) the Hernquist potential (has no core, and $r_\mathrm{h}=2.41b$)
\begin{align}
\label{HernPot}
\Phi_\mathrm{Hern}(r) = -GM/(b+r),
\end{align}
(iv) the NFW potential (has no core and has a divergent mass profile)
\begin{align}
\label{NFWPot}
\Phi_\mathrm{NFW}(r) = -GMr^{-1}\ln(1+r/b).
\end{align}  
The NFW potential arises from a density distribution
\begin{align}
\rho(r) \propto \left(\frac{r}{b} \right)^{-1}\left(1+\frac{r}{b}\right)^{-2}.
\label{eq:rhoNFW}
\end{align} 
In equation \eqref{NFWPot} the quantity $M$ is \textit{not} the mass of the model (which is formally infinite), just a constant with units of mass.

For illustration, in Figure \ref{convplots2} (left panels) we show three examples of numerically integrated orbits in some of these potentials. (Each orbit was integrated for $100$ azimuthal periods; the first few periods are highlighted in red). The first two (`I' and `II') orbit the isochrone potential \eqref{IsoPot}, which has a constant density core for $r\lesssim b$.  The third (`III') orbits the Hernquist potential \eqref{HernPot}, which is coreless.    In Table \ref{IIItable} we list the peri/apocentre $r_\mathrm{p/a}$, semi-major axis $a_\mathrm{g}$, eccentricity $e_\mathrm{g}$, azimuthal period $T_\phi$, and the values of $A^*$ and $\Gamma$ calculated using equation \eqref{2Davgeqn}. We also provide values of $A^*_\mathrm{num}, \Gamma_\mathrm{num}$ obtained by direct averaging of $\Phi_{\alpha\beta}$ along each numerically integrated outer orbit (see Appendix \ref{Numerical_Time_Averages}), to which we will return when discussing the validity of the axisymmetric averaging approximation in \S\ref{numveravg}.

%%%%%%%%%%%%%%%%%%%%%%%%%%%%%%%%%%%%%%%%%%

\subsubsection{Behavior of $A^*$}	
\label{ASph}

In Figure \ref{ABPlots} we plot $\log_{10}A^*$ in the $(a_\mathrm{g},e_\mathrm{g})$ plane for the potentials (\ref{IsoPot})-(\ref{NFWPot}). We see that $A^*$ is a strong function of $a_\mathrm{g}$ but a weaker function of $e_\mathrm{g}$. The dependence on $e_\mathrm{g}$ emerges predominantly for orbits with $e_\mathrm{g}\gtrsim 0.5$; it is rather weak at all $e_\mathrm{g}$ for orbits with $a_\mathrm{g}\lesssim b$,  where $b$ is the scale radius of the potential in question. The difference in radial $A^*$ behavior between different potentials is most pronounced for orbits with $a_\mathrm{g} \lesssim b$. In this region there is a sharp increase in $A^*$ in the uncored (Hernquist and NFW) potentials, but a much shallower gradient in the cored potentials (isochrone and Plummer).  

We can make the comparison more quantitative by examining the radial profile of $A^*$ for circular outer orbits (of radius $a_\mathrm{g}$ and eccentricity $e_\mathrm{g}=0$). Then $A^*$ is a function of $a_\mathrm{g}/b$ only, and is plotted in Figure \ref{ACirc}. We see that $A^*(a_\mathrm{g}/b)$ becomes significantly larger than $1$ for $a_\mathrm{g} \ll b$ in the case of non-cored potentials, but reaches a maximum of $0.5$ in the isochrone case. For those potentials with finite total mass $M$ we can construct the density-weighted average 
\begin{align}
\langle A^* \rangle_\rho = \frac{1}{M}\int_0^\infty 4\pi r^2 \rho(r) A^*(r) \,\md r,
\end{align} 
still assuming a circular outer orbit.  We find $\langle A^* \rangle_\rho = 0.0617, 0.4234$ and $0.65$ for isochrone, Plummer and Hernquist potentials respectively. The isochrone model has by far the smallest $\langle A^* \rangle_\rho$.

%%%%%%%%%%%%%%%%%%%%%%%%%%%%%%%%%%%%%%%%%%

\subsubsection{Behavior of $\Gamma$}	
\label{GamSph}

Figure \ref{GammaPlots} shows $\Gamma$ in the $(a_\mathrm{g},e_\mathrm{g})$ plane for the same four potentials. We see that $\Gamma \to 0$ for $a_\mathrm{g} \ll b$ in cored potentials. For the coreless potentials $\Gamma$ is always positive, as expected. We see that for $a_\mathrm{g}\gtrsim b$, the value of $\Gamma$ is quite sensitive to the outer orbit eccentricity $e_\mathrm{g}$ in all four potentials. At fixed $a_\mathrm{g}$, increasing $e_\mathrm{g}$ corresponds to a decrease in $\Gamma$. For the cored potentials this is because high-$e_\mathrm{g}$ orbits start probing the cluster core where the potential is roughly spherical harmonic (for which $\Gamma$ is effectively zero, see \S\ref{harmonicpot} and \S\ref{sect:gen}), which tends to lower $\Gamma$. 

Meanwhile, increasing $a_\mathrm{g}$ at fixed $e_\mathrm{g}$ tends to increase $\Gamma$.  At large $a_\mathrm{g}$ all finite mass potentials reduce to a Keplerian potential for which $\Gamma =1$. 
%For those potentials that are Keplerian at large radii we expect $\Gamma$ to converge to $1$ for $a_\mathrm{g} \gg b$. 
In the NFW potential, the $\Gamma$ profile is shallow because the potential decays slowly, namely as $\Phi_\mathrm{NFW} \sim r^{-1}\ln(r/b)$ for $r\gg b$. Hence it never becomes sufficiently Keplerian to reach $\Gamma \sim 1$.  

To better illustrate this convergence at large $a_\mathrm{g}/b$, in Figure \ref{GammaConvergence} we show $\Gamma(a_\mathrm{g}/b)$ for circular outer orbits $(e_\mathrm{g}= 0)$ in the same four potentials as in Figure \ref{GammaPlots}.  We see that the $\Gamma \to 1$ convergence does occur for all potentials that are asymptotically Keplerian as $r\to \infty$, although in some cases one has to go to radii $a_\mathrm{g} \gtrsim 50b$ to observe it satisfactorily.  Additionally, at very small radii the NFW density profile can be approximated as a power-law cusp, $\rho \propto r^{-1}$, see equation (\ref{eq:rhoNFW}).  Using the result listed in \S\ref{sect:gen} (and derived in Appendix \ref{SphericalABGamma}) we expect to find $\Gamma = 1/9$ as $a_\mathrm{g}\to 0$, and indeed this is reflected in Figure \ref{GammaConvergence}.
\\
\\
We note that some of the orbits in the potentials (\ref{IsoPot})-(\ref{NFWPot}) will have  commensurable (or almost commensurable) radial and azimuthal frequencies. For these orbits, i.e. at some points in $(a_\mathrm{g},e_\mathrm{g})$ space, equation \eqref{2Davgeqn} is not valid, because its derivation relies upon orbits densely filling an axisymmetric annulus, see \S \ref{Non-Com}.  This is particularly true of potentials with a core at small $a_\mathrm{g}$, where the potential is close to harmonic (c.f. \S \ref{numveravg}). Nevertheless, Figures \ref{ABPlots} and \ref{GammaPlots} give a good idea of how $A$ and $\Gamma$ change as we consider different orbits within the cluster.

%%%%%%%%%%%%%%%%%%%%%%%%%%%%%%%%%%%%%%%%%%
\begin{figure*}
\centering
\includegraphics[width=0.4\linewidth,height=0.41\linewidth]{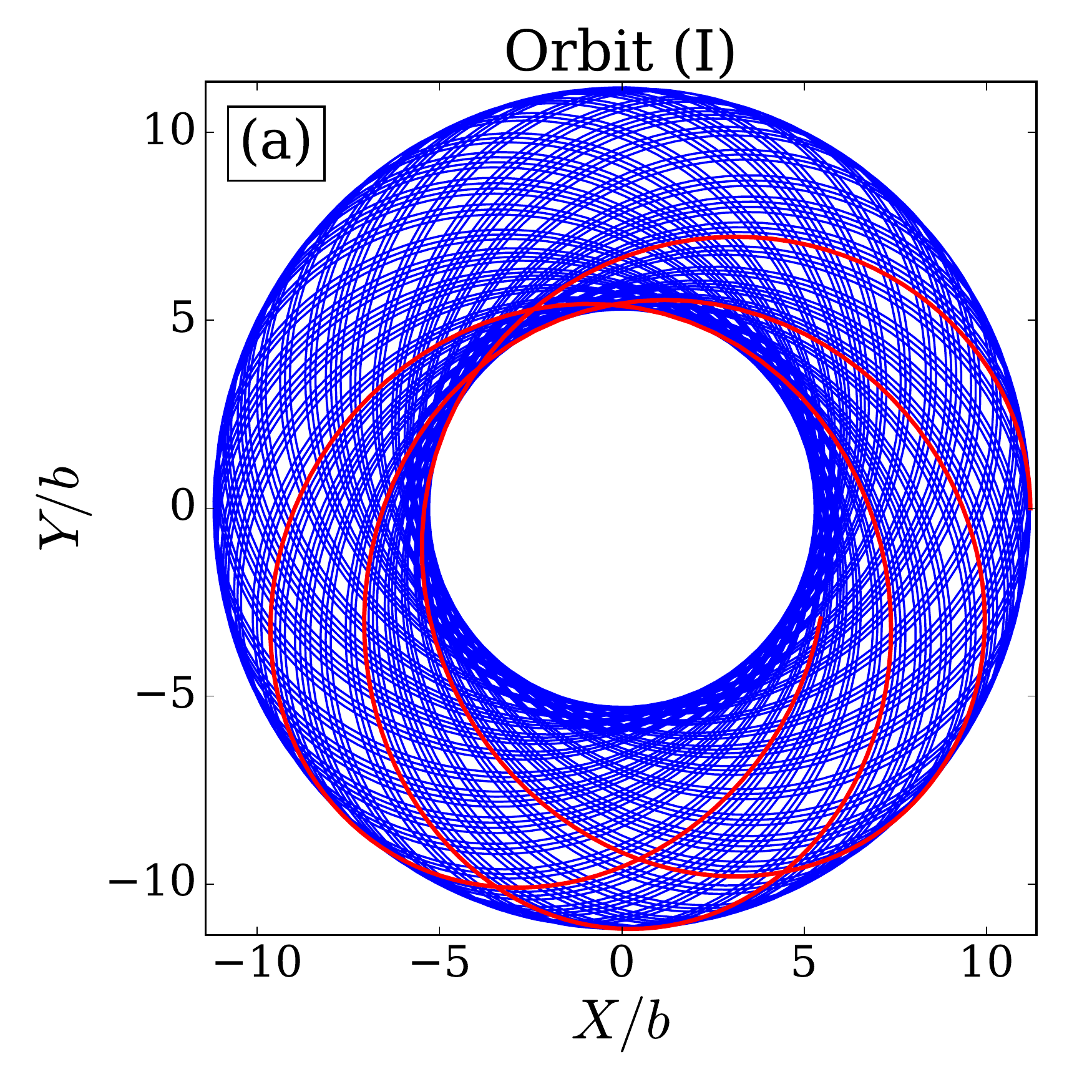}
\raisebox{0.01cm}{\includegraphics[width=0.42\linewidth,height=0.4\linewidth]{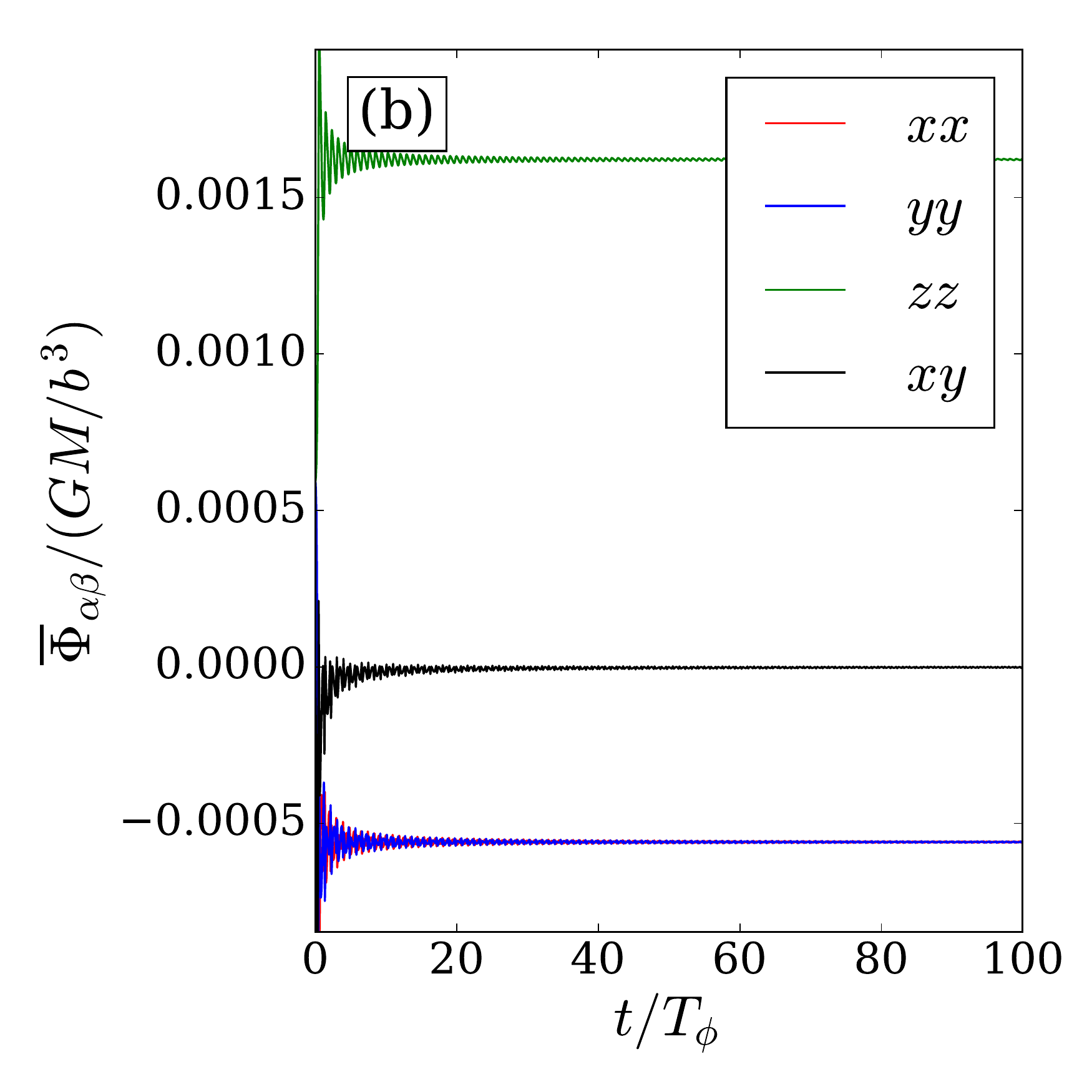}}
\includegraphics[width=0.4\linewidth,height=0.41\linewidth]{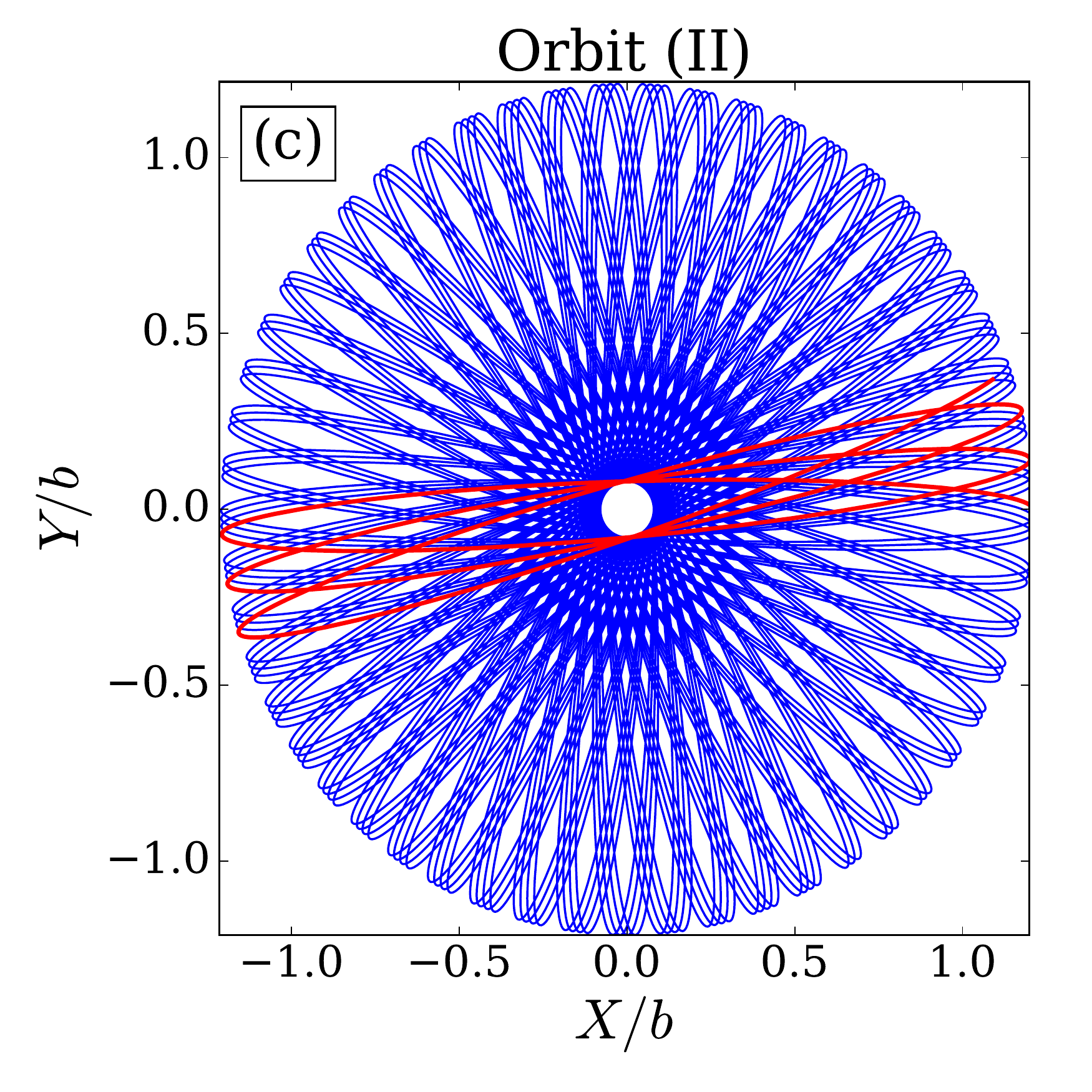}
\raisebox{0.01cm}{\includegraphics[width=0.41\linewidth,height=0.4\linewidth]{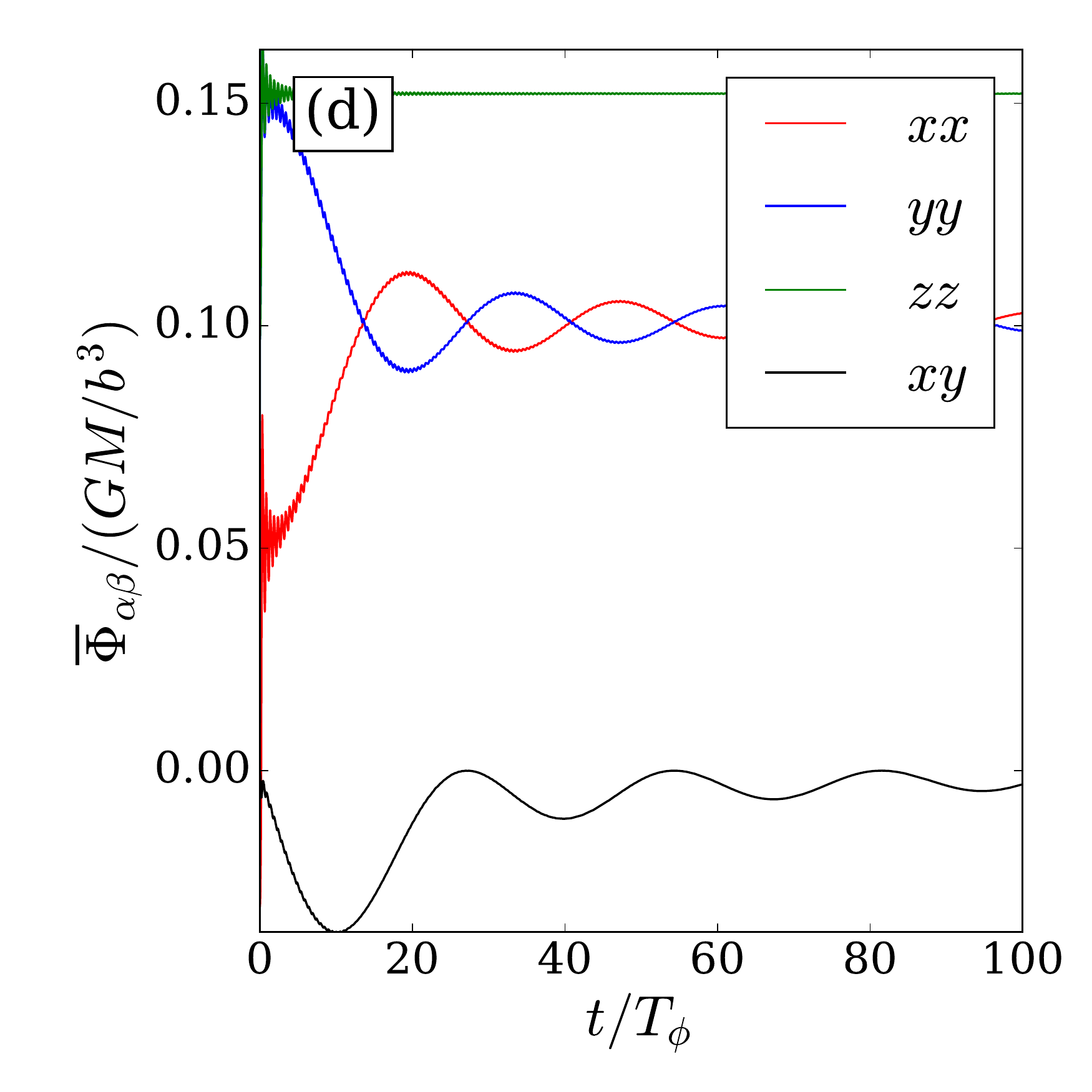}}
\includegraphics[width=0.4\linewidth,height=0.41\linewidth]{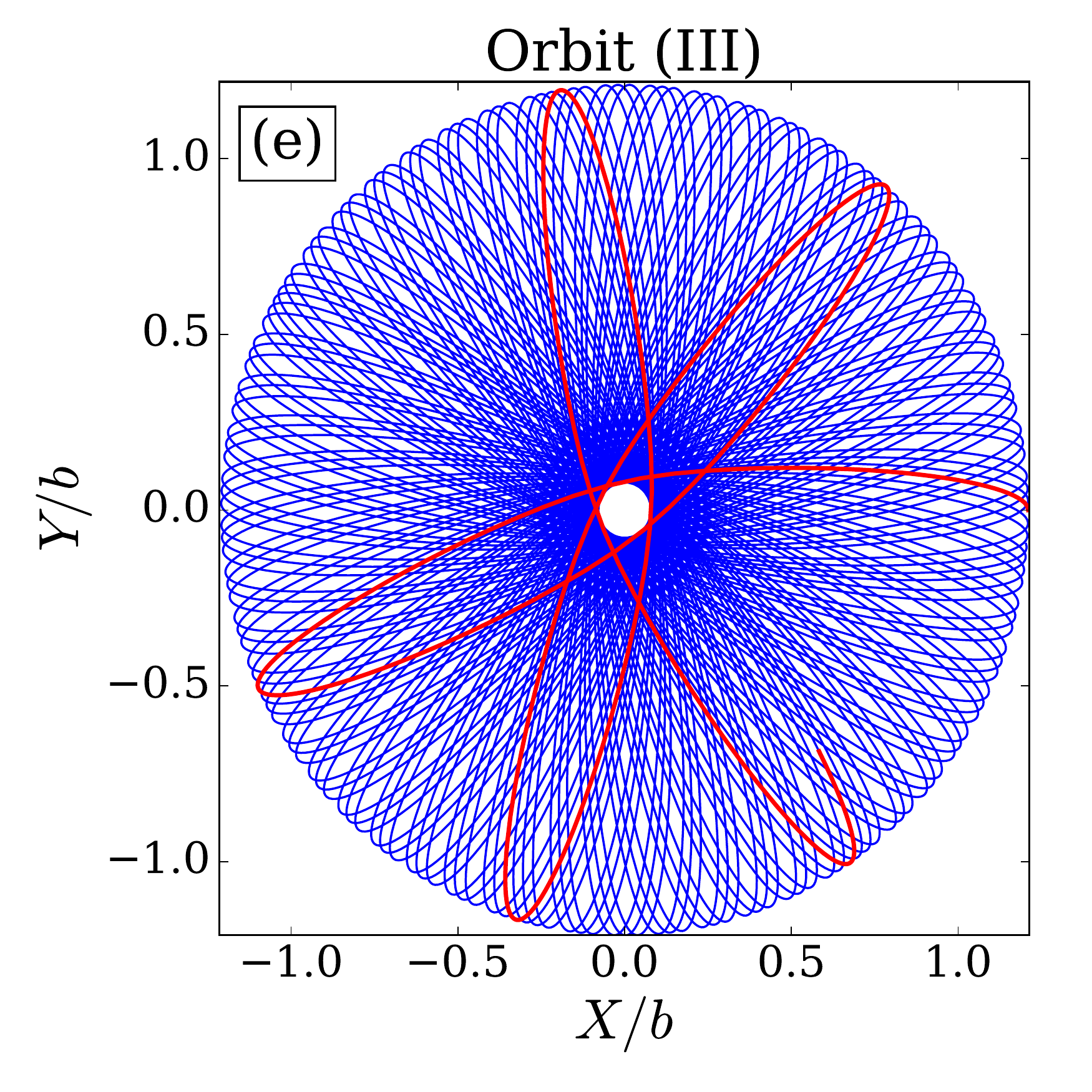}
\raisebox{0.01cm}{\includegraphics[width=0.41\linewidth,height=0.4\linewidth]{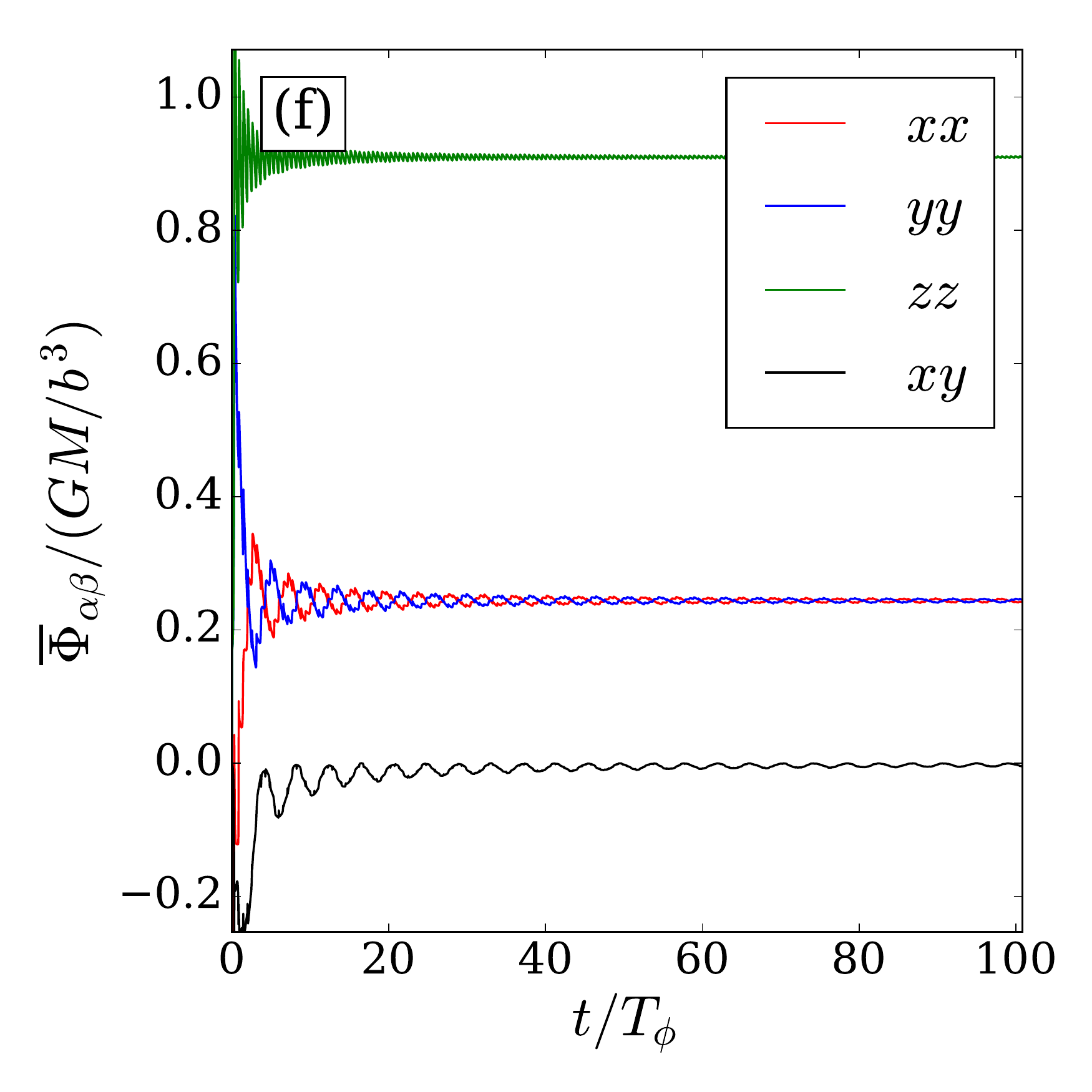}}
\caption{Orbits in spherical potentials (see Table \ref{IIItable}) used for demonstrating convergence to the `axisymmetric annulus' approximation. Left panels show the outer orbit $\Rg$ in the $(X,Y)$ plane, while right panels demonstrate the convergence of time-averaged coefficients $\overline{\Phi}_{\alpha\beta}$ (insets illustrate the color scheme for each $\alpha\beta$ coordinate pair). Orbits (I) and (II) were integrated in the spherical isochrone potential \eqref{IsoPot}, while Orbit (III) is in the Hernquist potential \eqref{HernPot}. All panels show 100 azimuthal periods' worth of data.  In the left panels we highlight the first few azimuthal periods of the orbit in red.  Convergence is much slower for Orbit (II) because it spends most of its time in a constant density core, so radial and azimuthal frequencies are almost commensurable, leaving large unfilled gaps in the annulus even after $100T_\phi$.}
\label{convplots2}
\end{figure*}
%%%%%%%%%%%%%%%%%%%%%%%%%%%%%%%%%%%%%%%%%%

%%%%%%%%%%%%%%%%%%%%%%%%%%%%%%%%%%%%%%%%%%

\subsection{Behavior of Hamiltonian characteristics in axisymmetric potentials}	
\label{AGammaAxi}

%%%%%%%%%%%%%%%%%%%%%%%%%%%%%%%%%%%%%%%%%%

For axisymmetric potentials it is difficult to make rigorous mathematical statements about $A$ and $\Gamma$.  In Appendix \ref{AxisymmetricABGamma} we show that in this case, in principle, $\Gamma$ can take any value $\in (-\infty,\infty)$; however, to achieve extreme negative values of $\Gamma$, or $\Gamma > 1$, may require very unusual orbits. (Two such examples are given in Appendix \ref{AxisymmetricABGamma}). Meanwhile, in this section we focus on the most typical orbits in axisymmetric potentials via numerical examples. The $A^*$ and $\Gamma$ values in this section are calculated numerically using the procedure outlined in Appendix \ref{Numerical_Time_Averages}, and are therefore denoted $A^*_\mathrm{num},\Gamma_\mathrm{num}$.

We will use two axisymmetric potentials in our numerical examples. The first is the flattened power-law potential \citep{Evans1994}: 
\begin{align} 
\Phi_\mathrm{FPL}(R,Z) = -\Phi_0\frac{b^\beta}{\left(R^2 + (Z/q)^2 + b^2\right)^{\beta/2}}, 
\label{FPot} 
\end{align}  
where $-\Phi_0$ is the central potential, $b$ is a core radius and $q$ is the oblateness parameter: $q<1$ corresponds to an oblate potential which can be used to model elliptical galaxies and galactic bulges \citep{Evans1994}.  The natural definition of $A^*$ in this case is $A^* \equiv A/(\vert \Phi_0 \vert /b^2)$.  We choose $\beta=1/2$ and $q=0.94$, meaning that this potential is only slightly flattened. One can derive a number of useful analytical results in such weakly non-spherical potentials; we defer this investigation to a future study. Here we simply demonstrate that even in the case of a weakly flattened potential large departures from the behaviour typical for purely spherical potentials described in \S\ref{AGamSph} become possible. 

%%%%%%%%%%%%%%%%%%%%%%%%%%%%%%%%%%%%%%%%%%
\begin{figure}
%\centering
\includegraphics[width=0.99\linewidth]{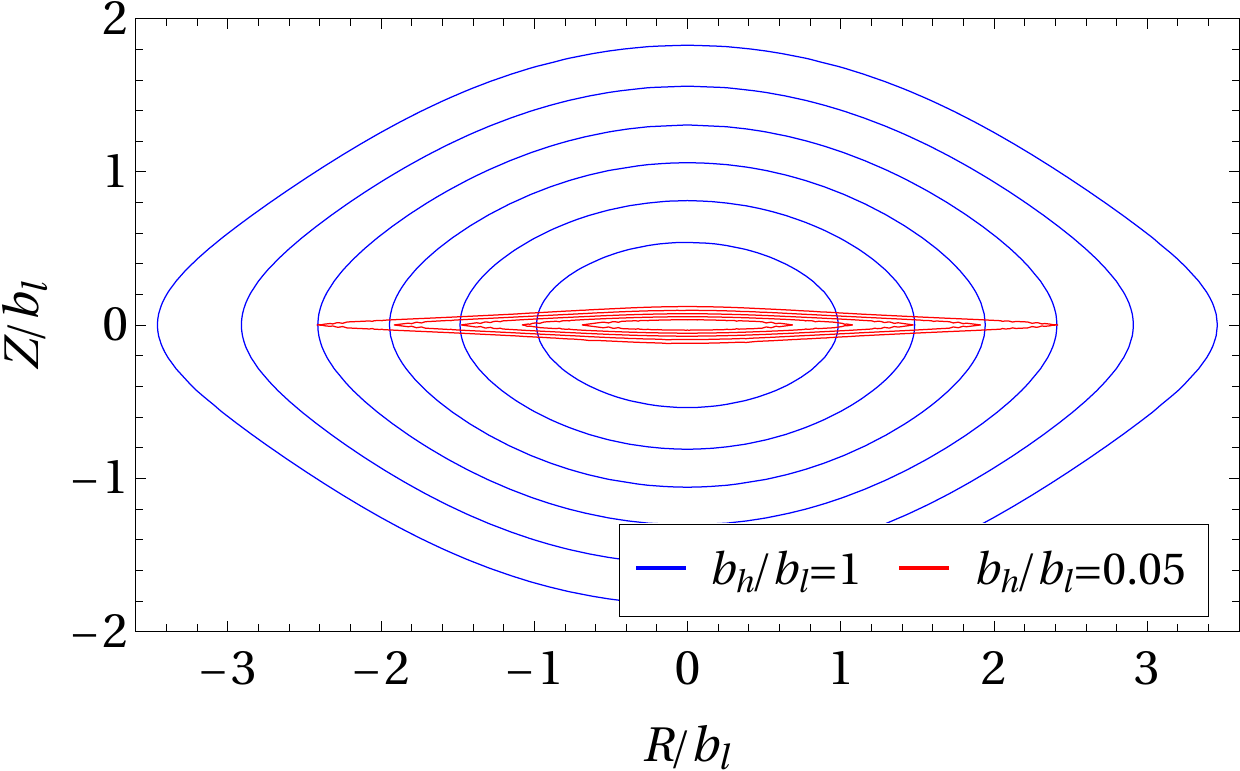}
\caption{Contours of constant $\log_{10}(\rho_\mathrm{MN}/\rho_{\mathrm{MN},0})$ where $\rho_\mathrm{MN} \equiv \nabla^2\Phi_\mathrm{MN}/(4\pi G)$ is the density distribution corresponding to the Miyamoto-Nagai potential \eqref{MNPot}, and $\rho_{\mathrm{MN},0}$ is the central density at $R=Z=0$.  We show the cases $b_h/b_\ell = 1$ (blue) and $b_h/b_\ell = 0.05$ (red).  Contours are spaced linearly from $-1.5$ to $0$.}
\label{MNDensityPlots}
\end{figure} 
%%%%%%%%%%%%%%%%%%%%%%%%%%%%%%%%%%%%%%%%%%

The other potential we will use is the Miyamoto-Nagai potential \citep{Miyamoto1975}: 
\begin{equation}
\Phi_\mathrm{MN}(R,Z)=-\frac{GM}{\sqrt{R^2+\left(b_\ell+\sqrt{Z^2+b_h^2}\right)^2}}, 
\label{MNPot}
\end{equation} 
where $b_\ell$ is the scale length and $b_h$ is the scale height. As one changes the value of $b_h/b_\ell$, the Miyamoto-Nagai potential smoothly transitions from the Kuzmin potential of a razor thin disk ($b_h \ll b_\ell$) to the spherical Plummer potential frequently used to model globular clusters ($b_h\gg b_\ell$) \citep{Binney2008}.  In Figure \ref{MNDensityPlots} we plot contours of constant density $\rho_\mathrm{MN} \equiv \nabla^2\Phi_\mathrm{MN}/(4\pi G)$ in the $(R,Z)$ plane for two Miyamoto-Nagai models used in this paper, namely $b_h/b_\ell = 1$ and $b_h/b_\ell = 0.05$. The natural definition of $A^*$ in this potential is $A^* \equiv A/(GM/b_\ell^3)$.

%%%%%%%%%%%%%%%%%%%%%%%%%%%%%%%%%%%%%%%%%%
\subsubsection{Orbits in the midplane of an axisymmetric potential} 
\label{PlaneAxi}
		
The simplest non-spherical case to consider is when the binary's outer orbit $\Rg$ is confined to the $(X,Y)$ midplane of an axisymmetric potential.  Then $\Rg$ still traces a planar rosette with a fixed peri/apocentre $(r_\mathrm{p},r_\mathrm{a})$ just as in the spherical case, so we can easily compute $A$ and $\Gamma$ as in \S\ref{tapspherical}.  In Appendix \ref{RecoverHT} we show how to compute $A,B,\Gamma$ in the case of epicyclic outer orbits in a disk-like potential. We find  $A=B=\nu^2$, where $\nu$ is the vertical epicyclic frequency at the guiding radius; therefore $\Gamma = 1/3$.
	
%%%%%%%%%%%%%%%%%%%%%%%%%%%%%%%%%%%%%%%%%%
\begin{figure*}
\centering
\includegraphics[width=0.33 \linewidth]{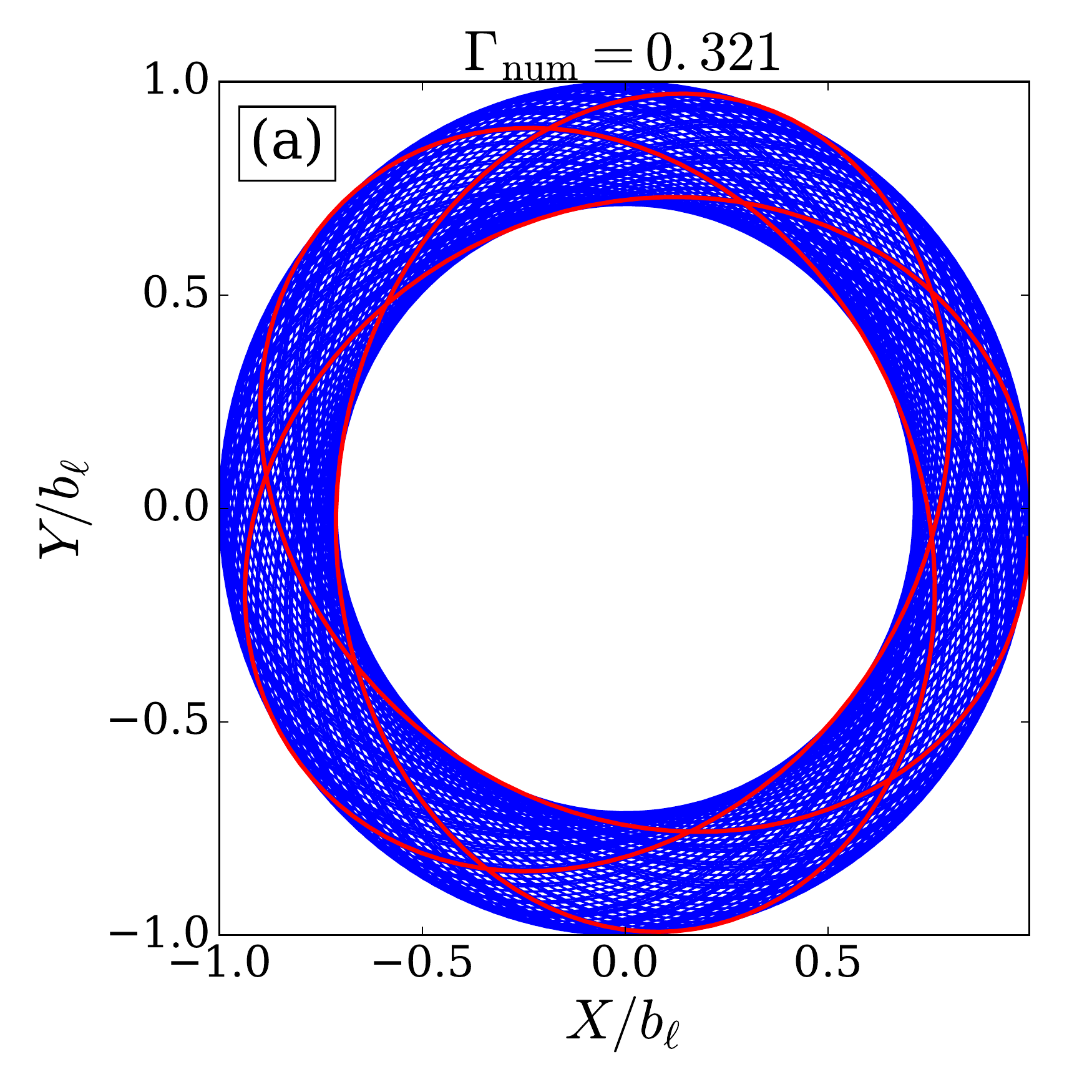}
\raisebox{0.14cm}{\includegraphics[width=0.315 \linewidth]{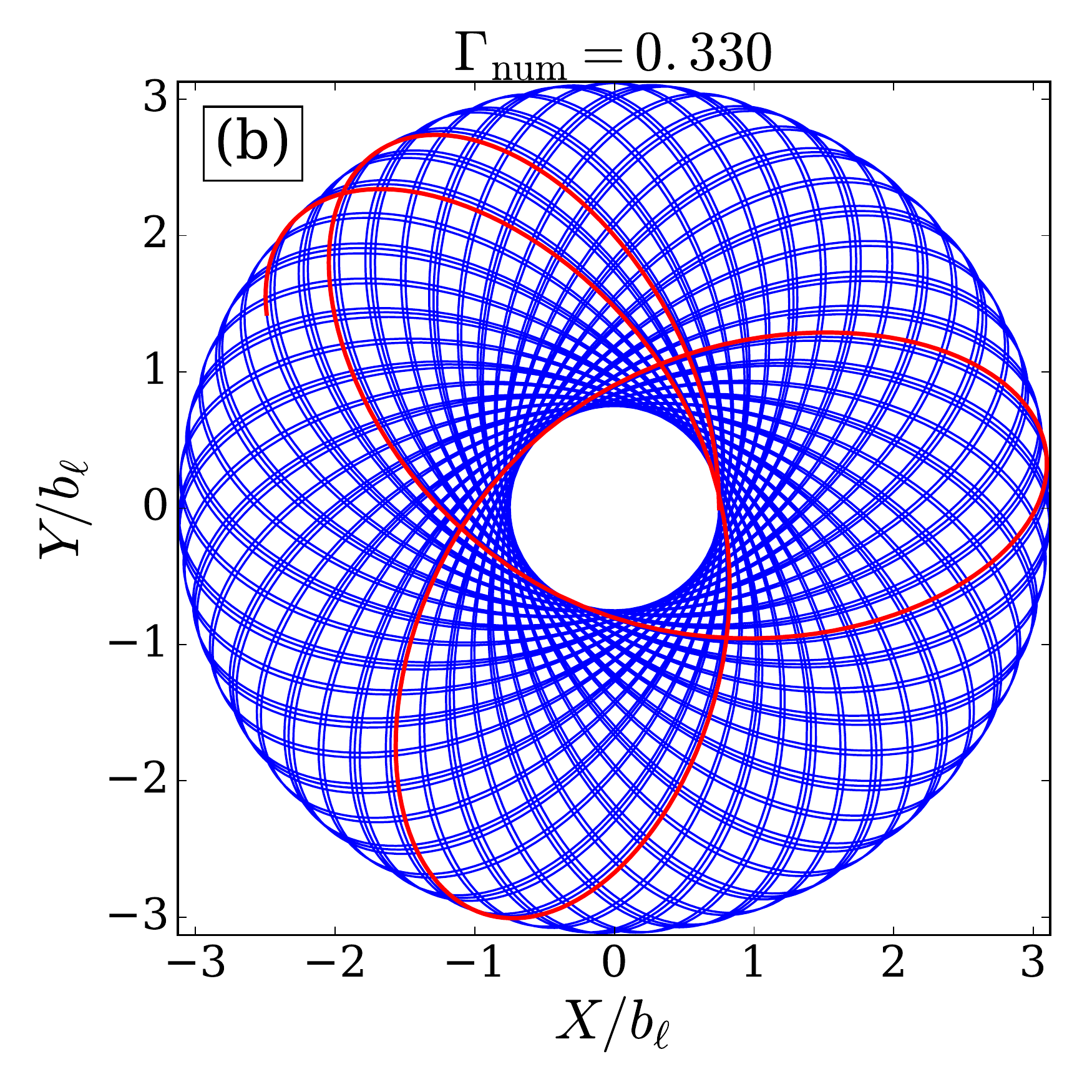}}
\includegraphics[width=0.33 \linewidth]{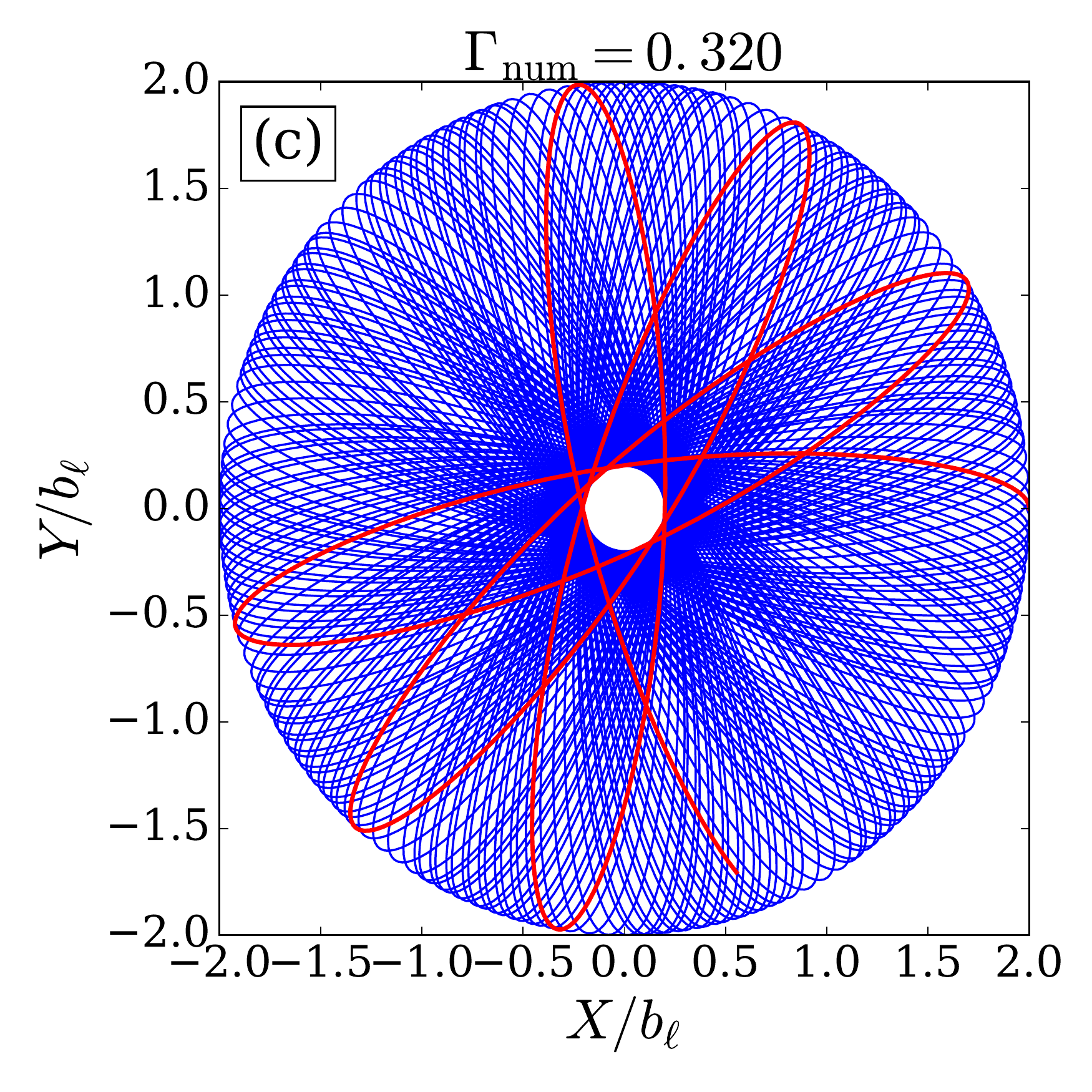}
\caption{Examples of orbits in the midplane of the thin disk represented by the Miyamoto-Nagai potential \eqref{MNPot} with $b_h/b_\ell = 0.05$. We integrate the orbits for $100T_\phi$ (the first few $T_\phi$ are highlighted red). The resulting numerically determined $\Gamma_\mathrm{num}$ values are all very close to the value $\Gamma=1/3$ that was predicted simply on the basis of the disk being very thin, so that the vertical curvature of the potential is much larger than the radial curvature. See \S \ref{PlaneAxi} for details.}
\label{DiskPlots}
\end{figure*}
%%%%%%%%%%%%%%%%%%%%%%%%%%%%%%%%%%%%%%%%%%
	
In fact, we already deduced in \S\ref{sect:gen} that $\Gamma=1/3$ will hold for \textit{any} orbit $\Rg$ which is confined to the plane of a very thin axisymmetric disk.  This follows from the fact that the curvature of the potential is by far greatest in the $Z$ direction at any given position in the disk, so that $\Phi_{zz}(R,0) \gg \Phi_{xx}(R,0) \,\, \forall \,\,R$.  Then $A\approx B\approx\overline{\Phi}_{zz}$ and so $\Gamma\approx 1/3$.  In Figure \ref{DiskPlots} we confirm this prediction using three very different orbits in the $(X,Y)$ midplane of a thin ($b_h/b_\ell = 0.05$) Miyamoto-Nagai potential.  The $\Gamma_\mathrm{num}$ values are (a) $0.321$, (b) $0.330$ and (c) $0.320$, all very close to $\Gamma=1/3$.

%%%%%%%%%%%%%%%%%%%%%%%%%%%%%%%%%%%%%%%%%%
\subsubsection{Orbits that are far from coplanar}	
	
As we show in Appendix \ref{sect:AGamma}, we always have $\Gamma \geq 0$ in realistic, finite-mass spherical potentials. For $\Gamma$ to fall below zero the potential must be non-spherical, but also, according to definitions (\ref{ABGamDef}), the outer orbit of the binary must have $\vert \overline{\Phi}_{xx} \vert > \vert \overline{\Phi}_{zz} \vert$. Qualitatively, this implies that the average `radial curvature' of the potential over the orbit needs to be greater than the average `vertical curvature'. This is not going to be the case while the orbit is confined near a single plane, as we have just seen. However, this situation is naturally realised in potentials that are highly prolate in the $Z$ direction (asymptotically `cylindrical', with $\Phi(R,Z)=\Phi(R)$). In such potentials $\Gamma\approx -1/3$ (see \S \ref{sect:gen}). Also, to probe the negative $\Gamma$ regime we can consider orbits in non-spherical potentials that make large excursions `out of the plane', i.e. in the $Z$ direction.  
	
\begin{figure*}
\centering
\includegraphics[width=0.32\linewidth,height=0.32\linewidth]{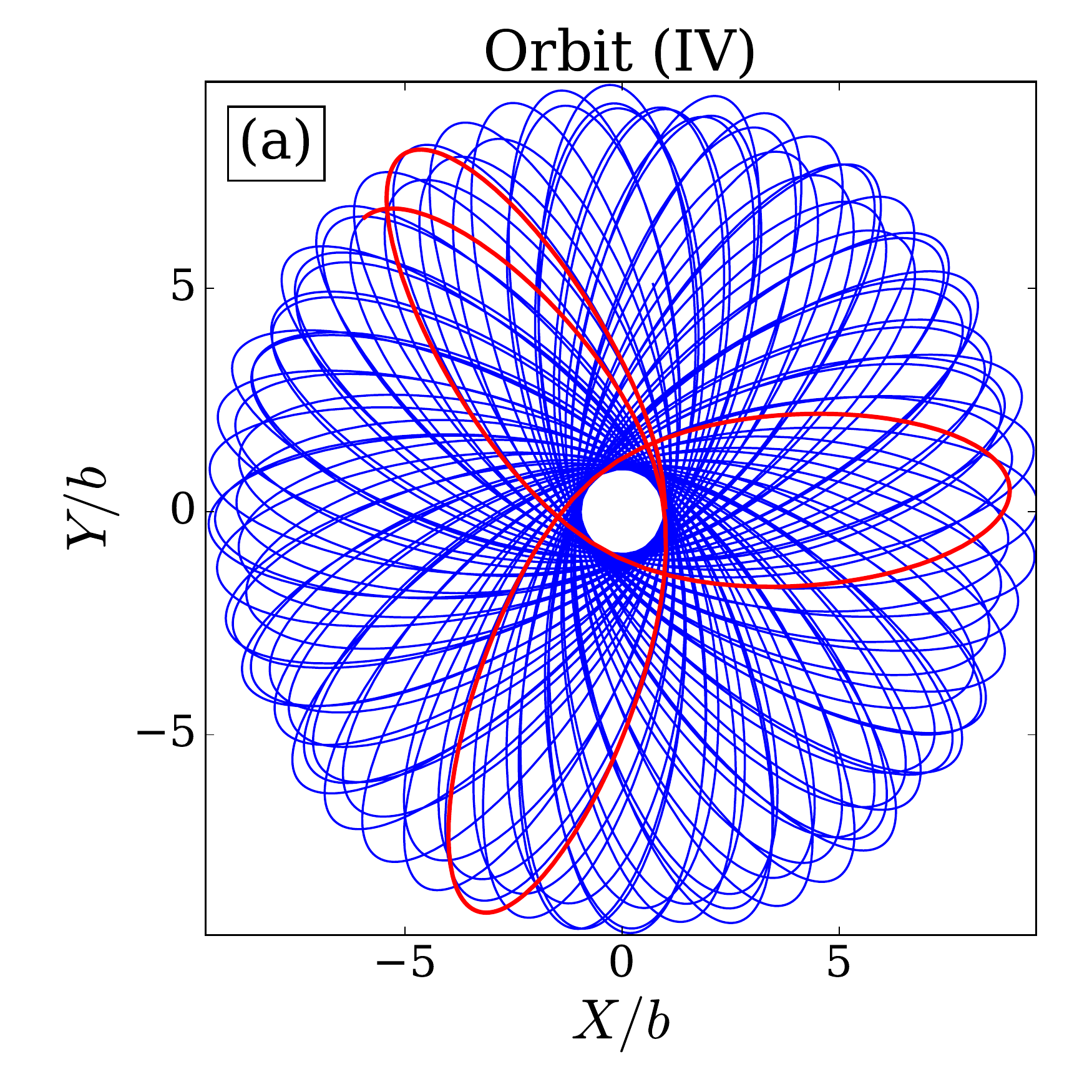}
\includegraphics[width=0.32\linewidth,height=0.32\linewidth]{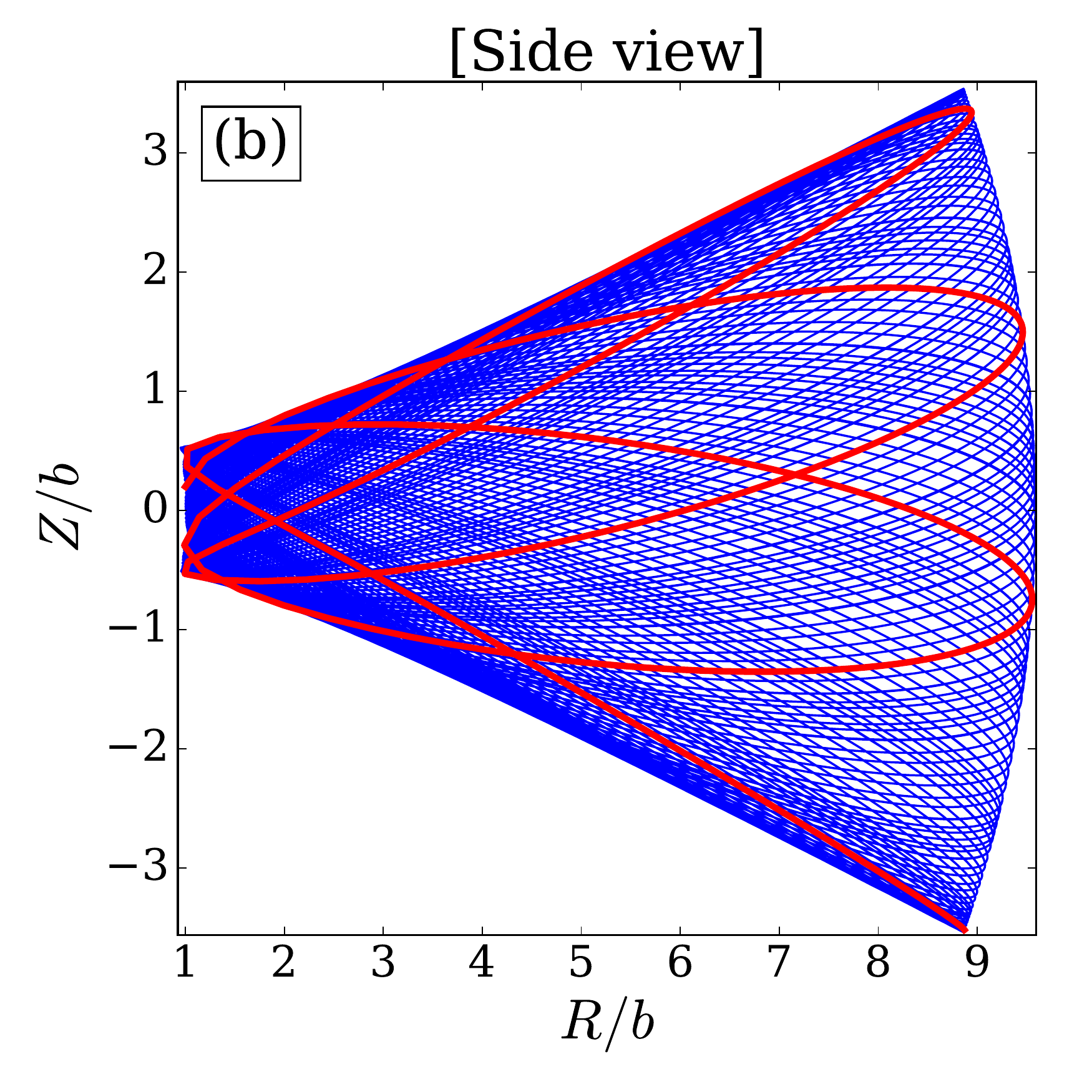}
\includegraphics[width=0.32\linewidth,height=0.32\linewidth]{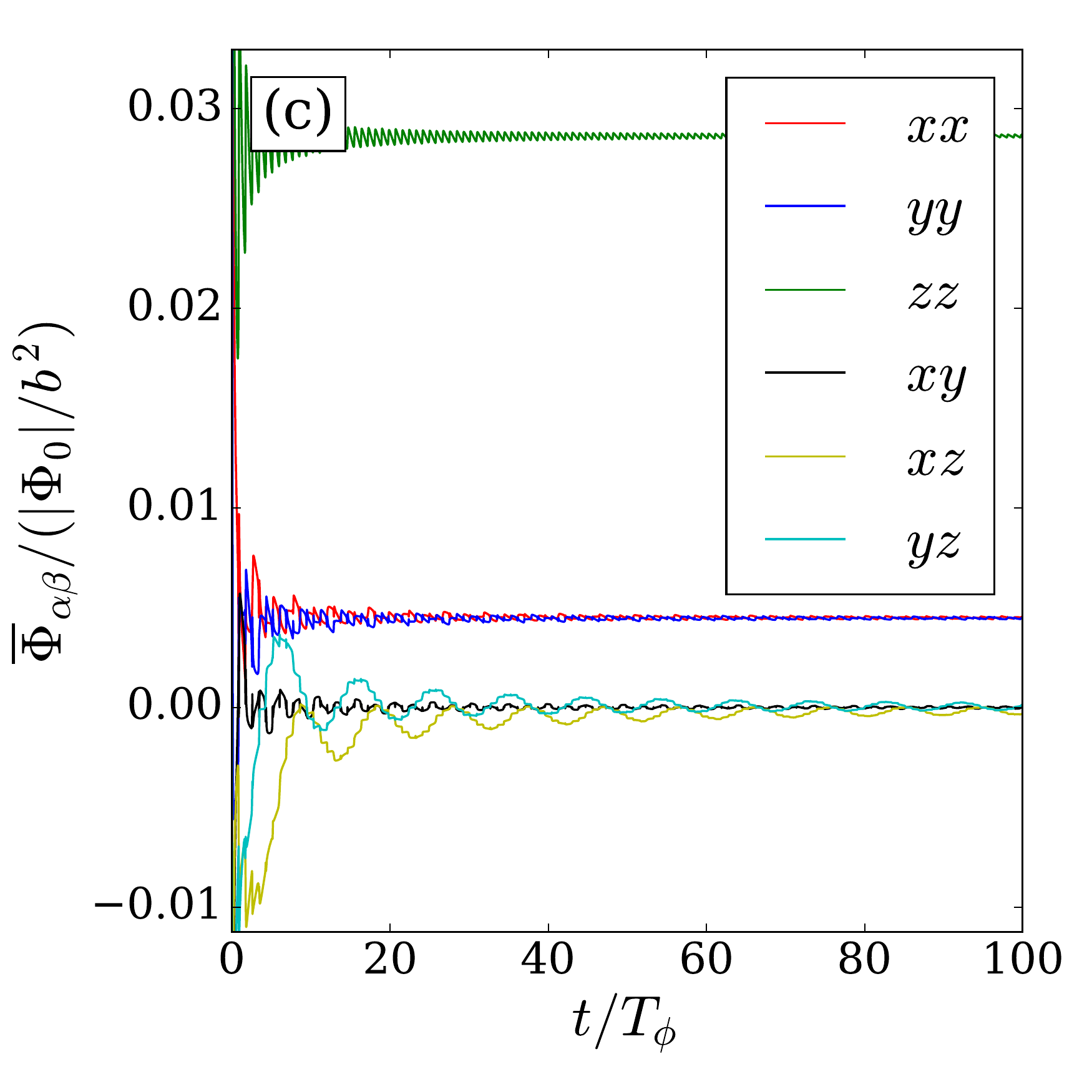}
\includegraphics[width=0.32\linewidth,height=0.32\linewidth]{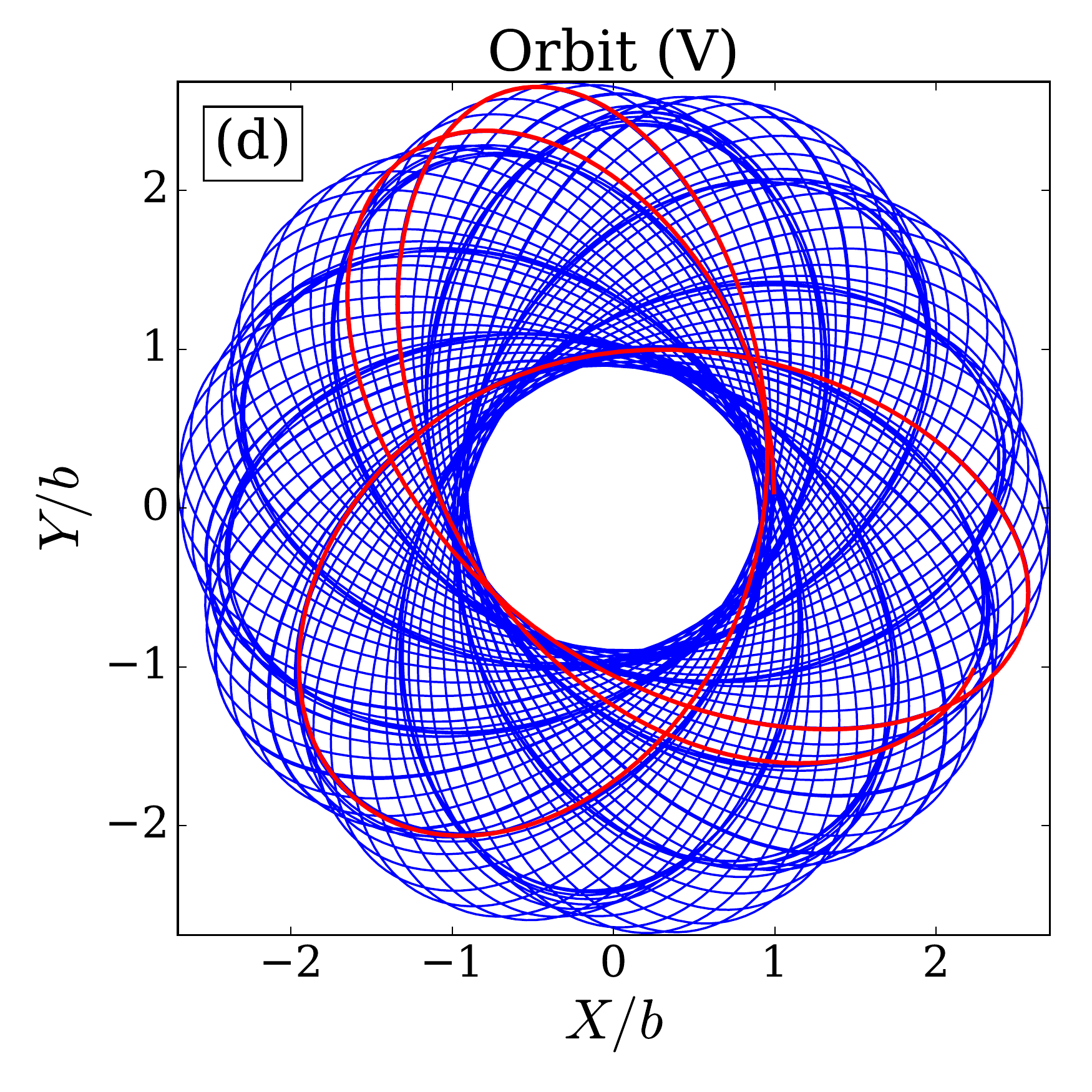}
\includegraphics[width=0.32\linewidth,height=0.32\linewidth]{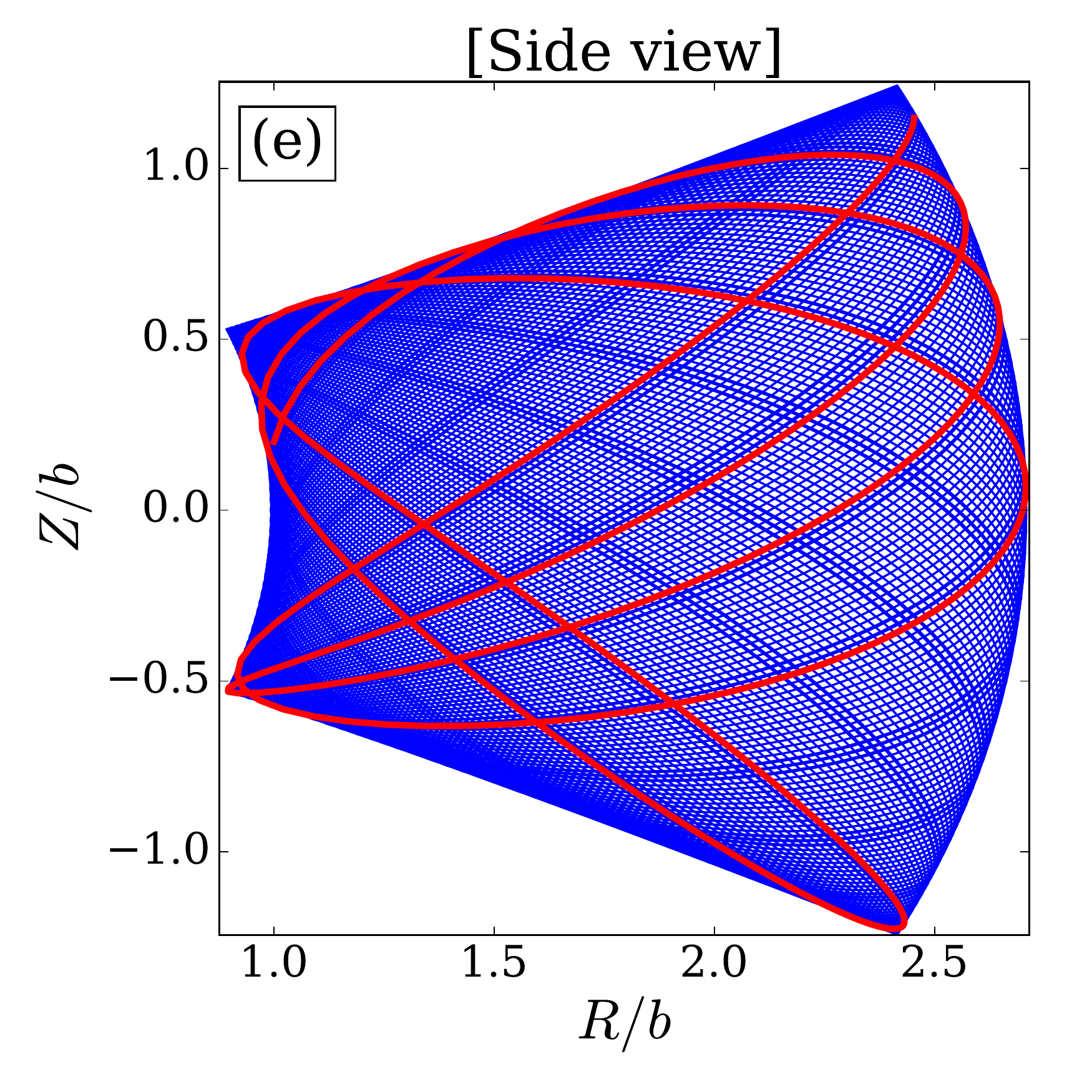}
\includegraphics[width=0.32\linewidth,height=0.32\linewidth]{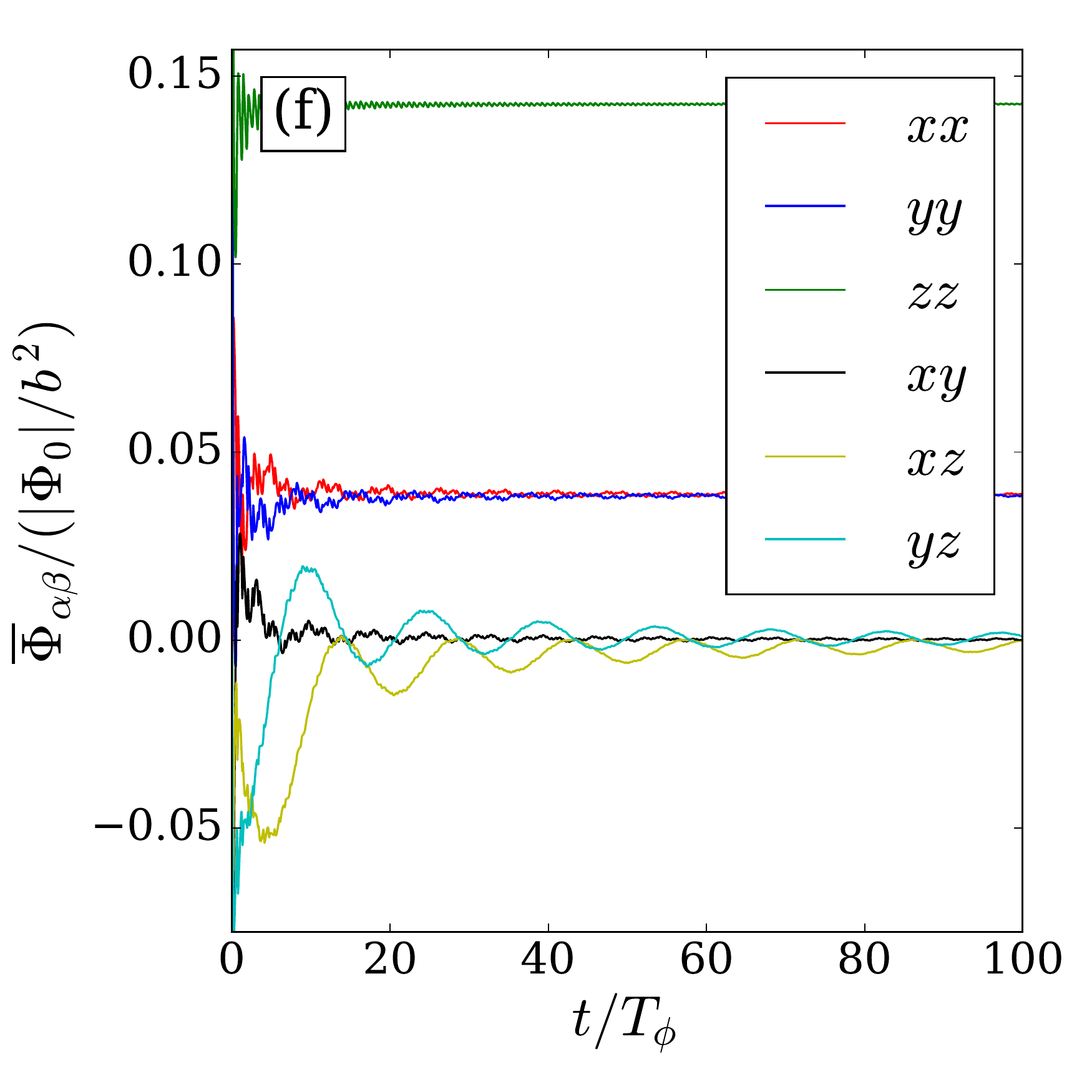}
\includegraphics[width=0.32\linewidth,height=0.32\linewidth]{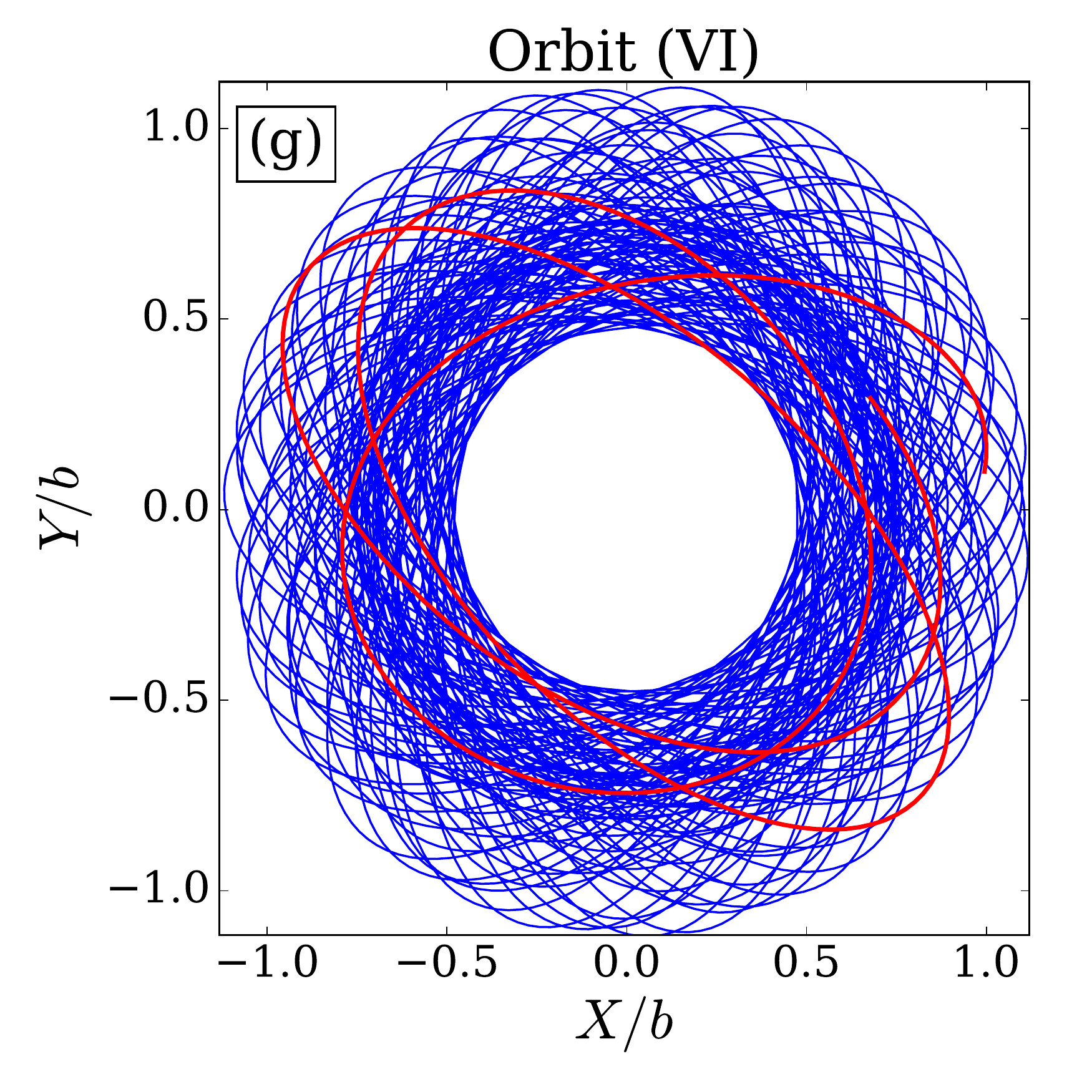}
\includegraphics[width=0.32\linewidth,height=0.32\linewidth]{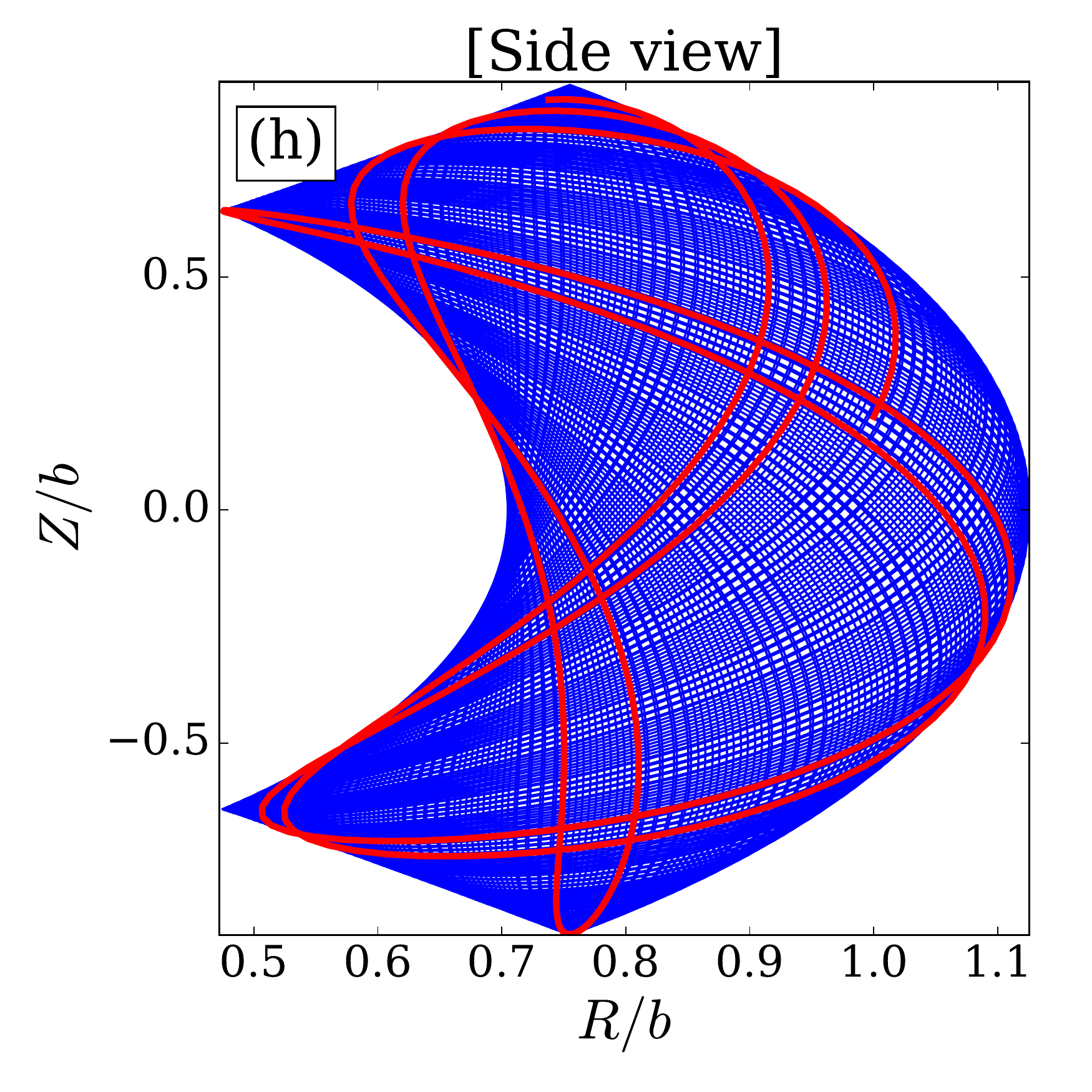}
\includegraphics[width=0.32\linewidth,height=0.32\linewidth]{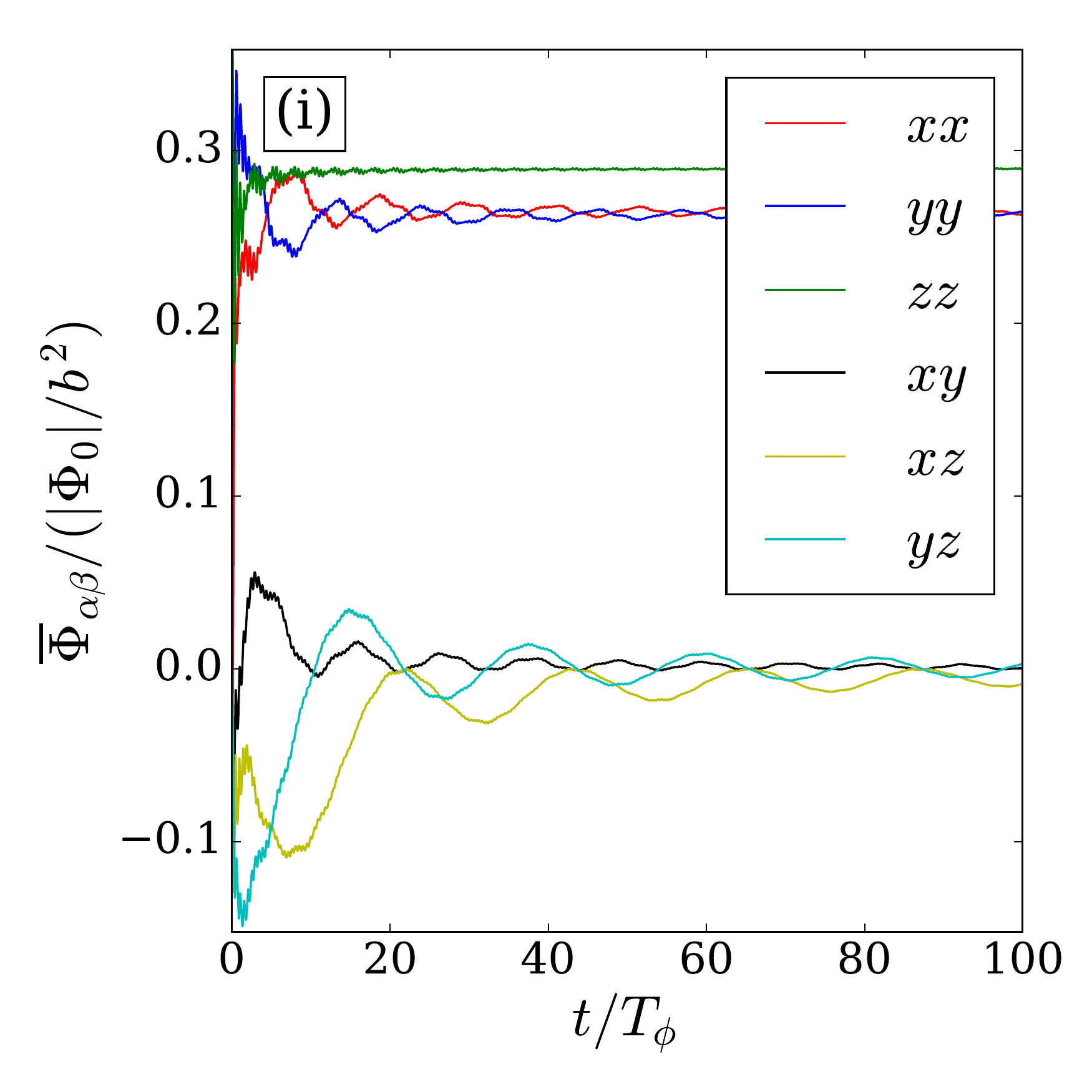}
\includegraphics[width=0.32\linewidth,height=0.32\linewidth]{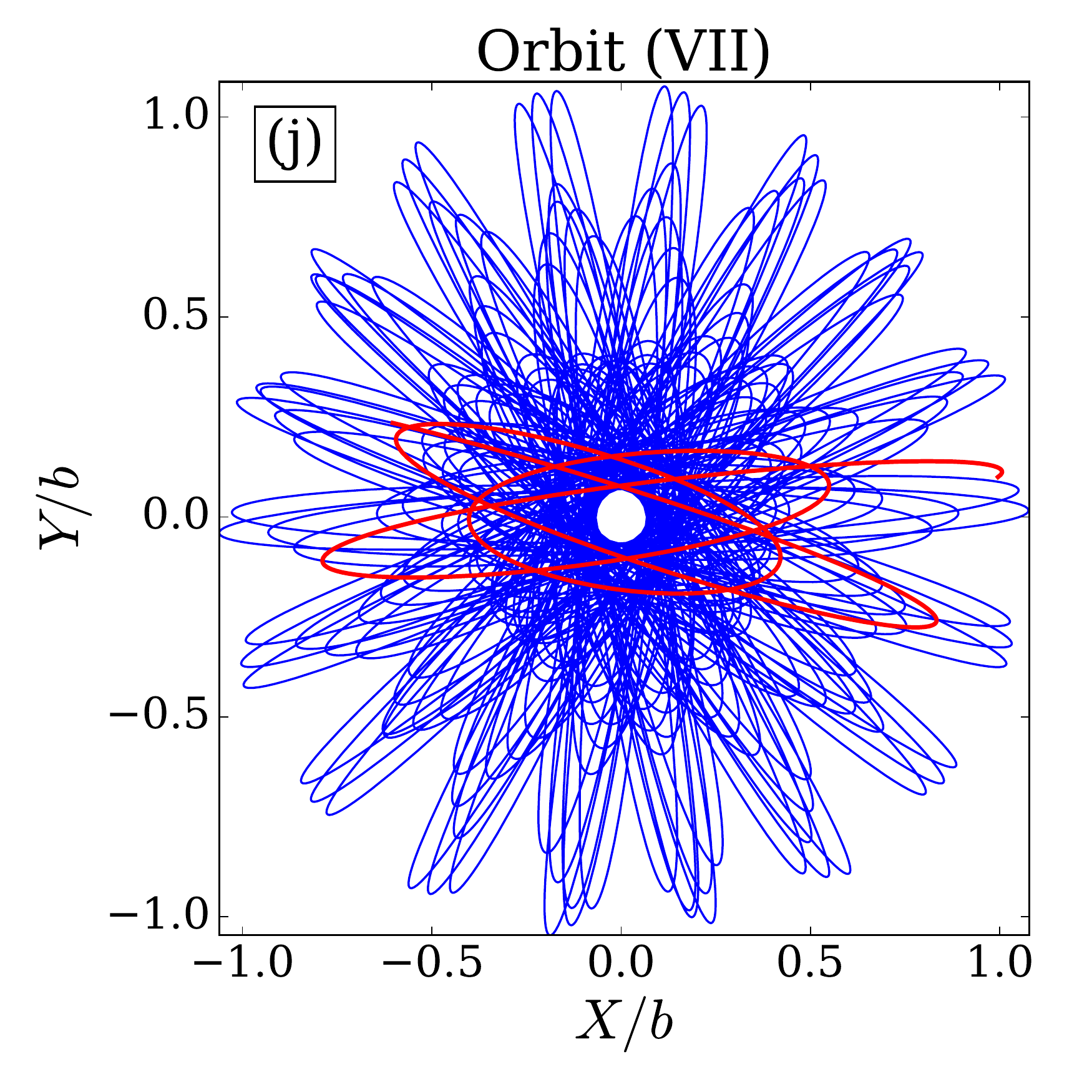}
\includegraphics[width=0.32\linewidth,height=0.32\linewidth]{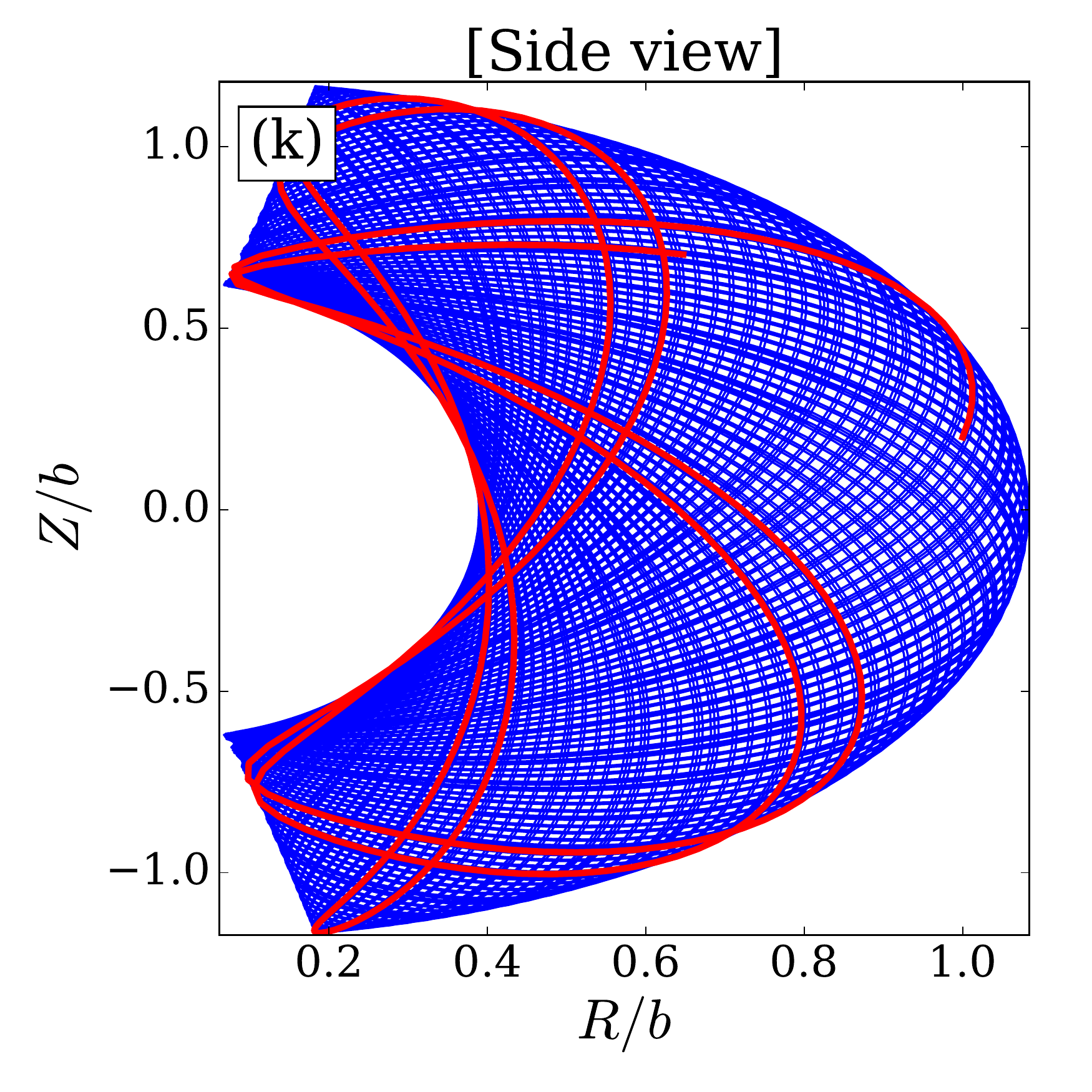}
\includegraphics[width=0.32\linewidth,height=0.32\linewidth]{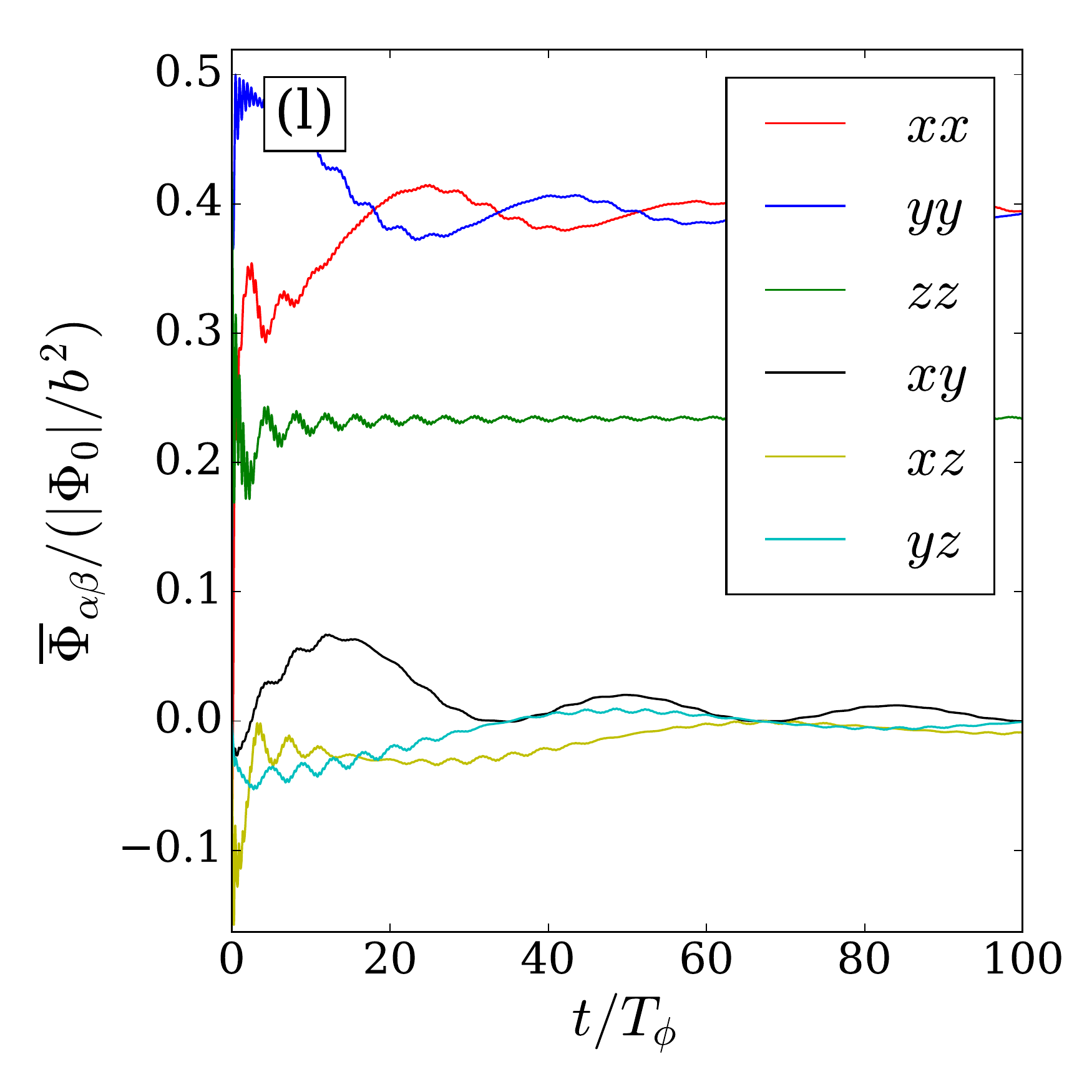}
\caption{Similar to Fig. \ref{convplots2}, but for Orbits (IV)-(VII) demonstrating convergence to the `axisymmetric torus' approximation in non-spherical potentials. Middle panels show the meridional projection of ${\bf R}_g(t)$. Also, the right panels show the convergence of $\Phi_{xz}$ and $\Phi_{yz}$, which are not identically zero for non-coplanar orbits (they should vanish only upon outer orbit averaging). These Orbits with properties listed in Table \ref{FPLtable} were integrated for 100 azimuthal periods in a flattened power law potential \eqref{FPot} with $\beta=1/2$ and $q=0.94$. Initially, the Orbits differ only in their azimuthal velocity $v_{\phi}$, see Table \ref{FPLtable}.
(Note the different axis scales for different orbits). 
}
\label{OPlotsF}
\end{figure*}

This is demonstrated in Figure \ref{OPlotsF}, in which we plot four Orbits (`IV'-`VII') in the flattened power-law potential \eqref{FPot} with $q=0.94$ and $\beta=1/2$.  These four Orbits are initiated with exactly the same initial conditions except for their initial azimuthal velocity $v_\phi$; the full details of the initial conditions, as well as the resulting $A^*_\mathrm{num}$ and $\Gamma_\mathrm{num}$ values, are given in Table \ref{FPLtable}. In the top row of Figure \ref{OPlotsF} we have Orbit (IV), with initial $v_\phi = 1.35\sqrt{GM/b}$.  Orbit (IV) is certainly not planar but the typical excursions in  $Z$ are fairly small compared to the excursions in $R$.  As a result the $\Gamma_\mathrm{num}=0.243$ value is less than $1/3$ but still significantly greater than zero.  As we move down the page we decrease the initial azimuthal velocity each time, so that Orbits (V)-(VII) initially have $v_\phi/\sqrt{GM/b}= 0.95,\, 0.35,\, 0.05$ respectively (while keeping all other initial conditions the same). The radial excursions decrease as the initial azimuthal velocity decreases, until they become comparable to the vertical excursions. Eventually $\Gamma$ moves below zero in Orbit (VII), see Table \ref{FPLtable}.  The $A^*$ values grow as we move down the page since the binary samples a stronger potential when it is closer to the origin.

%%%%%%%%%%%%%%%%%%%%%%%%%%%%%%%%%%%%%%%%%%
%%%%%%%%%%%%%%%%%%%%%%%%%%%%%%%%%%%%%%%%%%

%%%%%%%%%%%%%%%%%%%%%%%%%%%%%%%%%%%%%%%%%%
\begin{table*}
\caption{
\label{IIItable}
Details of the outer orbits $\Rg$ used for numerical verification of the `axisymmetric annulus' approximation in \S\ref{AGamSph}, \S\ref{sect:sph_validity} and Figure \ref{convplots2}. An orbit in a spherical potential (with mass $M$ and scale radius $b$) is uniquely specified by its peri/apocentre distances $r_\mathrm{p/a}$, or equivalently by its generalised semi-major axis $a_\mathrm{g}$ and eccentricity $e_\mathrm{g}$.  We also provide the orbit's azimuthal period $T_\phi$ around the cluster, its analytical $A^*$ and $\Gamma$ values, and the corresponding numerically computed values $A^*_\mathrm{num}, \Gamma_\mathrm{num}$, all to 3 significant figures.
}
\begin{tabular}{| l | l | l | l | l | l | l | l | l | l | l |}
    \hline
    Orbit of $\Rg$ & Potential & $(r_\mathrm{p}/b,r_\mathrm{a}/b)$  & $a_\mathrm{g}/b$ & $e_\mathrm{g}$ & $T_\phi\sqrt{\frac{GM}{b^3}}$ & $A^*$ & $A^*_\mathrm{num}$ & $\Gamma$ & $\Gamma_\mathrm{num}$ \\ \hline
  (I) & $\Phi_\mathrm{iso}$ & $(5.29,11.2)$  & $8.2$  & $0.36$  & $171$ & $0.00106$ & $0.00106$ &  $0.685$ & $0.686$
   \\ \hline
     (II)& $\Phi_\mathrm{iso}$ & $(0.08,1.21)$&$0.65$&$0.88$ & $18.0 $ & $0.253$ & $0.255$ & $0.0676$ & $0.0650$ \\
    \hline
  (III)& $\Phi_\mathrm{Hern}$ & $(0.08 ,1.21 )$&$0.65$&$0.88$ & $11.0 $& $1.15$ & $1.15$ & $0.192$ & $0.195$ \\
    \hline
    \end{tabular}
\end{table*}
%%%%%%%%%%%%%%%%%%%%%%%%%%%%%%%%%%%%%%%%%%

%%%%%%%%%%%%%%%%%%%%%%%%%%%%%%%%%%%%%%%%%%
\begin{table}
   \caption{Properties of Orbits (IV)-(VII) integrated in the flattened power-law potential \eqref{FPot} with $\beta=1/2$ and $q=0.94$ (c.f. Figure \ref{OPlotsF}). The initial conditions of Orbits (IV)-(VII) are identical except for the initial azimuthal velocity $v_\phi$.  We take initial $(R,v_R,\phi,Z,v_Z)=(b,\,0.1\sqrt{GM/b},\,0.1,\,0.2b,\,0.5\sqrt{GM/b})$, and initial $v_\phi$ is given below.}
   \label{FPLtable}
    \begin{tabular}{| l | l | l | l | l | l | l | l | l | l | l |}
    \hline
    Orbit  & Potential & $v_{\phi}/\sqrt{GM/b}$ & $A^*_\mathrm{num}$ & $\Gamma_\mathrm{num}$ 
    \\ \hline
  (IV) & $\Phi_\mathrm{FPL}$ & $1.35$  & $0.0332$ & $0.243$ &  
      \\ \hline
  (V) & $\Phi_\mathrm{FPL}$ & $0.95$  & $0.182$ & $0.192$ & 
      \\ \hline
  (VI) & $\Phi_\mathrm{FPL}$ & $0.35$  & $0.552$ & $0.016$ &  
      \\ \hline
  (VII) & $\Phi_\mathrm{FPL}$ & $0.05$  & $0.626$ & $-0.085$ & 
        \\ \hline
    \end{tabular}
\end{table}
%%%%%%%%%%%%%%%%%%%%%%%%%%%%%%%%%%%%%%%%%%

%%%%%%%%%%%%%%%%%%%%%%%%%%%%%%%%%%%%%%%%%%
\begin{table}
   \caption{Properties of Orbits (VIII)-(XI) integrated in the Miyamoto-Nagai potential \eqref{MNPot} with $b_h/b_\ell=1$ (c.f. Figure \ref{OPlotsMN}). The initial conditions of Orbits (VII)-(XI) are identical except for the initial vertical coordinate $Z$.  We take initial $(R,v_R,\phi,v_\phi,v_Z)=(b_\ell,\,0,\,0,\,0.25\sqrt{GM/b_\ell},\,0)$, and initial $Z$ is given below.}
    \label{MNtable}
    \begin{tabular}{| l | l | l | l | l | l | l | l | l | l | l |}
    \hline
    Orbit  & Potential & $Z/b_\ell$ & $A^*_\mathrm{num}$ & $\Gamma_\mathrm{num}$ 
    \\ \hline
  (VIII) & $\Phi_\mathrm{MN}, \, b_h/b_\ell=1$ & $0.1$  & $0.256$ & $0.153$ &  
    \\ \hline
  (IX) & $\Phi_\mathrm{MN}, \, b_h/b_\ell=1$ & $1.0$  & $0.122$ & $0.042$ & 
    \\ \hline
      (X) & $\Phi_\mathrm{MN}, \, b_h/b_\ell=1$ & $2.0$  & $0.0392$ & $-0.163$ &  
    \\ \hline
  (XI) & $\Phi_\mathrm{MN}, \, b_h/b_\ell=1$ & $3.0$  & $0.0140$ & $-0.384$ & 
    \\ \hline
\end{tabular}
\end{table}
%%%%%%%%%%%%%%%%%%%%%%%%%%%%%%%%%%%%%%%%%%

%%%%%%%%%%%%%%%%%%%%%%%%%%%%%%%%%%%%%%%%%%
\begin{figure*}
\centering
\includegraphics[width=0.32\linewidth,height=0.32\linewidth]{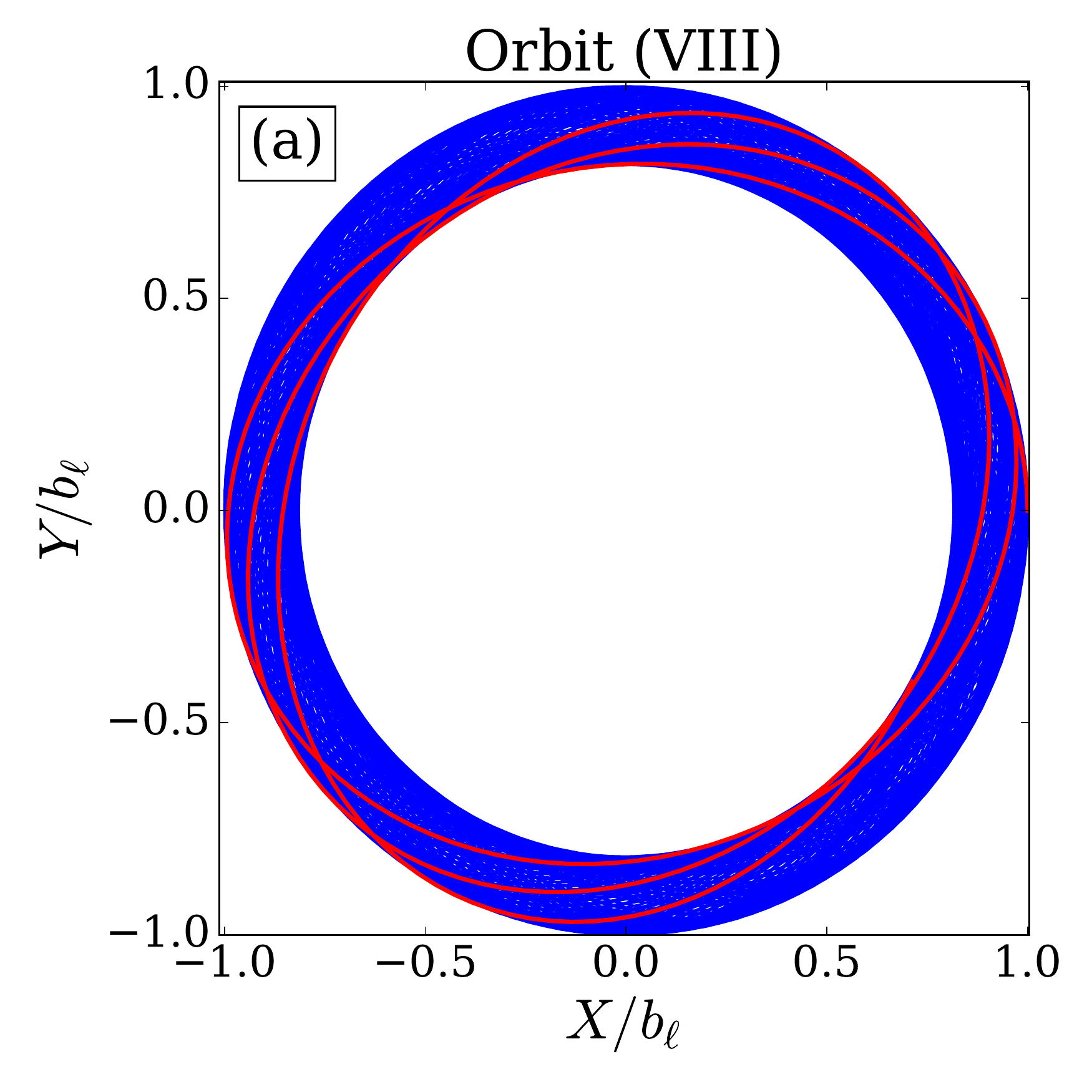}
\raisebox{0.1cm}{\includegraphics[width=0.32\linewidth,height=0.31\linewidth]{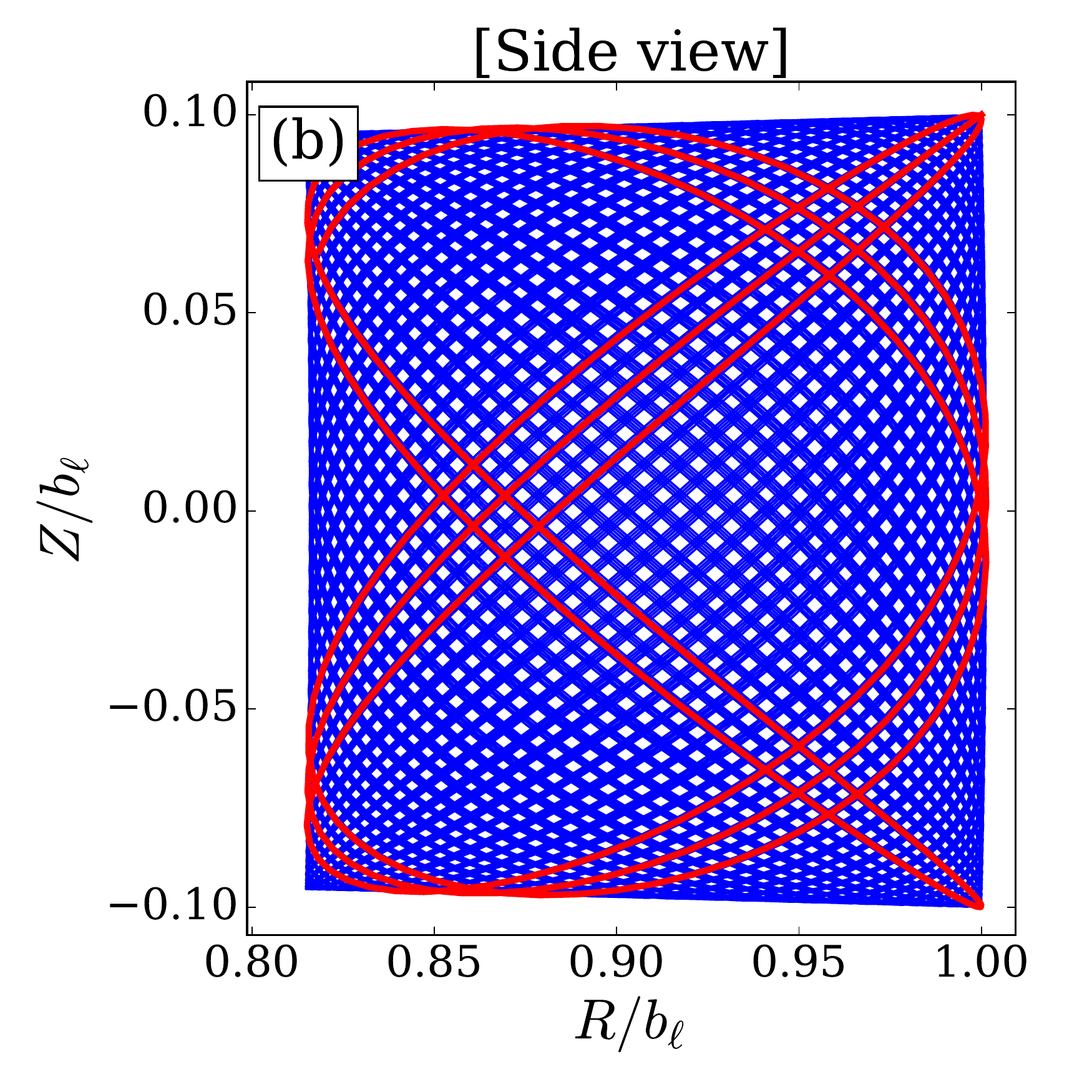}}
\raisebox{0.1cm}{\includegraphics[width=0.32\linewidth,height=0.3\linewidth]{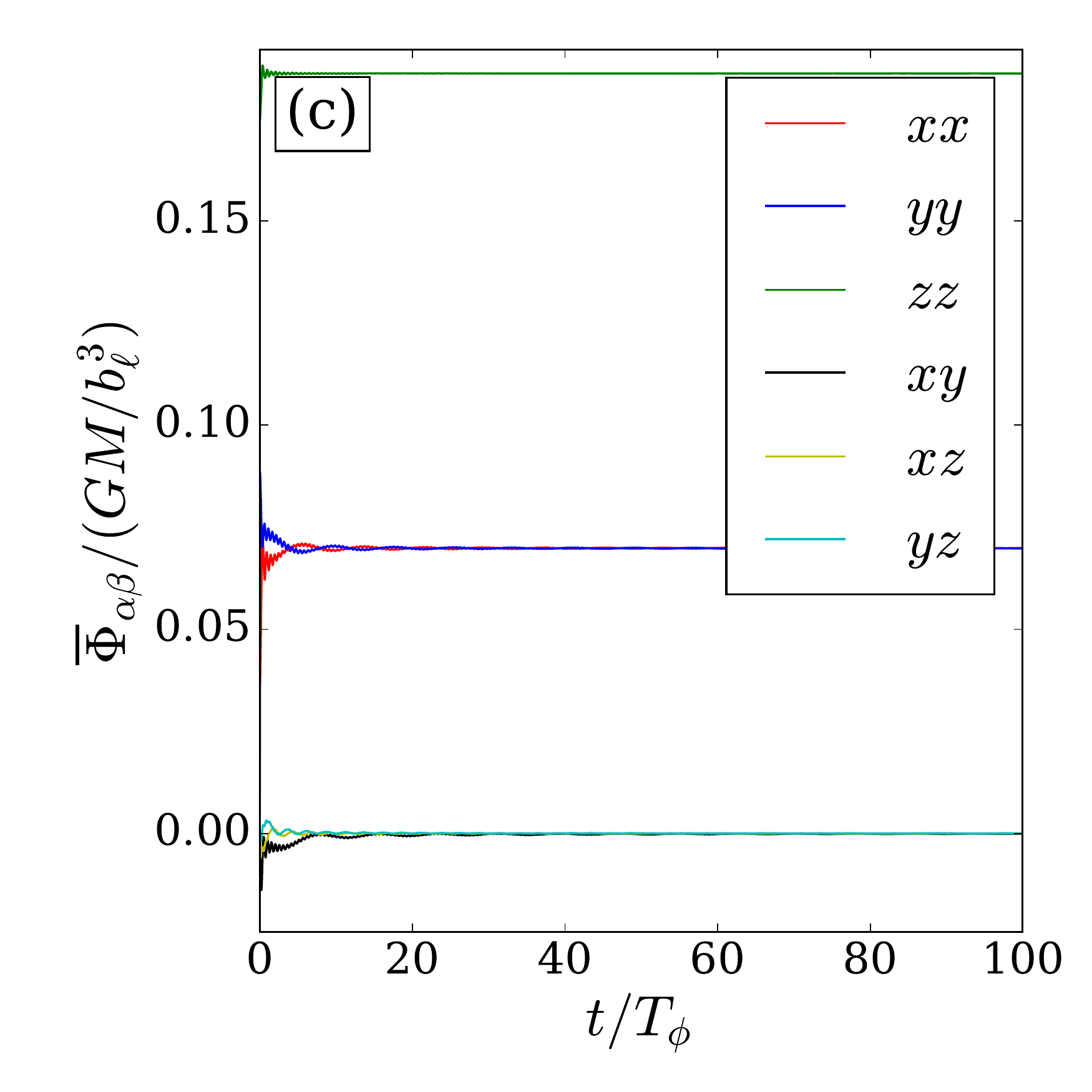}}
\includegraphics[width=0.32\linewidth,height=0.32\linewidth]{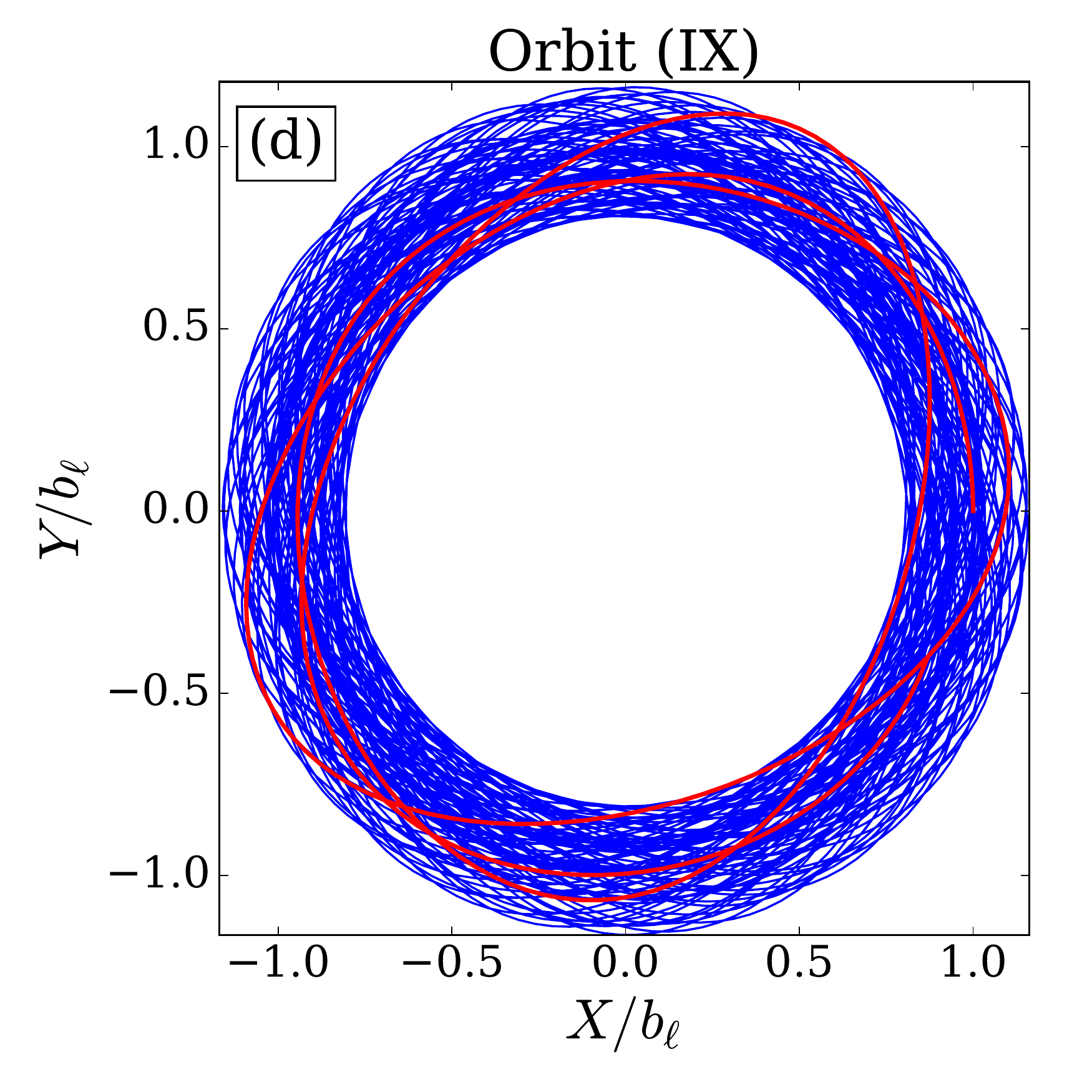}
\raisebox{0.1cm}{\includegraphics[width=0.32\linewidth,height=0.31\linewidth]{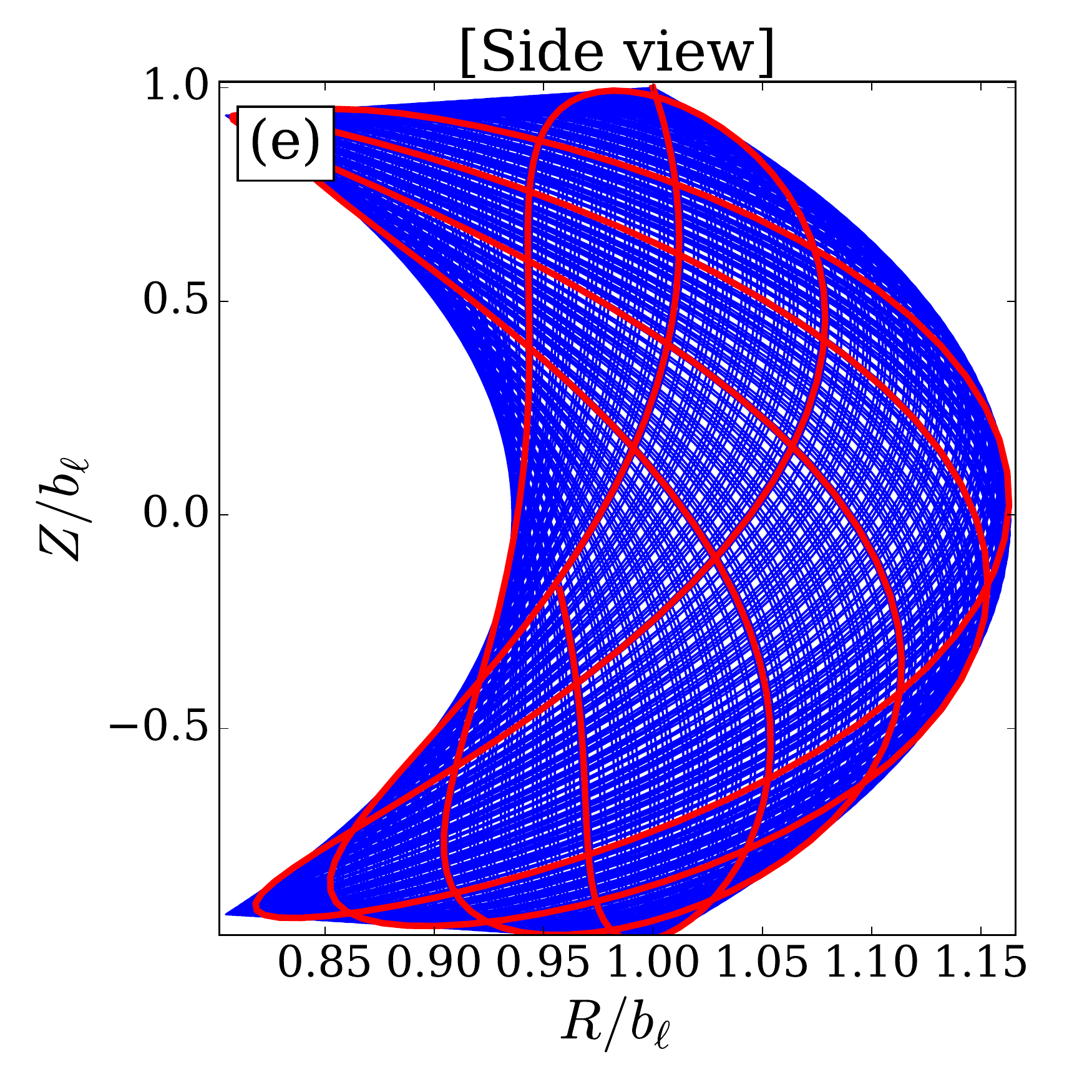}}
\raisebox{0.1cm}{\includegraphics[width=0.32\linewidth,height=0.3\linewidth]{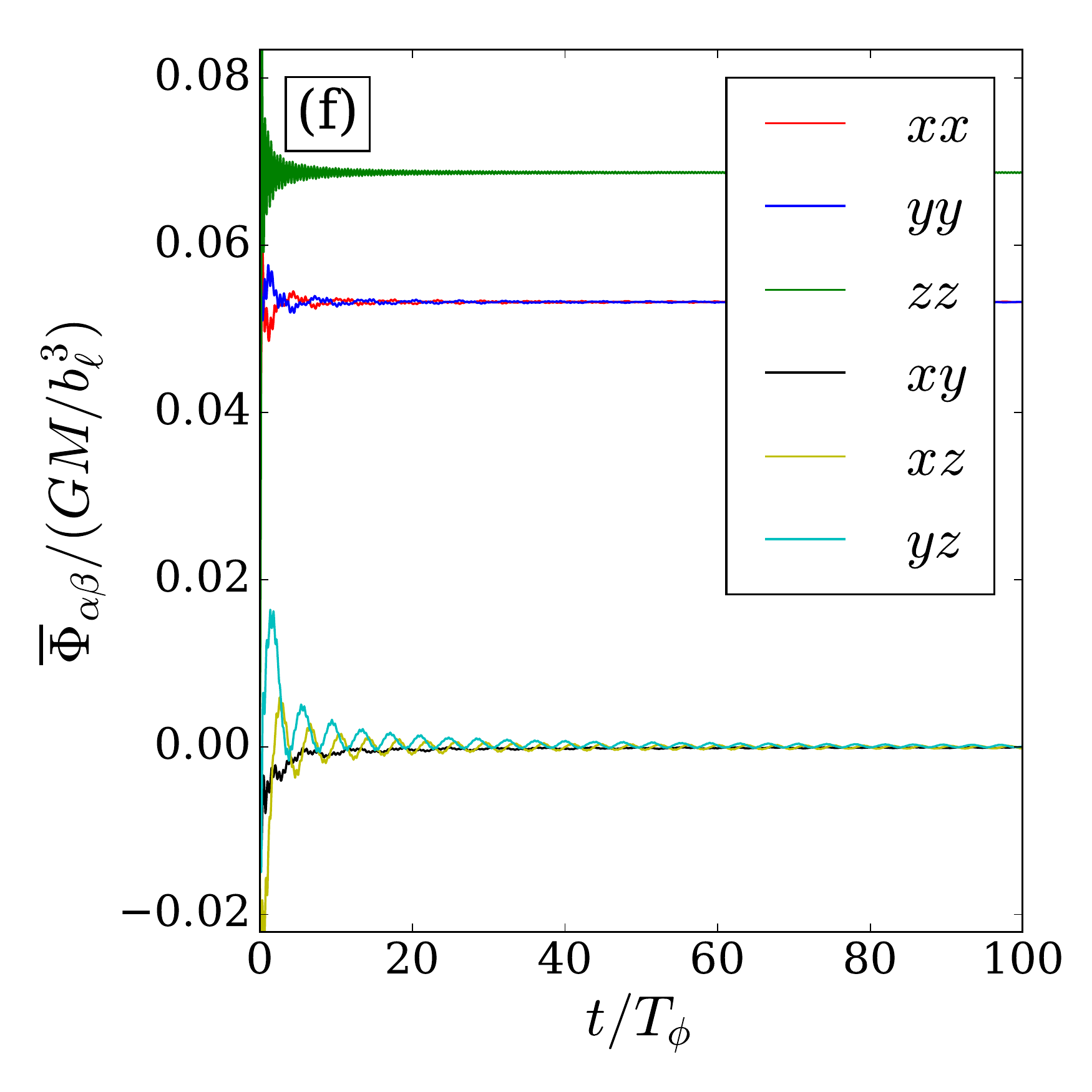}}
\includegraphics[width=0.32\linewidth,height=0.32\linewidth]{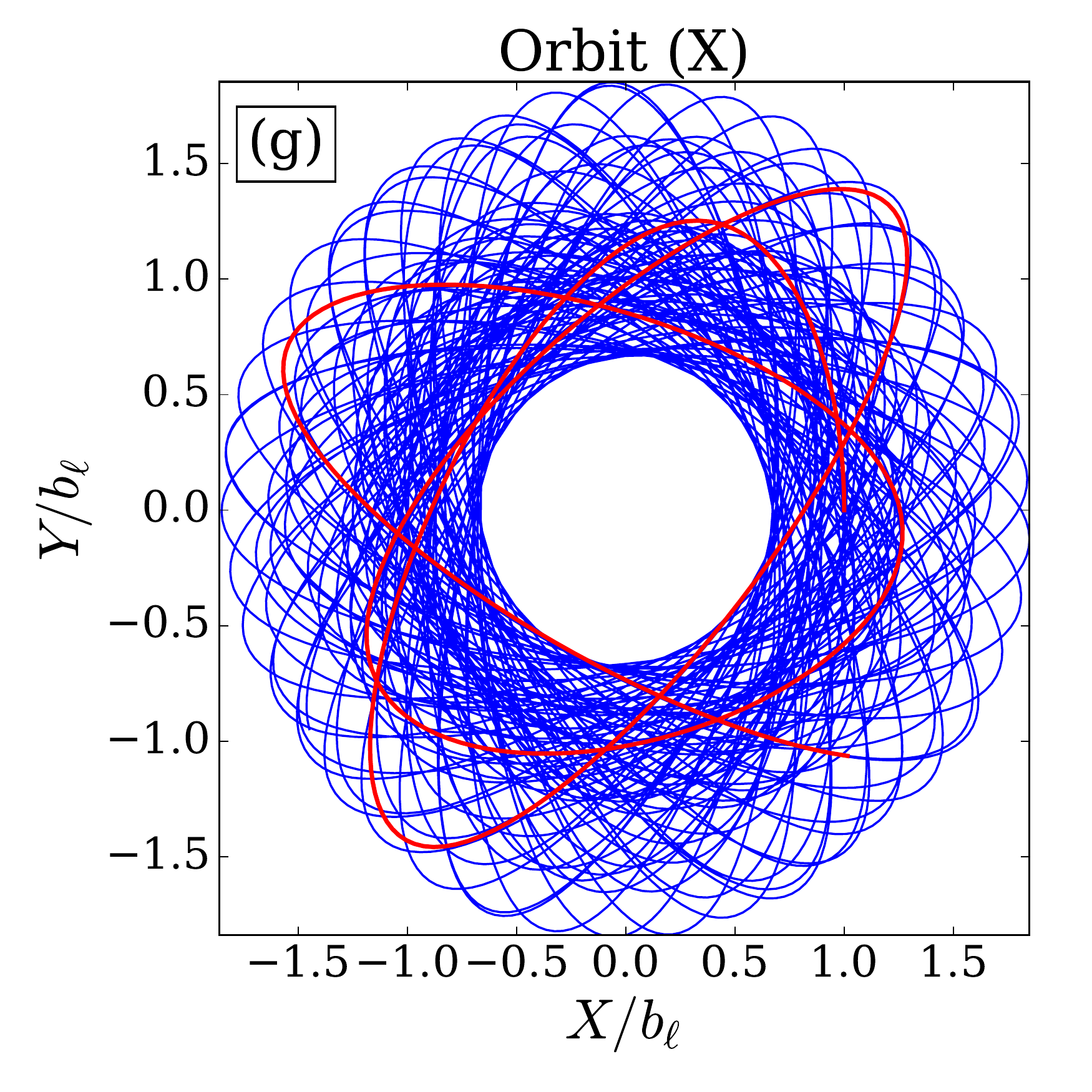}
\raisebox{0.1cm}{\includegraphics[width=0.31\linewidth,height=0.31\linewidth]{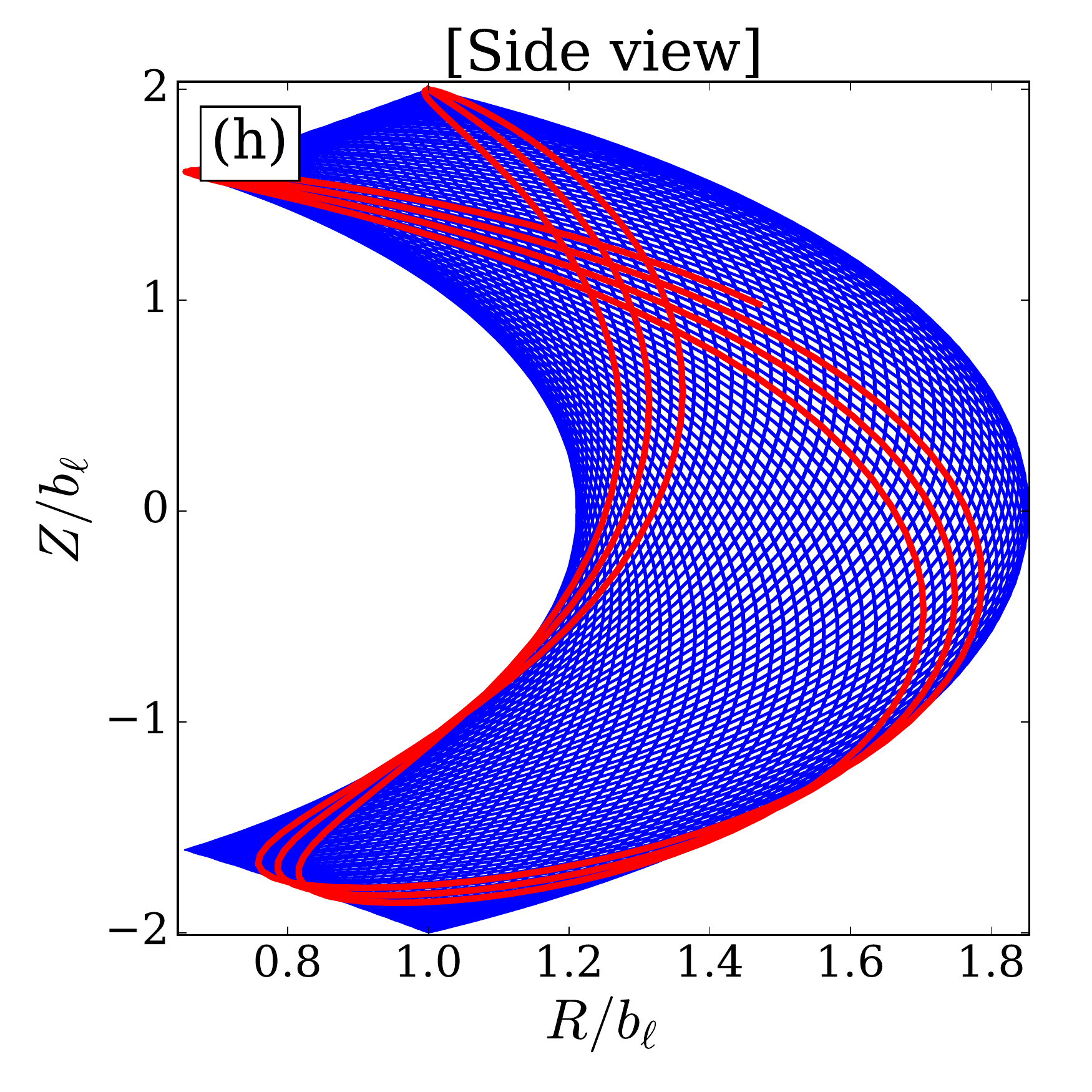}}
\raisebox{0.1cm}{\includegraphics[width=0.32\linewidth,height=0.3\linewidth]{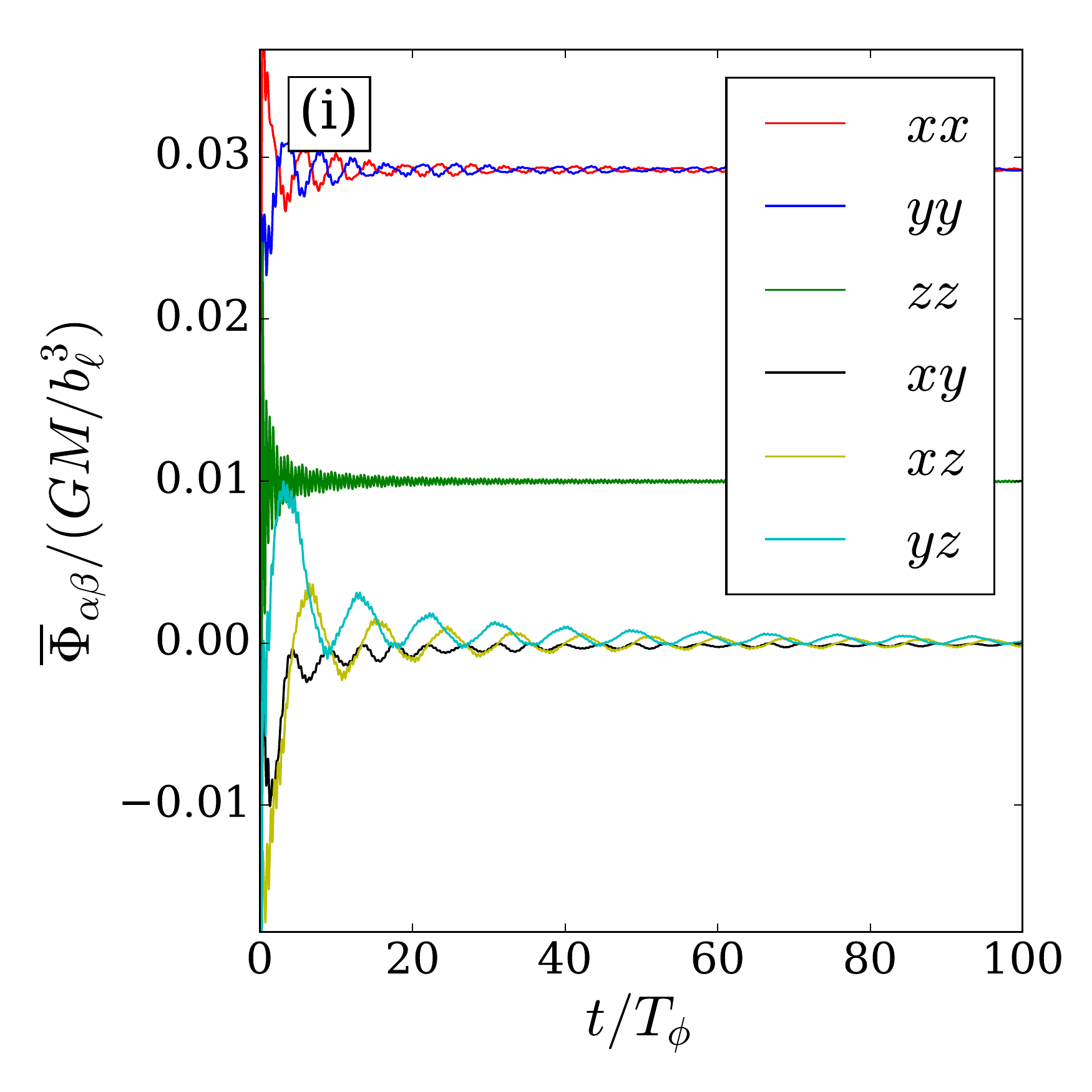}}
\includegraphics[width=0.32\linewidth,height=0.32\linewidth]{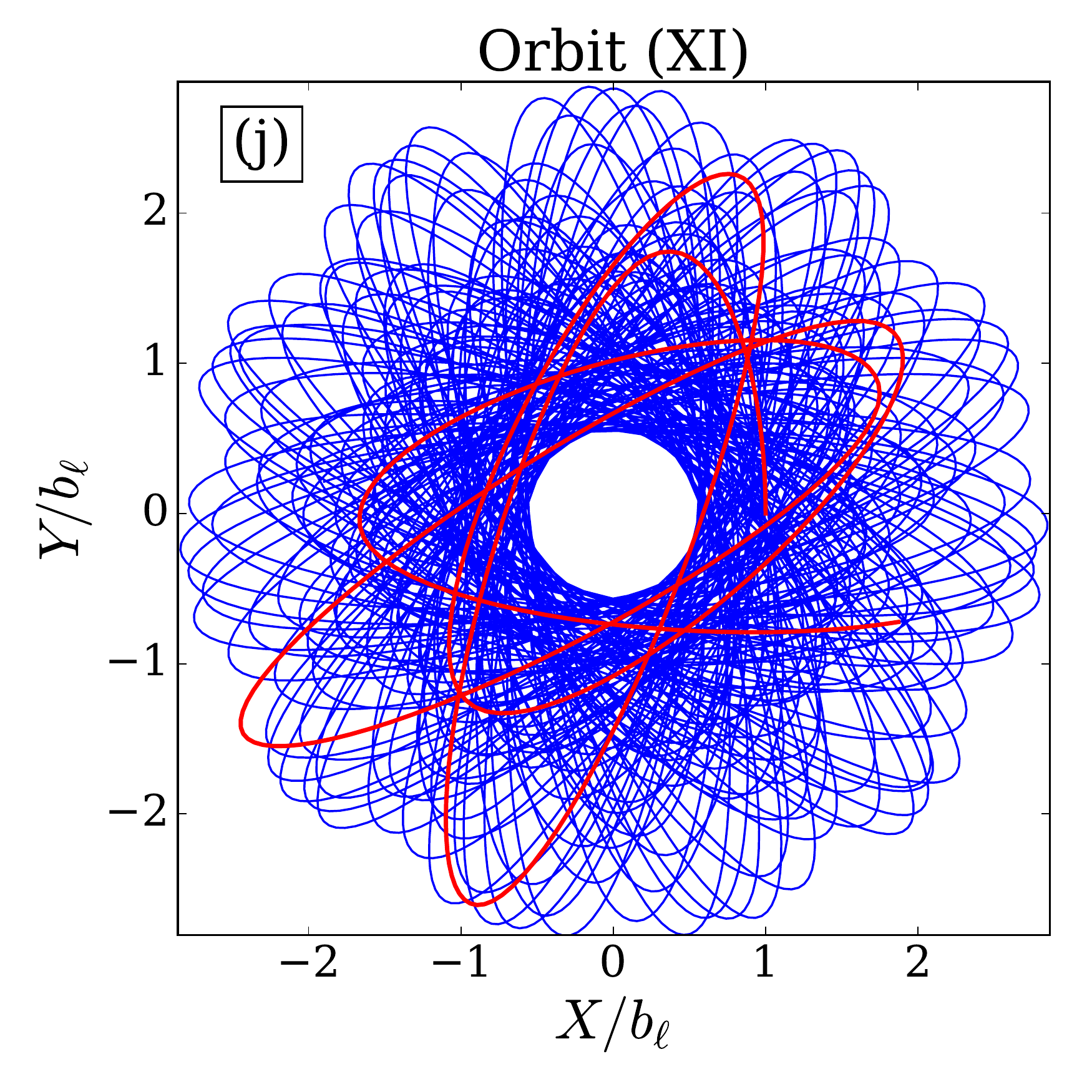}
\raisebox{0.1cm}{\includegraphics[width=0.3\linewidth,height=0.31\linewidth]{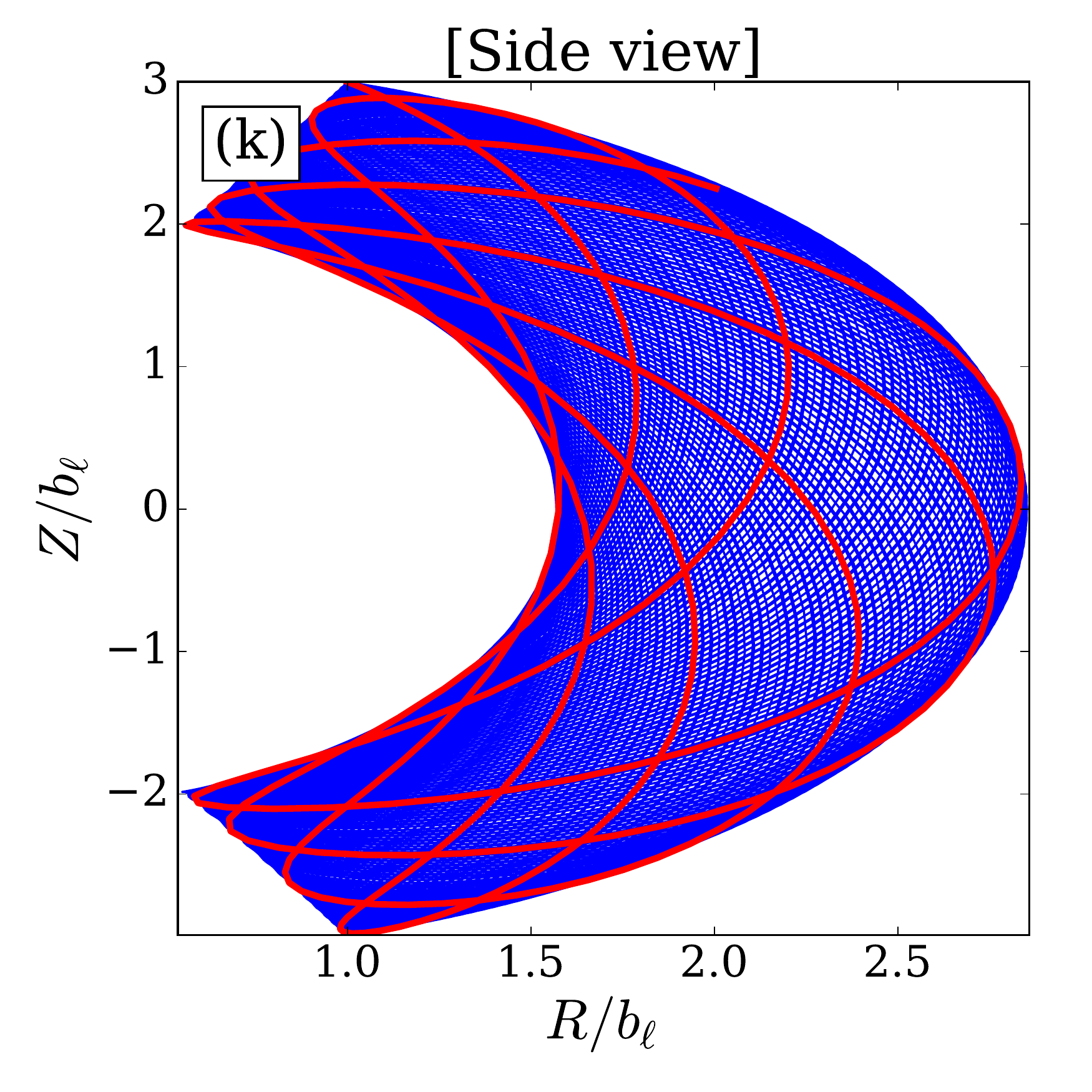}}
\raisebox{0.1cm}{\includegraphics[width=0.32\linewidth,height=0.3\linewidth]{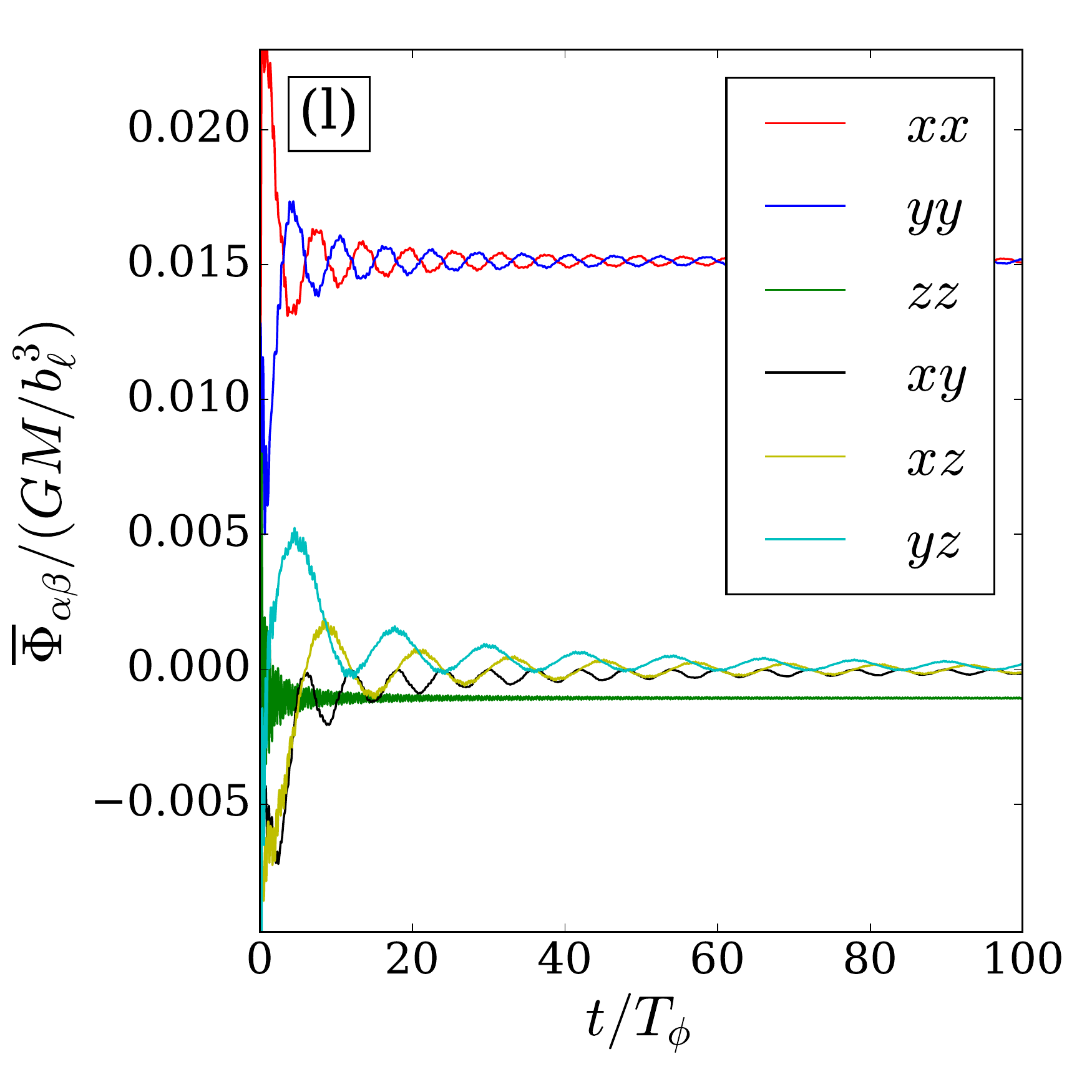}}
\caption{Similar to Figure \ref{OPlotsF}, but now for Orbits (VIII)-(XI) with characteristics given in Table \ref{MNtable} integrated in a Miyamoto-Nagai potential \eqref{MNPot} with $b_h/b_\ell=1$. These orbits differ only in their initial vertical coordinate $Z$, with initial $Z/b_\ell = 0.1, 1.0, 2.0, 3.0$ respectively.  All panels show 100 azimuthal periods' worth of data. (Note the drastically changing vertical scale in the central column, particularly panel (b)).}
\label{OPlotsMN}
\end{figure*}
%%%%%%%%%%%%%%%%%%%%%%%%%%%%%%%%%%%%%%%%%%
	
For our final set of examples we compare four more Orbits (`VIII'-`XI') in the Miyamoto-Nagai potential \eqref{MNPot} with $b_h/b_\ell=1$, the intial conditions for which are given alongside their resulting $A^*_\mathrm{num}, \Gamma_\mathrm{num}$ values in Table \ref{MNtable}.  At large distances this potential is significantly flatter than the $q=0.94$ flattened power-law potential. Orbits (VIII)-(XI) are plotted in the left hand column of Figure \ref{OPlotsMN}.  Each Orbit has exactly the same initial conditions except we change the initial vertical coordinate $Z$, using $Z = 0.1b_\ell,\, b_\ell,\, 2b_\ell$ and $3b_\ell$ respectively.  Increasing the initial $Z$ thickens the orbit.  Orbit (VIII) (top row, initial $Z=0.1b_\ell$) is almost coplanar and has $\Gamma_\mathrm{num}=0.153$.  The $\Gamma_\mathrm{num}$ value decreases as move down the page, reaching a minimum value of $\Gamma_\mathrm{num}=-0.384$ for Orbit (XI) (bottom row, $Z=3b_\ell$) which is thicker vertically than it is radially.  Meanwhile, as we move from top to bottom the $A_\mathrm{num}^*$ value decreases because the orbit spends more time away from the midplane where the tidal potential is strongest. An even more extreme orbit in this potential, with the same initial conditions except $Z=4 b_\ell$ (resulting in $\Gamma_\mathrm{num}=-1.4$), is presented in Figure \ref{fig:App_D_Ex1}. 

Finally, note that Orbit (VIII) is very similar in appearance to the orbit in Figure \ref{DiskPlots}a: both are roughly epicyclic, so they should obey equations \eqref{epiA} and \eqref{epiGam}. The difference between them is that in the case of Orbit (VIII) the potential felt by the binary is significantly less flattened, since this Orbit resides predominantly in the quasi-spherical core of the potential ($b_h/b_\ell=1$). As a result, $\Phi_{zz}$ does not dominate over $\Phi_{xx}$ and hence $\Gamma$ ends up being significantly less than $1/3$ (unlike the orbit in Figure \ref{DiskPlots}a which explores much more flattened version of the Miyamoto-Nagai potential with $b_h/b_\ell=0.05$).

%%%%%%%%%%%%%%%%%%%%%%%%%%%%%%%%%%%%%%%%%%
%%%%%%%%%%%%%%%%%%%%%%%%%%%%%%%%%%%%%%%%%%
\section{Validity of secular Hamiltonian}	 
\label{numveravg}

%%%%%%%%%%%%%%%%%%%%%%%%%%%%%%%%%%%%%%%%%%

In \S \ref{abgamma} we focused on understanding the typical values of $A, \Gamma$ for various types of orbit in different potentials.  However the Hamiltonian \eqref{H1Mt} is only valid if the symmetry conditions (\ref{phixxyy})-(\ref{phiyzxz}) for the time-averages $\overline{\Phi}_{\alpha\beta}$ are satisfied.  In addition it is reasonable to require that $\overline{\Phi}_{\alpha\beta}$ converge to fixed values (say to within a few percent) on timescales significantly shorter than the timescale for secular evolution (which will be derived in Paper II).  In this section we check the validity of these assumptions numerically.  The procedure for calculating $\overline{\Phi}_{\alpha\beta}$ numerically is given in Appendix \ref{Numerical_Time_Averages}.

%%%%%%%%%%%%%%%%%%%%%%%%%%%%%%%%%%%%%%%%%%

\subsection{Spherical potentials}
\label{sect:sph_validity}

%%%%%%%%%%%%%%%%%%%%%%%%%%%%%%%%%%%%%%%%%%

In a spherical potential, orbits $\Rg(t)$ that have non-commensurable frequencies densely fill an axisymmetric annulus.  If this is true then the time-averaged coefficients $\overline{\Phi}_{\alpha \beta}$ should obey the symmetry properties (\ref{phixxyy}), (\ref{phixy}). 

To verify this we use Orbits (I)-(III).  Their initial conditions are given alongside their $A^*, A^*_\mathrm{num}, \Gamma, \Gamma_\mathrm{num}$ values in Table \ref{IIItable}.  The right panels in Figure \ref{convplots2} show the corresponding running average (from $t=0$ to current time) of numerically computed $\overline{\Phi}_{\alpha \beta}$.  As the number of completed orbits grows, the time-averaged derivatives of the potential tend to converge towards fixed values.  

Orbit (I) has rather large semi-major axis and small eccentricity, so that it stays far from the core at all times, filling its annulus densely.  The `axisymmetric annulus' approximation works very well in this case, so the (semi-)analytic and numerically computed values agree: $A^* = A^*_\mathrm{num}$ and $\Gamma = \Gamma_\mathrm{num}$ to within $1\%$ accuracy. 

We have picked rather extreme examples in Orbits (II) and (III) in order to demonstrate behaviour of orbits $\Rg$ that are both very radial and tightly bound near the centre of the cluster.  Orbit (II) spends a lot of time in the isochrone potential's constant density core where its frequencies are almost commensurable ($\Omega_R \approx 2\Omega_\phi$); as a result it precesses slowly, so that there are unfilled gaps left in its annulus even after $t=100 T_\phi$. This issue does not arise in the uncored Hernquist potential, so Orbit (III) fills its annulus more efficiently than Orbit (II). Nevertheless, the axisymmetric approximation is still very successful in both cases, with a maximum discrepacy of $\sim4\%$ arising between the $\Gamma$ and $\Gamma_\mathrm{num}$ values of Orbit (II).  

However, we notice that while the converged symmetry properties of the $\overline{\Phi}_{\alpha \beta}$ (see equations (\ref{phixxyy}), (\ref{phixy})) are well established after $\gtrsim 15T_\phi$ for Orbits (I) and (III), they are less well established for Orbit (II) even at $t\gtrsim 45T_\phi$.  Again this is because Orbit (II) does not fill its annulus efficiently.  This can be problematic because if the $\overline{\Phi}_{\alpha \beta}$ fail to converge on a timescale shorter than the secular evolution timescale, the doubly-averaged theory can break down, as we will see in Paper II.

%%%%%%%%%%%%%%%%%%%%%%%%%%%%%%%%%%%%%%%%%%

\subsection{Axisymmetric potentials}

%%%%%%%%%%%%%%%%%%%%%%%%%%%%%%%%%%%%%%%%%%

In a (non-spherical) axisymmetric potential, orbits $\Rg(t)$ that have non-commensurable frequencies densely fill an axisymmetric torus and so the time-averaged coefficients $\overline{\Phi}_{\alpha \beta}$ should obey the symmetry properties (\ref{phixxyy}), (\ref{phixy}), and (\ref{phiyzxz}). To verify this numerically we use Orbits (IV)-(VII) in the flattened power-law potential \eqref{FPot} with $q=0.94$ and $\beta=1/2$ (see Table \ref{FPLtable} and Figure \ref{OPlotsF}), and Orbits (VIII)-(XI) in the Miyamoto-Nagai potential \eqref{MNPot} with $b_h/b_\ell=1$ (see Table \ref{MNtable} and Figure \ref{OPlotsMN}).

Some features of the $\overline{\Phi}_{\alpha\beta}$ convergence plots are similar to the spherical case.  For example, in Figure \ref{OPlotsF} the derivatives $\overline{\Phi}_{\alpha\beta}$ converge rather slowly in the bottom panel because the Orbit (VII) fills its torus rather sparsely, owing to the large fraction of time it spends in the almost-harmonic potential of the core.  

Note that the rightmost columns in Figures \ref{OPlotsF} and \ref{OPlotsMN} also show the convergence of $\Phi_{xz}$ and $\Phi_{yz}$. This is different from Figure \ref{convplots2} since now we are dealing with non-planar orbits so that these derivatives are no longer identically zero. Although the corresponding time-averages $\overline{\Phi}_{xz}, \overline{\Phi}_{yz}$ do indeed converge to zero in all of our axisymmetric examples as expected, in most cases their convergence takes significantly longer than that of the other $\overline{\Phi}_{\alpha\beta}$ coefficients.  This is what we would expect from looking at the $\phi_\mathrm{g}$ dependence of equations \eqref{phixxnon}-\eqref{phiyznon}: the derivatives $\Phi_{xx}$, $\Phi_{yy}$ and $\Phi_{xy}$ fluctuate twice as rapidly with respect to $\phi_\mathrm{g}$ compared to $\Phi_{xz}, \Phi_{yz}$. Slower convergence of $\overline{\Phi}_{xz}$ and $\overline{\Phi}_{yz}$ seems to be especially apparent for strongly non-coplanar orbits, i.e. orbits which make large excursions in the $Z$ direction.  Orbits that inefficiently fill their torus (i.e. on timescales longer than the secular evolution timescale) can render the doubly-averaged theory inaccurate, as discussed in detail in \S7 of Paper II.

%%%%%%%%%%%%%%%%%%%%%%%%%%%%%%%%%%%%%%%%%%
%%%%%%%%%%%%%%%%%%%%%%%%%%%%%%%%%%%%%%%%%%

\section{Discussion}	
\label{sect:disc}

%%%%%%%%%%%%%%%%%%%%%%%%%%%%%%%%%%%%%%%%%%
	
In deriving the secular Hamiltonian (equations \eqref{H1Mt}, \eqref{H1Star}) we relied on several approximations. First, we assumed that the outer orbit-averaging procedure used for computing potential derivatives $\overline{\Phi}_{\alpha\beta}$ converges rapidly when compared to the timescale for secular evolution. The rate of convergence of various $\overline{\Phi}_{\alpha\beta}$ components was explored in \S \ref{numveravg}. In Paper II we will use direct numerical integrations of binaries orbiting in stellar cluster potentials to study how well this double averaging procedure works in practice.

Second, we truncated our expansion of the tidal Hamiltonian in \S \ref{sect:tidalH} at the quadrupole order. However, studies of the LK mechanism have shown the importance of higher order --- `octupole' --- terms for the dynamics of triples in certain situations \citep{Lithwick2011,Li2014}. This raises a question of whether octupole terms can be important for the secular dynamics of binaries in external tidal fields. While we derive octupole-level corrections to our doubly-averaged Hamiltonian in Appendix \ref{Higher_Order}, in practice they are unlikely to be important. This is because in realistic situations the ratio of the semi-major axis of the inner binary orbit ($a\lesssim 100$ AU) is much smaller than the size of its outer orbit ($\vert \Rg \vert \sim 1$ pc, comparable to the cluster size), rendering the timescale on which octupole-level effects may manifest themselves too long (we will see in Paper II that a characteristic timescale of secular evolution driven by quadrupole terms in a typical globular cluster is at least tens of Myrs). 

Third, our calculation assumes a spatially smooth and time-invariant tidal potential. This approximation neglects the granularity and stochastic variability of the cluster potential caused by encounters with other stars, which are very important in dense environments of clusters \citep{Heggie1975,Hut1983a,Hut1983b,Heggie1996,Collins2008}. The cumulative effect of a large number of such encounters is what eventually contributes to the smooth tidal field of the cluster \citep{Collins2010}; thus, one hopes that in the long run our framework should provide a qualitatively accurate picture of binary evolution in clusters. Nevertheless, in the future we plan to extend our calculation by including the effects of individual stellar encounters on evolution of the binary inner orbit \citep{Weinberg1987}. We will also explore the role of strong encounters (responsible for the formation and disruption of  binaries in clusters, \citealt{Heggie1975}) in resetting the whole course of secular evolution of the binary.

An effect that can modify the binary's \textit{outer} orbit in a stellar cluster is resonant relaxation.  \citet{Rauch1996} showed that in quasi-Keplerian systems (such as nuclear clusters dominated by a central super-massive black hole) angular momentum is efficiently exchanged between stellar orbits that have commensurable frequencies. When applied to a binary in a quasi-Keplerian cluster, precession of the binary's outer orbit due to resonant interactions with other stars (so-called `vector resonant relaxation') can alter its inclination relative to the inner orbital plane, potentially bringing an initially low-inclination binary into a high-inclination regime and triggering LK oscillations \citep{Hamers2018a}.  While this effect has not been explored for binaries in non-Keplerian potentials, vector resonant relaxation can indeed operate in non-Keplerian systems such as globular clusters \citep{Meiron2018b}, as can more general forms of resonant relaxation allowing for exchange of both angular momentum and energy of the outer orbit \citep{Hamilton2018}.  

Additionally, the fact that a binary is typically heavier than the average star in a cluster means that it will tend to sink towards the centre of the cluster via dynamical friction \citep{Binney2008}. Moreover, the global properties of the cluster itself may evolve as a result of two-body relaxation leading to core collapse. All of these effects can change the values of $A$ and $\Gamma$ for a given binary over long time intervals.  We will explore their impact upon binary evolution in future work.

Finally, an important assumption that lies at the foundation of our time-averaging procedure is that different frequencies characterizing binary motion in the cluster are not commensurable with each other (see \S \ref{Non-Com}). If this condition is violated, the outer orbit $\Rg$ no longer fills an axisymmetric torus inside the cluster uniformly in azimuth, rendering the equations (\ref{phixxyy})-(\ref{phiyzxz}) invalid. This issue is addressed in more detail next.

%%%%%%%%%%%%%%%%%%%%%%%%%%%%%%%%%%%%%%%%%%

\subsection{Commensurable frequencies} 

%%%%%%%%%%%%%%%%%%%%%%%%%%%%%%%%%%%%%%%%%%

Orbits in realistic spherical potentials obey the following relation between the radial ($\Omega_R$) and azimuthal ($\Omega_\phi$) frequencies \citep{Binney2008}:
\begin{align} 
1/2 \leq \Omega_\phi/\Omega_R \leq 1.
\end{align}
Thus, in spherical potentials there are infinitely many rational values of $\Omega_\phi/\Omega_R$ in the interval $(1/2, 1)$. However, even the orbits with rational values of $\Omega_\phi/\Omega_R = p/q$ should still be described (at least roughly) by the `filled annulus' approximation as long as the integers $p,q\gg 1$. For these systems, equation \eqref{H1Mt} should be approximately valid. 

Of course, a small number of low-order resonant points in the $(a_\mathrm{g},e_\mathrm{g})$ plane will have  $\Omega_\phi/\Omega_R=p/q$ with integer $p,q \sim $ a few. Binaries on these outer orbits will not satisfy the axisymmetric averaging approximation. In particular, $\Omega_\phi/\Omega_R$ is always rational in the harmonic $(\Omega_\phi/\Omega_R \equiv 1/2)$ and Keplerian $(\Omega_\phi/\Omega_R \equiv 1)$ potentials.  If the potential is purely harmonic we can treat it as in \S \ref{harmonicpot}.  Moreover, we show in Appendix \ref{RecoverLK} that Keplerian potentials are perfectly described by our doubly-averaged formalism with $\Gamma=1$.  However it should be stressed that we have \textit{not} used the axisymmetric averaging approximation in either of these cases: the harmonic potential just happens to be effectively axisymmetric after single-averaging, and the Keplerian potential is known to be axisymmetric under double averaging to quadrupole order.  In neither case do orbits `fill their annulus'\footnote{The Keplerian potential is not axisymmetric to octupole order, see Appendix \ref{Higher_Order}.}.

Another problematic case is when the binary experiences a potential that is \textit{almost} harmonic or \textit{almost} Keplerian, so that the outer orbit precesses apsidally, but not quickly enough to fill a circular annulus on a secular timescale and thereby qualify for an axisymmetric treatment. For example, orbits that spend a lot of time in a constant-density core of the cluster potential experience an \textit{almost} harmonic potential and so tend to fill their annulus very slowly (see Figure \ref{convplots2}c).

%%%%%%%%%%%%%%%%%%%%%%%%%%%%%%%%%%%%%%%%%%
 
\subsection{Relation to previous work}

%%%%%%%%%%%%%%%%%%%%%%%%%%%%%%%%%%%%%%%%%%

Many previous studies have looked at secular evolution of binaries perturbed by external potentials. The effect of an arbitrary quadrupole perturbation upon a binary has been briefly considered by \citet{Mikkola2006}.  In particular, their equation (20) gives the quadrupole potential experienced by a binary in a star cluster consisting of a large number of point masses $m_k$. Our perturbing Hamiltonian $H_1$ is recovered from their result in the mean-field limit (i.e. by replacing the exact potential of the cluster, $-G\sum_k m_k/\vert \mathbf{R}-\mathbf{R}_k\vert$, with the smooth potential $\Phi(\mathbf{R})$).  However, \citet{Mikkola2006} did not explicitly convert to orbital elements, perform any averaging, or develop any secular theory as we do here. In a similar vein, a short paper by \citet{Katz2011b} considered the secular dynamics of a binary perturbed by a generic quadratic potential and included axisymmetric potentials as a special case.  They \textit{did} convert to orbital elements but did not go much further; in particular they did not provide any prescription for computing the coefficients of the averaged perturbing potential.

Studies of tidal effects of the Galactic disk on wide binaries \citep{Heisler1986,Byl1986,Yabushita1989} represent an important limit ($\Gamma\to 1/3$) of our general theory, see \S \ref{sect:epi} and Appendix \ref{RecoverHT}. Since HT86, Galactic tides have been included in many studies of cometary orbits (e.g. \citealt{Matese1989,Matese1996,Breiter1996,Wiegert1999,Brasser2001,Fouchard2004,
Fouchard2005,Fouchard2006,Breiter2007}), as well as planetesimal orbits (e.g. \citealt{Higuchi2007}). 

\citet{Veras2013a} considered a very general form of the perturbed two-body problem, allowing for both position- and velocity-dependent tidal forces to act upon the binary.  Their equations (25)-(29) are more general versions of our singly-averaged equations (c.f. our singly-averaged Hamiltonian \eqref{H1M}), and our equations are recovered if one sets the velocity-dependent forces to zero.  However they did not derive any analogues of our doubly-averaged equations. \citet{Veras2013b} noted that Galactic forces may impact the evolution of exoplanetary systems around stars near the bulge of the Galaxy where the Galactic tide is much stronger than it is in the Solar neighbourhood. 

Another interesting and obvious limit of our theory, $\Gamma=1$ --- which is, however, rather distinct, see \S \ref{linkLK} and Appendix \ref{RecoverLK} --- has been explored in numerous studies of Lidov-Kozai dynamics \citep{Lidov1962,Kozai1962,Fabrycky2007,Naoz2016} and its extensions. One interesting extension to the LK problem was made by \citet{Petrovich2017}, who considered the effect of an axisymmetric (non-spherical) nuclear cluster potential on compact-object binaries that are themselves orbiting a central super-massive black hole (SMBH). The non-spherical part of the cluster potential was considered to drive nodal precession of the binary's quasi-Keplerian outer orbit around the SMBH (continuously changing the relative inclination in the triple system composed of the binary and SMBH, which is important for the operation of LK cycles in this sub-system), while the dominant spherical part drove apsidal precession of the outer orbit. Our doubly-averaged formalism covers this problem in the case where the characteristic timescales for nodal and apsidal precession of the outer orbit are much shorter than the secular timescale, so that the outer orbit fills its torus. Our singly-averaged equations cover it in all cases. However, unlike \citet{Petrovich2017}, we also account for the {\it direct} effect of the tidal torque due to the potential of the cluster on the orbital elements of the inner orbit of the binary.

Several authors have considered the problem of a star in orbit around a SMBH in a nuclear cluster (e.g. \citealt{Sridhar1999,Ivanov2005,Lockmann2008,Subr2009, Chang2009,Haas2011,Merritt2013,Li2015,Iwasa2016,Iwasa2017}). The SMBH-star system effectively forms a binary.  The binary's Keplerian orbital elements may then evolve on secular timescales due to some combination of (i) the mean field nuclear cluster potential, (ii) GR pericentre precession, (iii) an infalling massive black hole on a slowly decaying circular orbit, (iv) a circumnuclear ring of material, etc.  While this class of problems is reminiscent of our work, they are not quite the same because in the SMBH-star case the barycentre of the binary does not move, and so there is no clean separation between single- and double-averaging. In some cases, e.g. for a binary that sits at the centre of a spherical cusp, there is even no well-defined tidal expansion of the potential. Instead, averaging of the potential is incorporated into the averaging over the stellar Keplerian orbit around the SMBH, which is different from our approach.

One of the most interesting recent applications of secular dynamics has been the possibility of substantial shrinking of binary orbits by LK cycles with dissipative effects. Such applications include the origin of hot Jupiters \citep{Fabrycky2007,Naoz2011,Petrovich2015,Hamers2017a}, formation of blue stragglers \citep{Perets2009,Knigge2009}, white-dwarf mergers \citep{Thompson2011,Katz2011,Toonen2018}, and compact-object binary mergers in globular or nuclear star clusters \citep{Antognini2014,Rodriguez2015,Naoz2016,Silsbee2017,Petrovich2017,Leigh2018}. Binary evolution driven by cluster tides explored in our work represents a different evolutionary scenario that may lead to similar outcomes (without invoking a nearby third companion). This possibility is explored in a separate study \citep{Hamilton2019c}.

%%%%%%%%%%%%%%%%%%%%%%%%%%%%%%%%%%%%%%%%%%
%%%%%%%%%%%%%%%%%%%%%%%%%%%%%%%%%%%%%%%%%%

\section{Summary} 
\label{conclusions}

%%%%%%%%%%%%%%%%%%%%%%%%%%%%%%%%%%%%%%%%%%

This work explores secular evolution of binary systems orbiting in axisymmetric stellar clusters. We derive a Hamiltonian describing this evolution for an arbitrary form of the smooth cluster potential, average it over the (inner) orbital motion of the binary, and then average it again over the (outer) orbit of the binary around the cluster assuming the potential is axisymmetric. Our results can be summarized as follows.
\begin{itemize}

\item When the doubly-averaged Hamiltonian is cast in  dimensionless form, all the information about the tidal potential is contained in a single dimensionless parameter $\Gamma$, which depends on the background potential $\Phi$ and the orbit of the binary in the cluster. The value of this parameter determines the phase portrait of the binary evolution, which is explored in a companion study (Paper II).

\item The timescale of secular evolution is set by another (dimensional) parameter $A$, which, like $\Gamma$, depends on the cluster potential and the binary's outer orbit.

\item In certain cases $A$ and $\Gamma$ can be calculated (semi-)analytically. Such cases include (a) orbits in spherical potentials, (b) orbits confined to the midplane of an axisymmetric potential, and (c) epicyclic orbits near the midplane of an axisymmetric potential. We demonstrate how our calculations reproduce the known results for Lidov-Kozai evolution, evolution of Oort Cloud comets due to the Galactic tide, and so on.

\item We map out the behavior of $A$ and $\Gamma$ in different spherically symmetric potentials as a function of size and radial elongation of the binary orbit. We find that $\Gamma$ is small in the central regions of clusters with cored potentials, but tends to unity in clusters with finite mass as the orbit size increases. In general, $0 \leq \Gamma \leq 1$ in realistic finite-mass spherical potentials.

\item In general axisymmetric potentials, $\Gamma$ can easily attain negative values, in particular for highly inclined (i.e. non-coplanar) orbits.  

\item The accuracy with which our doubly-averaged Hamiltonian characterizes binary evolution deteriorates for highly non-coplanar orbits in axisymmetric potentials. Commeurability of orbital frequencies in the cluster potential may also present a problem for application of our theory at a quantitative level. 

\end{itemize}

These results will be extensively used in Paper II, where we systematically explore the dynamics of binaries driven by the tidal field of clusters for different values of $\Gamma$. There we verify numerically the predictions of the secular theory based on our doubly-averaged Hamiltonian, and derive timescales for secular eccentricity oscillations. In future papers in this series, these calculations will be applied to understanding secular evolution of binaries in dense clusters and exploring their relevance for the formation of compact-object mergers \citep{Hamilton2019c}, blue stragglers, and so on.

\section*{Acknowledgements}

We thank John Magorrian for pointing out some literature and Eugene Vasiliev, Fabio Antonini and Almog Yalinewich for useful comments. The numerical orbit integration was done with \texttt{galpy} (http://github.com/jobovy/galpy, \citealt{Bovy2015}). 
CH is funded by a Science and Technology Facilities Council (STFC) studentship.

%%%%%%%%%%%%%%%%%%%%%%%%%%%%%%%%%%%%%%%%%%%%%%%%%%

%%%%%%%%%%%%%%%%%%%% REFERENCES %%%%%%%%%%%%%%%%%%

% The best way to enter references is to use BibTeX:

\bibliographystyle{mnras}
\bibliography{Bibliography} % if your bibtex file is called example.bib

% Alternatively you could enter them by hand, like this:
% This method is tedious and prone to error if you have lots of references
%\begin{thebibliography}{99}
%\bibitem[\protect\citeauthoryear{Author}{2012}]{Author2012}
%Author A.~N., 2013, Journal of Improbable Astronomy, 1, 1
%bibitem[\protect\citeauthoryear{Others}{2013}]{Others2013}
%Others S., 2012, Journal of Interesting Stuff, 17, 198
%\end{thebibliography}

%%%%%%%%%%%%%%%%%%%%%%%%%%%%%%%%%%%%%%%%%%%%%%%%%%

%%%%%%%%%%%%%%%%% APPENDICES %%%%%%%%%%%%%%%%%%%%%

\appendix

%%%%%%%%%%%%%%%%%%%%%%%%%%%%%%%%%%%%%%%%%%
%%%%%%%%%%%%%%%%%%%%%%%%%%%%%%%%%%%%%%%%%%

\section{The singly-averaged Hamiltonian in orbital elements} \label{algebraic_expressions}

%%%%%%%%%%%%%%%%%%%%%%%%%%%%%%%%%%%%%%%%%%
	
In this Appendix we give the full algebraic expressions in terms of orbital elements for the terms $\langle r_\alpha r_\beta \rangle_M$ that enter the singly-averaged Hamiltonian \eqref{H1M}.  They are 
    \begin{align} 
	\langle x^2 \rangle_M =\nn& \frac{a^2}{16}\left( \cos 2i (2+3e^2+5e^2\cos2\omega \cos2\Omega) \right. \\ \nn& \left. + 5e^2\cos2\omega (3\cos2\Omega+2\sin^2i) \right. \\ \nn&\left. + (2+3e^2)(3+2\cos2\Omega\sin^2 i) \right. \\ &\left.-20e^2\cos i \sin2\omega\sin2\Omega \right), 
	\label{eq:x2}
	\end{align}
	\begin{align} 
	\langle y^2 \rangle_M =\nn& \frac{a^2}{16}\left( \cos 2i (2+3e^2-10e^2\cos 2\omega \cos^2\Omega) \right. \\ \nn& \left. + 5e^2\cos2\omega (1-3\cos2\Omega) \right. \\ \nn&\left. + (2+3e^2)(3-2\cos2\Omega\sin^2 i) \right. \\ &\left.+20e^2\cos i \sin2\omega\sin2\Omega \right), 
	\label{eq:y2}
	\end{align}
	\begin{align} 
	\langle z^2 \rangle_M = \frac{a^2}{4}\sin^2i\left( 2+3e^2-5e^2\cos2\omega \right), 
	\label{eq:z2}
	\end{align}
	\begin{align} 
	\langle xy \rangle_M =\nn& \frac{a^2}{16}\left( 20e^2\cos i\cos2\Omega\sin2\omega \right. \\ &\left.+ (5e^2(3+\cos 2i)\cos 2\omega+2(2+3e^2)\sin^2i)\sin2\Omega \right), 
	\label{eq:xy}
	\end{align}
	\begin{align} \langle xz \rangle_M =\nn& \frac{a^2}{4}\sin i \left( 5e^2\cos\Omega\sin2\omega \right. \\ &\left.-\cos i (2+3e^2-5e^2\cos2\omega)\sin\Omega \right), 
	\label{eq:xz}
	\end{align}
	\begin{align} \langle yz \rangle_M =\nn& \frac{a^2}{4}\sin i \left( 5e^2\sin\Omega\sin2\omega \right. \\ &\left.+\cos i (2+3e^2-5e^2\cos2\omega)\cos\Omega \right).
	\label{eq:yz}
	\end{align}

%%%%%%%%%%%%%%%%%%%%%%%%%%%%%%%%%%%%%%%%%%
%%%%%%%%%%%%%%%%%%%%%%%%%%%%%%%%%%%%%%%%%%

\section{Recovering the Lidov-Kozai quadrupole Hamiltonian} \label{RecoverLK}

%%%%%%%%%%%%%%%%%%%%%%%%%%%%%%%%%%%%%%%%%%

To derive the LK Hamiltonian we take equation \eqref{H1M} and average it over time using $\Phi(r) = -GM/r$, and with $\Rg(t)$ describing a Keplerian ellipse with the focus at the origin.
		
First, it is obvious that for the this potential, $\Phi''(R_\mathrm{g}) \propto \Phi'(R_\mathrm{g})/R_\mathrm{g} \propto R_\mathrm{g}^{-3}$.  Then we must average the right hand sides of equations \eqref{phixxnon}-\eqref{phixynon} which requires that we average the quantities $R_\mathrm{g}^{-3}, R_\mathrm{g}^{-3}\cos2\phi_\mathrm{g}$ and $R_\mathrm{g}^{-3}\sin 2\phi_\mathrm{g}$.  
		
Without loss of generality we may choose the argument of pericentre $\omega_\mathrm{g}$ of the ellipse to be zero, so that $\phi_\mathrm{g} = f_\mathrm{g}$, the true anomaly.  Then  $R_\mathrm{g}=a_\mathrm{g}(1-e_\mathrm{g}\cos E_\mathrm{g})$ with $E_\mathrm{g}$ being the eccentric anomaly, so for an arbitrary function $\mathcal{S}(\phi_\mathrm{g})$ we can write 
\begin{align} 
\overline{R_\mathrm{g}^{-3} \mathcal{S}(\phi_\mathrm{g})} &= \frac{1}{2\pi} \int_0^{2\pi} \md E_\mathrm{g} a_\mathrm{g}^{-3}(1-e_\mathrm{g}\cos E_\mathrm{g})^{-2} \mathcal{S}(f_\mathrm{g}), \end{align} and we can convert between $f_\mathrm{g}$ and $E_\mathrm{g}$ using \begin{align} \cos f_\mathrm{g} = \frac{\cos E_\mathrm{g}-e_\mathrm{g}}{1-e_\mathrm{g}\cos E_\mathrm{g}}. \end{align} The answers are \begin{align} &\overline{ R_\mathrm{g}^{-3} } = \frac{4}{3}\overline{ R_\mathrm{g}^{-3} \cos 2\phi_\mathrm{g}}  = a_\mathrm{g}^{-3}(1-e_\mathrm{g}^2)^{-3/2} ,\\  &\overline{ R_\mathrm{g}^{-3}\sin 2\phi_\mathrm{g}} = 0.\end{align} Using these identities, we find a remarkably simple relation between the time-averaged coefficients: \begin{align} \overline{\Phi}_{xx} = \overline{\Phi}_{yy} = -\frac{1}{2}\overline{\Phi}_{zz}, \,\,\, \mathrm{with}  \, \,\,\overline{\Phi}_{zz} = GMa_\mathrm{g}^{-3}(1-e_\mathrm{g}^2)^{-3/2}, 
\end{align}  
and as expected $\overline{\Phi}_{xy}=\overline{\Phi}_{xz}=\overline{\Phi}_{yz}=0$.  Note that the regime $\overline{\Phi}_{xx} = \overline{\Phi}_{yy}$ is exactly that of an axisymmetric perturbing potential (see e.g. \cite{Katz2011}). The resulting perturbing Hamiltonian will therefore be the same as \eqref{H1Mt}, with the added simplification that $A = (GM/2)a_\mathrm{g}^{-3}(1-e_\mathrm{g}^2)^{-3/2}$ and $\Gamma=1$.  Making these substitutions we find \begin{align} \nn \overline{\langle H_1 \rangle}_{M} = -& \frac{GMa^2}{16a_\mathrm{g}^3(1-e_\mathrm{g}^2)^{3/2}} \\ &\times   [(2+3e^2)(3\cos^2 i-1)+15e^2\sin^2 i \cos 2\omega]. \label{LKHam} \end{align} This is precisely the dimensionless test particle quadrupole Lidov-Kozai Hamiltonian \citep{Lidov1962,Kozai1962,Kinoshita2007,Lithwick2011,Antognini2014}.
It describes the secular evolution of a hierarchical triple system in which the outer orbit dominates the angular momentum budget.

%%%%%%%%%%%%%%%%%%%%%%%%%%%%%%%%%%%%%%%%%%
%%%%%%%%%%%%%%%%%%%%%%%%%%%%%%%%%%%%%%%%%%

\section{Epicyclic orbits} 
\label{RecoverHT}

%%%%%%%%%%%%%%%%%%%%%%%%%%%%%%%%%%%%%%%%%%

In this Appendix we look at the behavior of $A$ and $\Gamma$ in the case of a binary performing epicyclic motion in an axisymmetric disk, to connect with the results of \citet{Heisler1986} (HT86), who calculated the secular effect of the Galactic tide on the Oort Cloud comets.

Let the guiding centre of the binary's orbit be a circle of radius $R_\mathrm{c}$ in the $Z=0$ plane of the potential.  The potential experienced by the binary can then be approximated as\footnote{We implicitly assume that the disk is symmetric about its midplane, $\Phi(R,Z)=\Phi(R,-Z)$, so that $\partial^2\Phi/\partial R\partial Z = 0$ at $Z=0$. Otherwise the binary would not remain in the midplane anyway.} \begin{align}
\Phi(R,Z)=& \nn\Phi(R_\mathrm{c},0)+\left(\frac{\partial\Phi}{\partial R}\right)_{(R_\mathrm{c},0)}(R-R_\mathrm{c})\\ &+\frac{1}{2}\left(\frac{\partial^2\Phi}{\partial R^2}\right)_{(R_\mathrm{c},0)}(R-R_\mathrm{c})^2 +\frac{1}{2}\left(\frac{\partial^2\Phi}{\partial Z^2}\right)_{(R_\mathrm{c},0)}Z^2.
\end{align} 
Using this expression and equations (\ref{phixxyy})-(\ref{phizz}) it is easy to show that 
\begin{align}
\overline{\Phi}_{xx} = \overline{\Phi}_{yy}=\frac{1}{2}(\kappa^2-2\Omega_\mathrm{c}^2); \,\,\,\,\,\,\, \overline{\Phi}_{zz} = \nu^2,
\end{align}
while all other $\overline{\Phi}_{\alpha\beta}=0$; here \begin{align}
\Omega_\mathrm{c}^2=\left(\frac{1}{R}\frac{\partial \Phi}{\partial R}\right)_{(R_\mathrm{c},0)},
\end{align} 
is the angular frequency of the guiding centre, while \begin{align} 
\kappa^2 &\equiv  \left(\frac{\partial^2\Phi}{\partial R^2} + \frac{3}{R}\frac{\partial \Phi}{\partial R} \right)_{(R_\mathrm{c},0)}, \,\,\,\,\,\,\,\,\,\nu^2 &\equiv \left(\frac{\partial^2\Phi}{\partial Z^2}\right)_{(R_\mathrm{c},0)},
\end{align}
are the radial and vertical epicyclic frequencies of $\Rg$ respectively. Hence
\begin{align}
A&=\nu^2 + \frac{1}{2}(\kappa^2-2\Omega_\mathrm{c}^2),
\label{epiA} \\ \Gamma &=\frac{\nu^2 - \frac{1}{2}(\kappa^2-2\Omega_\mathrm{c}^2)}{3[\nu^2 + \frac{1}{2}(\kappa^2-2\Omega_\mathrm{c}^2)]}. 
\label{epiGam}
\end{align}

Near the midplane of a galactic disk, and in particular in the HT86 case of the solar neighbourhood of the Milky Way, it is almost always the case that $\Omega_\mathrm{c}\sim\kappa \ll \nu$.  Thus to a very good approximation $A=\nu^2$ and $\Gamma=1/3$.  Plugging these results into our doubly-averaged Hamiltonian \eqref{H1Mt}, written in Delaunay variables, we find \begin{align}
\overline{\langle H_1 \rangle}_M &= \frac{\nu^2L^2}{4\mu^2J^2}(J^2-J_z^2)[J^2+5(L^2-J^2)\sin^2\omega] \nn \\ &= \frac{\pi G \rho_0 L^2}{\mu^2J^2}(J^2-J_z^2)[J^2+5(L^2-J^2)\sin^2\omega],
\end{align}
where we have eliminated $\nu$ in favour of the density in the Solar neighbourhood $\rho_0$ using Poisson's equation $4 \pi G \rho_0 =(\nabla^2\Phi)_{\Rg} \approx (\partial^2\Phi/\partial z^2)_{\Rg} \equiv \nu^2$.  This is precisely the Hamiltonian arrived at by HT86 (c.f. their equation (14)) when considering the effect of the Galactic tide on the Oort Cloud comets.

%%%%%%%%%%%%%%%%%%%%%%%%%%%%%%%%%%%%%%%%%%
%%%%%%%%%%%%%%%%%%%%%%%%%%%%%%%%%%%%%%%%%%

\section{Signs and sizes of \texorpdfstring{$A$}{} and \texorpdfstring{$\Gamma$}{}} 
\label{sect:AGamma}

%%%%%%%%%%%%%%%%%%%%%%%%%%%%%%%%%%%%%%%%%%

%%%%%%%%%%%%%%%%%%%%%%%%%%%%%%%%%%%%%%%%%%
\begin{table*}
   \caption{
   \label{GammaUTable}
   Summary of $U$ and $\Gamma(U)$ ranges that are possible for orbits $\Rg$ in different classes of potential $\Phi$.}
    \begin{tabular}{| l | l | l | l | l | l | l | l | l | l | l |}
    \hline
   \textbf{Type of the potential/orbit} $\Rg$ & $U$ \textbf{range} & $\Gamma$ \textbf{range}  & 
    \\ \hline
 General axisymmetric potential & $-\infty \leq U \leq \infty$ & $-\infty \leq \Gamma \leq \infty$ &
    \\ \hline
  Spherical potential (assuming $\md\rho/\md r \leq 0$ and finite mass) & $1/3 \leq U \leq \infty$ & $0 \leq \Gamma \leq 1$ &
      \\ \hline
  Midplane of a thin disk & $U=1$ & $\Gamma = 1/3$  & 
      \\ \hline
  Vertical cylindrical potential & $U=0$ & $\Gamma = -1/3$  & 
      \\ \hline
        Axisymmetric harmonic potential & $0\leq U \leq 1$ & $-1/3 \leq \Gamma \leq 1/3$  & 
      \\ \hline
        Spherical harmonic potential & $U=1/3$ & $\Gamma = 0$  & 
      \\ \hline
        Keplerian potential & $U \to \infty$ & $\Gamma \to 1$  & 
      \\ \hline
              Spherical cusp potential (density $\rho \propto r^{-\beta}$) & $U = 1/(3-\beta)$ & $\Gamma =\beta/[3(4-\beta)]$  & 
      \\ \hline
    \end{tabular}
\end{table*}
%%%%%%%%%%%%%%%%%%%%%%%%%%%%%%%%%%%%%%%%%%

Here we provide some technical details about the statements on the signs and values of $A$ and $\Gamma$ made in \S \ref{sect:gen}. Also, in Table \ref{GammaUTable} we summarize some information about these coefficients for certain potentials.  We provide two examples of orbits with extreme values of $\Gamma$.

%%%%%%%%%%%%%%%%%%%%%%%%%%%%%%%%%%%%%%%%%%

\subsection{Spherical potentials} 
\label{SphericalABGamma}

%%%%%%%%%%%%%%%%%%%%%%%%%%%%%%%%%%%%%%%%%%

Consider a spherically symmetric potential $\Phi=\Phi(r)$, where $r\equiv\sqrt{R^2+Z^2}$ is the sphericl radius. According to our convention, the outer orbit of the binary always lies in $Z=0$ plane of the associated cylindrical $(R,\phi,Z)$ coordinate system. Then it is a simple matter to show that 
\begin{align}
\left(\frac{\partial^2\Phi}{\partial Z^2}\right)_{\Rg} &= \left(\frac{1}{R}\frac{\partial \Phi}{\partial R}\right)_{\Rg} = \left(\frac{1}{r}\frac{\md \Phi}{\md r}\right)_{\Rg},\\
\left(\frac{\partial^2\Phi}{\partial R^2}\right)_{\Rg} &=\left(\frac{\md^2\Phi}{\md r^2}\right)_{\Rg}.
\end{align}
Using these conversions as well as equations \eqref{phixxyy}, \eqref{phizz}, \eqref{ABGamDef} we find 
\begin{align} 
A &= \frac{1}{2}\left[\overline{\left(\frac{\md^2\Phi}{\md r^2}\right)}_{\Rg} + 3 \overline{\left(\frac{1}{r}\frac{\md\Phi}{\md r}\right)}_{\Rg} \right], \label{eq:A}\\ 
B &= -\frac{1}{2}\left[\overline{\left(\frac{\md^2\Phi}{\md r^2}\right)}_{\Rg} - \overline{\left(\frac{1}{r}\frac{\md\Phi}{\md r}\right)}_{\Rg} \right]. \label{eq:B}
\end{align}

We can now prove that $A>0$  for (almost) any realistic spherical potential, and thereby show that for such systems $0\leq \Gamma \leq 1$. In a spherical potential $\Phi(r)$ we have
\begin{align} 
\label{masseqn}
\frac{\md \Phi}{\md r} = \frac{GM(r)}{r^2}, 
\end{align} 
where $M(r)$ is the cluster mass enclosed inside radius $r$. Also, Poisson's equation reads
\begin{align}
\frac{\md^2 \Phi}{\md r^2} + \frac{2}{r}\frac{\md \Phi}{\md r} = 4\pi G \rho,
\end{align}   
allowing us to rewrite equation (\ref{eq:A}) as
\begin{align}  
A =\frac{1}{2} \left[4\pi G\overline{\rho(\Rg)} + \overline{\left(\frac{GM(r)}{r^3}\right)}_{\Rg} \right]. 
\label{impliesA} 
\end{align}
Since $\rho > 0$ and $M>0$ at all radii, this inevitably results in $A>0$.  

Finally, since 
\begin{align} 
\frac{\md^2\Phi}{\md r^2} - \frac{1}{r}\frac{\md \Phi}{\md r} = r \frac{\md}{\md r} \left(\frac{1}{r} \frac{\md \Phi}{\md r} \right),
\end{align}  
equation (\ref{eq:B}) can be rewritten as
\begin{align}
B =-\frac{1}{2} \overline{\left[r\frac{\md}{\md r} \left(\frac{GM(r)}{r^3}\right) \right]}_{\Rg}. 
\label{impliesB} 
\end{align}
For any spherical system in which the density is a non-increasing function of radius, $\md (M/r^3)/\md r \leq 0$ for any $r$ and hence $B \geq 0$.

Let us now focus on spherical systems with $\md\rho/\md r\leq 0$.  If the cluster has a constant density core, then $M(r)\sim r^3$ as $r \to 0$ and so $B\to 0$ (equation \eqref{impliesB}).  Hence if $\Rg$ orbits entirely inside the constant density region, $\Gamma = 0$.  A potential without a core will always have a non-zero value of $\Gamma$ for orbits at small radii.  

At the other extreme, as $r\to\infty$ we have $\rho \to 0$, and usually the enclosed mass $M(r) \to$ const (although see below for potentials arising from power-law cusp density profiles).  Hence $\md (M(r)/r^3) / \md r \to -3M/r^4$ where $M$ is the total mass of the cluster, and in this limit we get $B=3A$. Thus an orbit $\Rg$ that spends its time exclusively at very large radii $r$ compared to the scale radius of the cluster will have $\Gamma \to 1$.  This is precisely the Lidov-Kozai limit: for potentials that are Keplerian as $r\to \infty$ (i.e. those with finite mass), orbiting far from the core is equivalent to orbiting a point mass at the origin. 

Finally, for any orbit in a spherical cluster with a power-law density cusp $\rho(r) \propto r^{-\beta}$ with $0<\beta<3$ (so that the mass is finite at the centre) one naturally has $M(r)=4\pi\rho r^3/(3-\beta)$. Then we find from equations \eqref{impliesA}, \eqref{impliesB} that 
\begin{align} 
\label{eq:cuspAB} 
A = 2\pi G\overline{\rho}~\frac{4-\beta}{3-\beta}, \,\,\,\,\, 
B = 2\pi G\overline{\rho}~\frac{\beta}{3-\beta}, 
\end{align} 
and so $U=1/(3-\beta)$ and $\Gamma = \beta/[3(4-\beta)]$.  

%%%%%%%%%%%%%%%%%%%%%%%%%%%%%%%%%%%%%%%%%%

\subsection{Axisymmetric potentials}  
\label{AxisymmetricABGamma}

%%%%%%%%%%%%%%%%%%%%%%%%%%%%%%%%%%%%%%%%%%

In a general axisymmetric potential there is no constraint on how negative the parameter $U$ (defined by equation (\ref{Udef}) and plotted in Figure \ref{GammaOfU}) can be. Non-spherical potentials naturally feature regions with $\Phi_{zz} < 0$, especially near the poles; choosing a highly inclined (with respect to the equatorial plane of the potential) orbit with large radius so that $\overline{\rho}$ is vanishingly small, one can drive strongly negative $U$, thereby achieving extreme (positive or negative) values of $\Gamma$. 

\begin{figure*}
\centering
\includegraphics[width=0.315\linewidth]{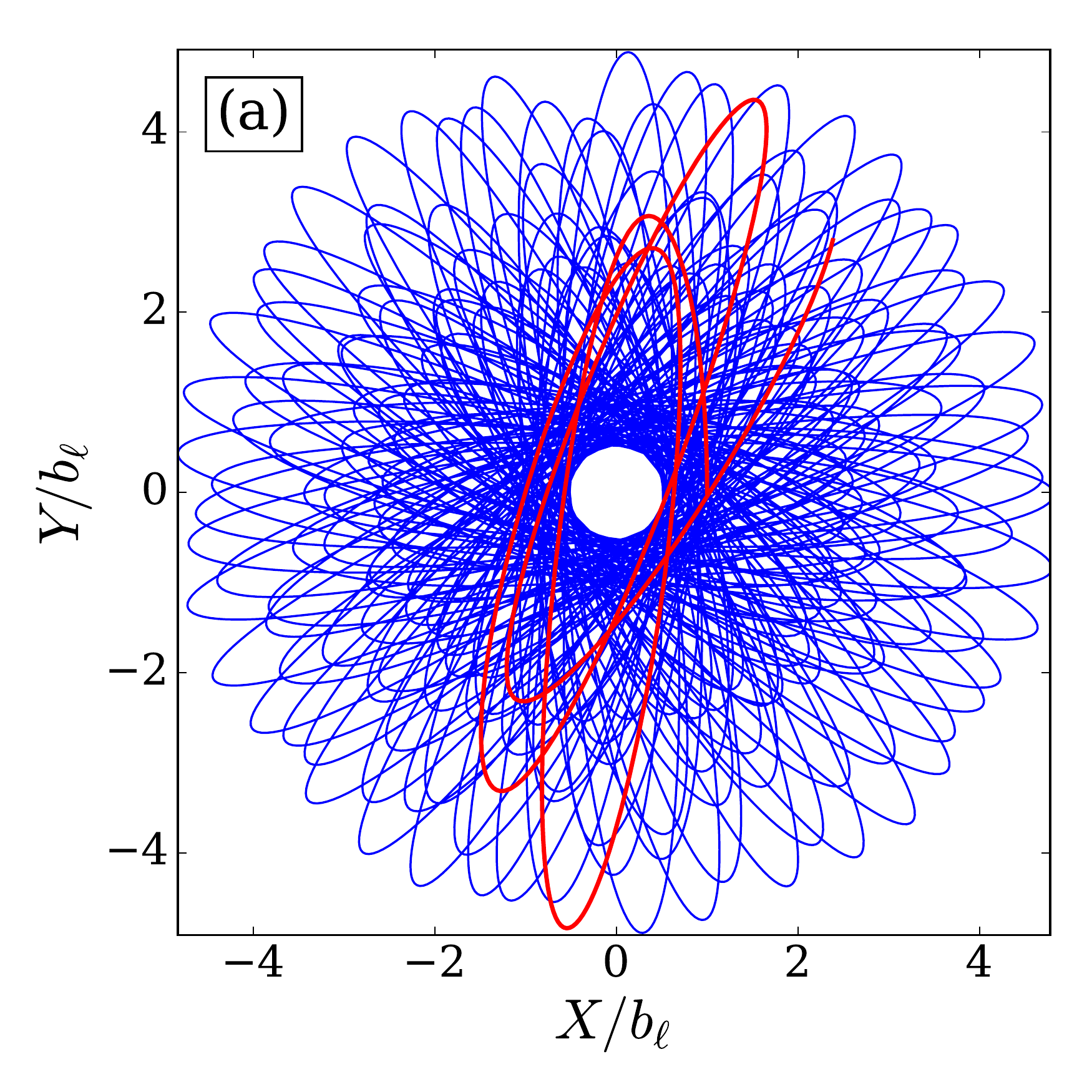}
\includegraphics[width=0.33\linewidth]{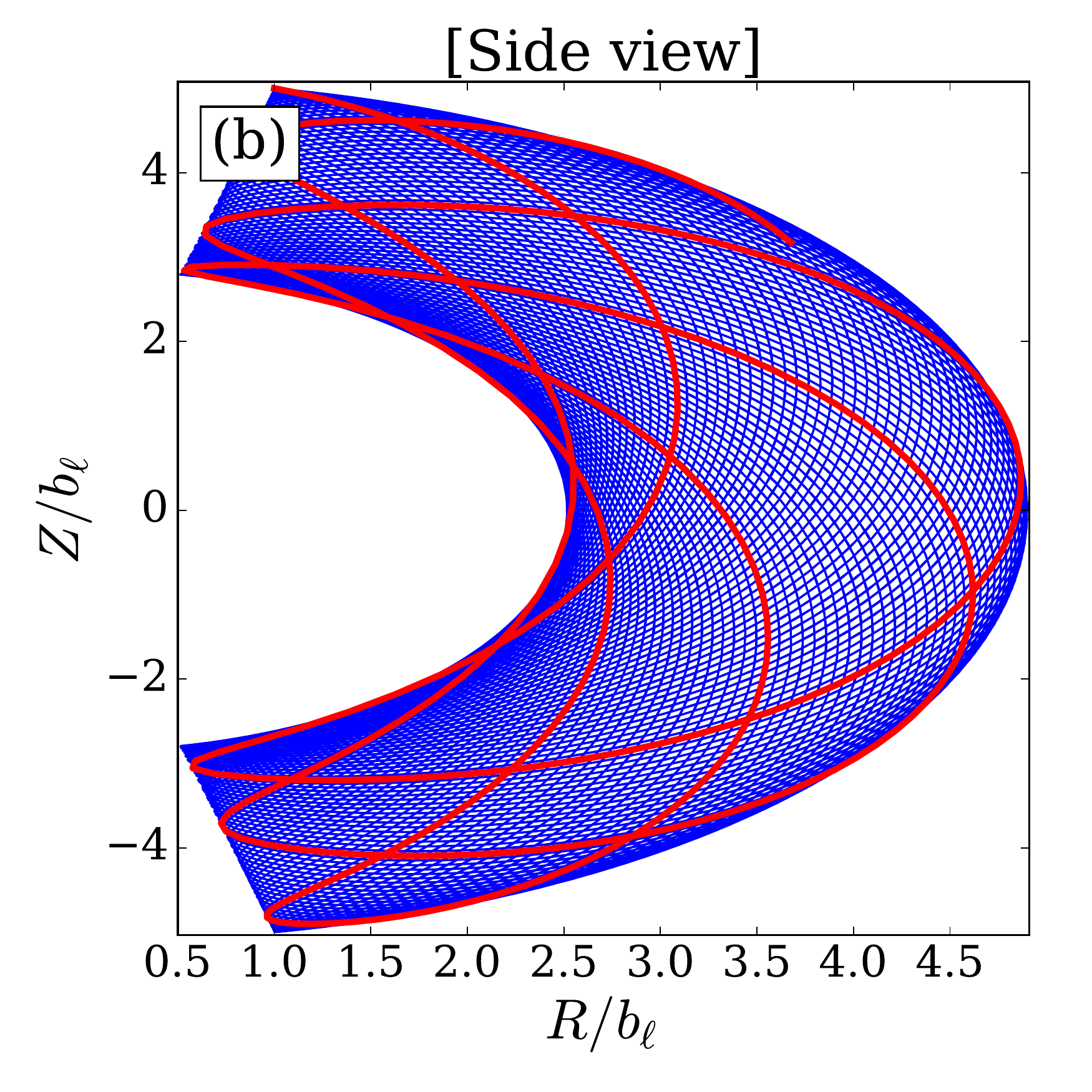}
\includegraphics[width=0.32\linewidth]{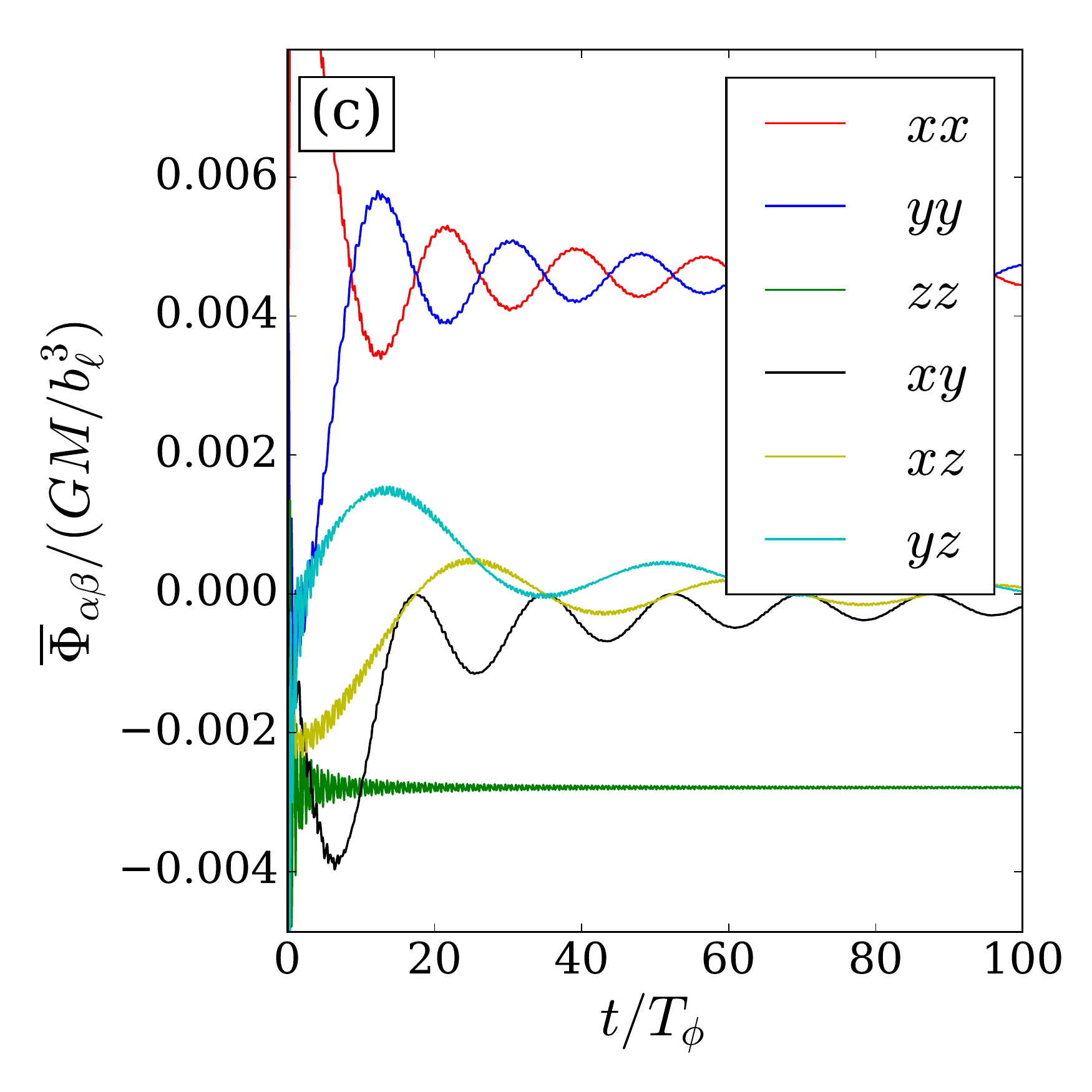}
\caption{Example of an orbit with $\Gamma_\mathrm{num}=-1.4$.  We use the Miyamoto-Nagai potential with $b_h/b_\ell = 1$ and the same initial conditions as Orbits (VIII)-(XI) in the main text, except the initial $Z$ value is $5.0 b_\ell$.}
\label{fig:App_D_Ex1}
\end{figure*}

In Figure \ref{fig:App_D_Ex1} we give an example of an orbit with $\Gamma_\mathrm{num}=-1.4$. We use the Miyamoto-Nagai potential with $b_h/b_\ell=1$ and precisely the same initial conditions as Orbits (VIII)-(XI) in the main text (see Table \ref{MNtable}), except that we now take the initial $Z$ coordinate to be $5.0b_\ell$. All three panels show $100T_\phi$ of data. The large initial value of $Z$ means that the orbit spends a lot of time near the poles of the potential where $\Phi_{zz} < 0$.

\begin{figure*}
\centering
\includegraphics[width=0.4\linewidth]{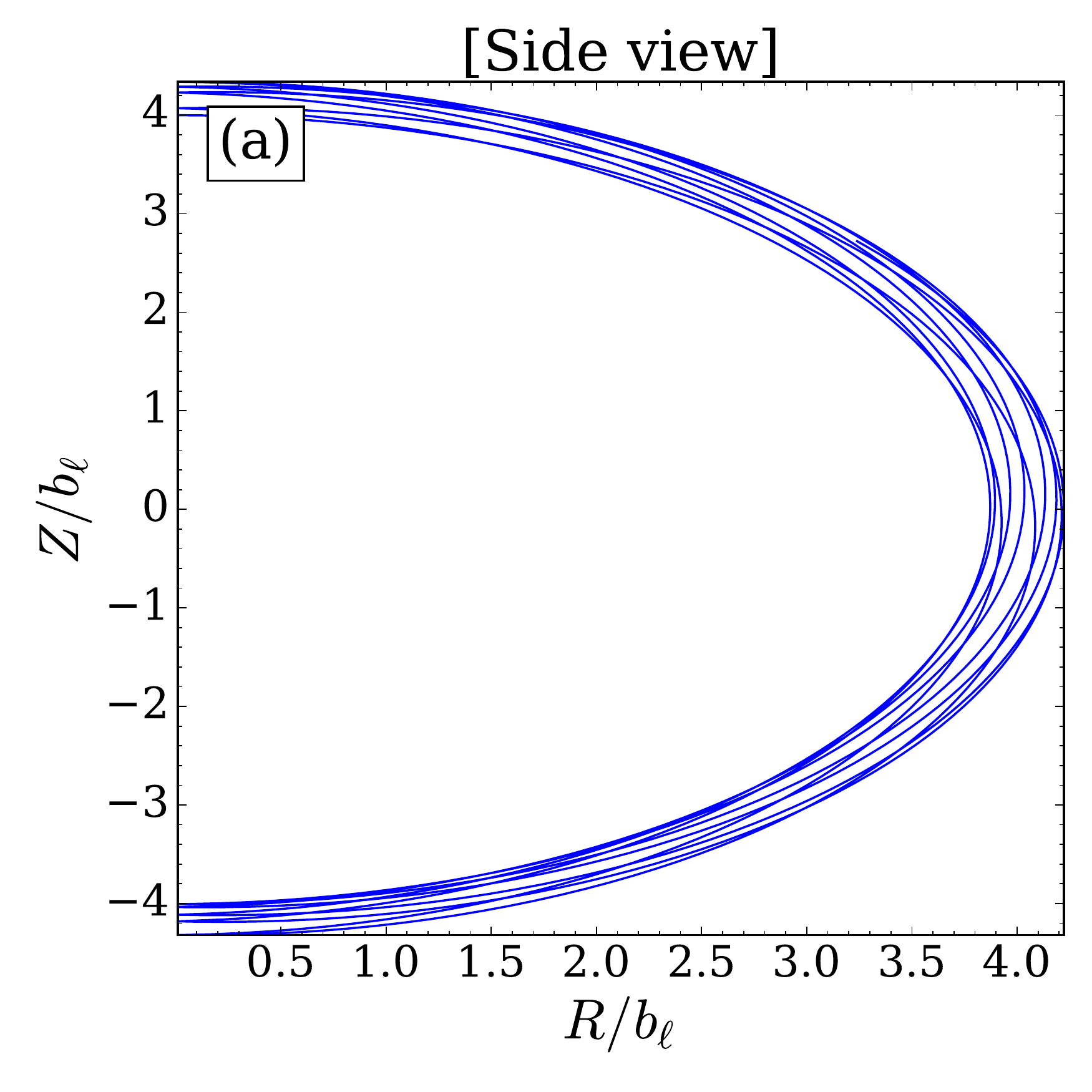}
\includegraphics[width=0.435\linewidth]{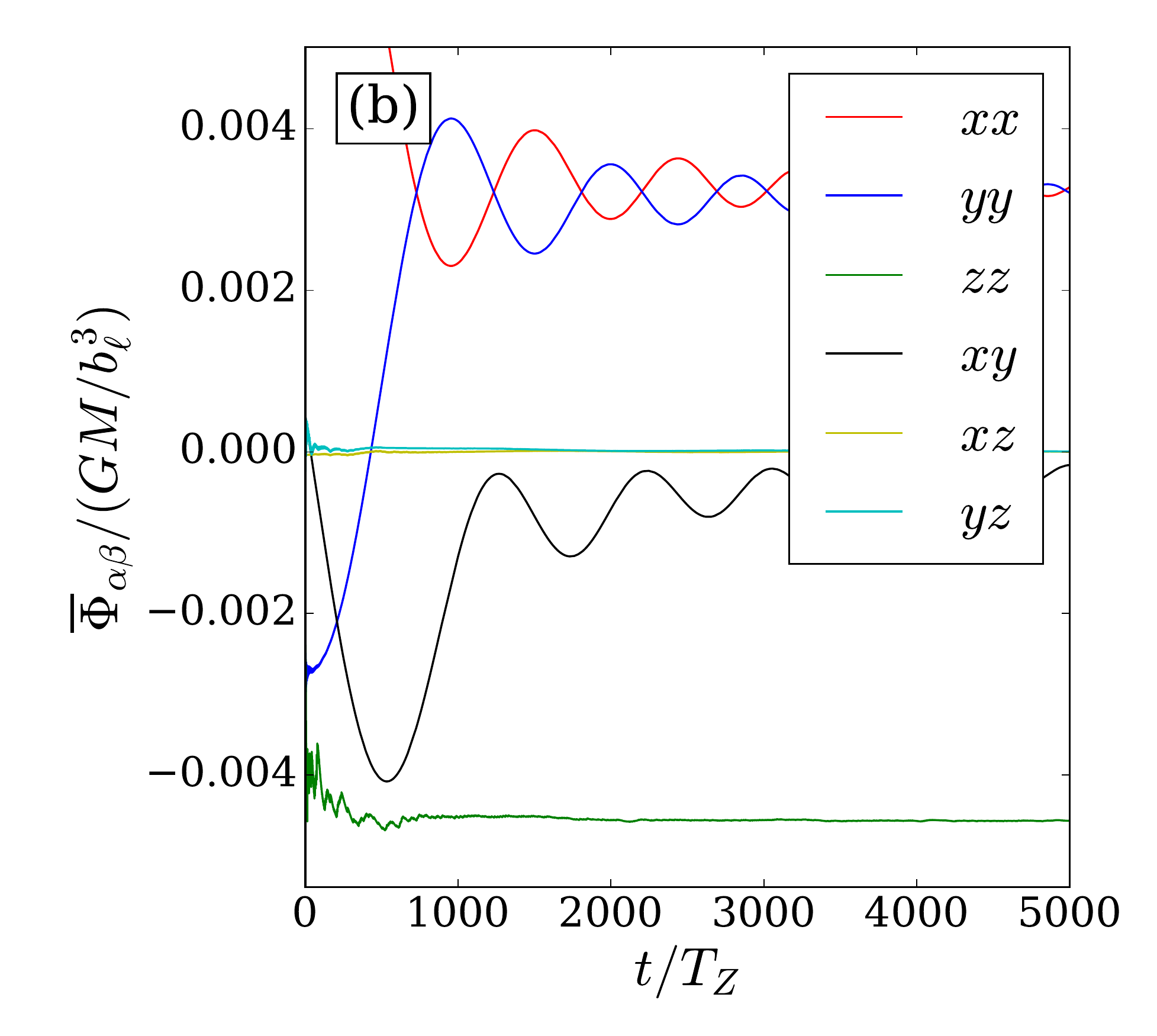}
\caption{Example of an orbit with $\Gamma_\mathrm{num}=2.0$.  We use the Miyamoto-Nagai potential with $b_h/b_\ell = 0.1$ and initial conditions $(R,v_R,\phi,v_\phi,Z,v_Z)=(0.011b_\ell,0,0.04,0.4\sqrt{GM/b_\ell},4b_\ell,0)$. In panel (a) we show the meridional $(R,Z)$ plane for only the first $5$ vertical periods $T_Z$ of the integration. The orbit precesses very slowly in azimuth, filling its torus after several thousand $T_Z$.}
\label{fig:App_D_Ex2}
\end{figure*}

In Figure \ref{fig:App_D_Ex2} we provide an example of a very polar orbit resulting in $\Gamma_\mathrm{num} = 2.0$ ($U=-2.3$).  We use the Miyamoto-Nagai potential $b_h/b_\ell = 0.1$, and initial conditions \begin{align} (R,v_R,\phi,v_\phi,Z,v_Z)=(0.011b_\ell,0,0.04,0.4\sqrt{GM/b_\ell},4.0b_\ell,0).\end{align}
In Figure \ref{fig:App_D_Ex2}a we display only the first $5$ vertical periods $T_Z$ of the orbit in the meridional $(R,Z)$ plane. When integrated for a long time, the orbit remains almost polar but precesses very slowly around the $Z$ axis until it eventually fills an axisymmetric torus after a few thousand $T_Z$. Figure \ref{fig:App_D_Ex2}a shows that the convergence of the $\overline{\Phi}_{\alpha\beta}$ coefficients in this case is very slow and takes $\gtrsim 5000T_Z$. In practice, unless the binary is very tight, secular theory is unlikely to work well for such an orbit. Indeed, for a relatively wide binary the secular evolution timescale is likely to be much shorter than $5000T_Z$, meaning that our assumption that the binary fills its torus (and hence the $\Phi_{\alpha\beta}$ converge) in much less than a secular timescale is violated. We will explore this issue in more detail in Paper II, \S7. 

%%%%%%%%%%%%%%%%%%%%%%%%%%%%%%%%%%%%%%%%%%
%%%%%%%%%%%%%%%%%%%%%%%%%%%%%%%%%%%%%%%%%%

\section{Octupole Hamiltonian} 
\label{Higher_Order}

%%%%%%%%%%%%%%%%%%%%%%%%%%%%%%%%%%%%%%%%%%

The Hamiltonian derived in \S\ref{setting} is correct to quadratic order in $a/\vert \Rg \vert$, the so-called `quadrupole approximation'.  We can attempt to derive a more accurate Hamiltonian by keeping the higher order terms in the series expansion of equation \eqref{eq:expansion}. The next (`octupole') term that we would include is
 \begin{align} 
\frac{1}{2!}\sum_{\beta\gamma} \left(\frac{\partial^2\Phi}{\partial R_\alpha \partial R_\beta \partial R_\gamma}  \right)_{\Rg}r_{i,\beta}\, r_{i,\gamma}.
 \end{align}  
 We then use the fact that $\mathbf{r} \equiv \mathbf{r}_1 - \mathbf{r}_2$ and $m_1\mathbf{r}_1+m_2\mathbf{r}_2 = \mathbf{0}$ to write 
 \begin{align}
 r_{1,\alpha} = \frac{m_2}{m_1+m_2}r_\alpha; \,\,\,\,\,\,\, r_{2,\alpha} = -\frac{m_1}{m_1+m_2}r_\alpha.
 \end{align} 
 As a result, the equation for the relative motion $\md^2 \mathbf{r} / \md t^2$ (equation \eqref{d2rdt2}) can be written purely in terms of $\mathbf{r}$.  The next order (`octupole') correction to the Hamiltonian \eqref{H1Mt} is \begin{align} H_\mathrm{oct} = \frac{1}{3!}\left( \frac{m_2-m_1}{m_2+m_1} \right) \sum_{\alpha\beta\gamma} \Phi_{\alpha\beta\gamma} r_\alpha \, r_\beta \, r_\gamma. \end{align} Note that the octupole term vanishes for equal-mass binaries ($m_1=m_2$).
 
The corresponding doubly-averaged perturbing octupole term is then simply \begin{align} \overline{\langle H_\mathrm{oct} \rangle}_M = \frac{1}{3!}\left( \frac{m_2-m_1}{m_2+m_1} \right) \sum_{\alpha\beta\gamma} \overline{\Phi}_{\alpha\beta\gamma} \langle r_\alpha \, r_\beta \, r_\gamma \rangle_M. \label{H2Mt}\end{align}

%%%%%%%%%%%%%%%%%%%%%%%%%%%%%%%%%%%%%%%%%%

\subsection{Time-averaging over an axisymmetric torus}

%%%%%%%%%%%%%%%%%%%%%%%%%%%%%%%%%%%%%%%%%%

As in the quadrupole case, it turns out that when time-averaged over an axisymmetric torus the coefficents $\overline{\Phi}_{\alpha\beta\gamma}$ satisfy various symmetry properties.  After a little algebra one can show that \begin{align}
\overline{\Phi}_{xxz} = \overline{\Phi}_{yyz},
\end{align}
and all other $\overline{\Phi}_{\alpha\beta\gamma}=0$ except for $\overline{\Phi}_{zzz}$.  Hence
\begin{align} 
\overline{\langle H_\mathrm{oct} \rangle}_M =& \frac{1}{3!}\left( \frac{m_2-m_1}{m_2+m_1} \right) \nn \\ &\times \left[3\overline{\Phi}_{xxz} \langle (x^2+y^2)z \rangle_M + \overline{\Phi}_{zzz} \langle z^3 \rangle_M \right]. \label{H2MtAxi} 
\end{align}
Writing out the $\langle . \rangle_M$ factors in terms of orbital elements we have 
\begin{align}
\langle (x^2+y^2)z \rangle_M =& -\frac{5}{128} a^3 e \left[28 e^2 \sin^3i  \sin 3 \omega \nn \right. \\ & \left. +(4 + 3 e^2)(7\sin i + 3\sin 3i)\sin \omega\right],
\end{align}
\begin{align}
\langle z^3 \rangle_M =-\frac{5}{16} a^3 e  (6 + e^2 - 7 e^2 \cos 2 \omega) \sin^3 i \sin\omega.
\label{phiz3}
\end{align}

Equations (\ref{H2MtAxi})-(\ref{phiz3}) provide a general framework for accounting for the octupole contribution to the tidal Hamiltonian in an arbitrary axisymmetric potential. Note there is no $\Omega$ dependence in the octupole Hamiltonian, so $J_z$ is still conserved to octupole order.

%%%%%%%%%%%%%%%%%%%%%%%%%%%%%%%%%%%%%%%%%%

\subsection{Link to the test particle octupole LK Hamiltonian} 

%%%%%%%%%%%%%%%%%%%%%%%%%%%%%%%%%%%%%%%%%%

Note that one \textit{cannot} recover the test particle octupole term of the doubly-averaged Lidov-Kozai Hamiltonian by putting $\Phi=-GM/r$ in \eqref{H2MtAxi}, because equation \eqref{H2MtAxi} is derived under the axisymmetric approximation.  The time-averaged potential of a perturber on a Keplerian orbit is only axisymmetric at the \textit{quadrupole} level, and the symmetry is broken by octupole terms.  Instead one must integrate over the outer Keplerian orbit exactly, as in Appendix \ref{RecoverLK}.

In general there are $10$ independent time-averaged coefficients $\overline{\Phi}_{\alpha\beta\gamma}$ to consider.  We choose the outer orbit to be in the $Z=0$ plane so we can immediately eliminate four of these, $\Phi_{zzz}=\Phi_{xxz}=\Phi_{yyz}=\Phi_{xyz}=0$.  We can also choose the pericentre of the outer orbit to be on the $X$ axis without loss of generality, so that the ellipse traced by the outer orbit is symmetric under $Y\to -Y$.  Then all $\Phi_{\alpha\beta\gamma}$ that contain an odd number of $Y$ derivatives will be antisymmetric under $Y\to -Y$, so their time-averages over this ellipse will vanish: $\overline{\Phi}_{xxy}=\overline{\Phi}_{zzy}=\overline{\Phi}_{yyy}=0$.  This leaves us with only three non-zero terms in the doubly-averaged octupole LK Hamiltonian: 
\begin{align} \overline{\langle H_\mathrm{oct} \rangle}_M =& \frac{1}{3!}\left( \frac{m_2-m_1}{m_2+m_1} \right) \nn \\ &\times \left[\overline{\Phi}_{xxx} \langle x^3 \rangle_M + 3\overline{\Phi}_{yyx} \langle y^2x \rangle_M + 3\overline{\Phi}_{zzx} \langle z^2x \rangle_M \right]. \label{H2MtLK} \end{align}  For reference we now write down the terms that make up equation \eqref{H2MtLK}. First we write down the necessary $\Phi_{\alpha\beta\gamma}$ coefficients in terms of cylindrical coordinates $R_\mathrm{g}$ and $\phi_\mathrm{g}$:  
\begin{align} \Phi_{xxx} &= \frac{3GM}{R_\mathrm{g}^4}\cos \phi_\mathrm{g}(5\cos^2 \phi_\mathrm{g}-3), \\ \Phi_{yyx} &= \frac{3GM}{R_\mathrm{g}^4}\cos \phi_\mathrm{g}(5\sin^2 \phi_\mathrm{g}-1), \\ \Phi_{zzx} &= -\frac{3GM}{R_\mathrm{g}^4}\cos \phi_\mathrm{g}. \end{align} 
The time-averages of these coefficients are \begin{align} \label{eq:phixxxavg} \overline{\Phi}_{xxx} &= GMa_\mathrm{g}^{-4}(1-e^2_{\mathrm{g}})^{-5/2} \times (9e_\mathrm{g}/4), \\ \overline{\Phi}_{yyx} &= GMa_\mathrm{g}^{-4}(1-e^2_{\mathrm{g}})^{-5/2} \times (3e_\mathrm{g}/4), \\ \overline{\Phi}_{zzx} &= GMa_\mathrm{g}^{-4}(1-e^2_{\mathrm{g}})^{-5/2} \times (-3e_\mathrm{g}). \end{align}  The mean-anomaly averaged quantities are 
%%%%%%%%%%%%%%%%%%%%%%%%%%%%%%%%
\begin{align} \langle x^3 \rangle_M =& -(5/64) a^3 e (\cos \omega \cos \Omega - \cos i \sin \omega \sin \Omega) \nn \\ &\times \left[ \cos 2i (6+e^2 +7e^2 \cos 2\omega \cos 2\Omega ) \right. \nn \\ &\left. +7e^2 \cos 2\omega(3\cos 2\Omega + 2\sin^2 i) \right.  \nn \\ &\left. +  (6+e^2)(3+2\cos 2\Omega \sin^2 i) \right.  \nn \\ &\left. - 28e^2 \cos i \sin 2\omega \sin 2 \Omega \right], \end{align}
%%%%%%%%%%%%%%%%%%%%%%%%%%%%%%%%
\begin{align}
\langle y^2x \rangle_M =& (5/256) a^3 e \nn \\ & \times \left[ 2\cos 2i \cos \omega (-2-5e^2+7e^2\cos2\omega)\right. \nn \\ &\left. \times (\cos \Omega + 3\cos 3\Omega) \right. \nn \\ &\left. + 2\cos \omega(-7(2+e^2+e^2\cos2\omega)\cos\Omega \right. \nn \\ &\left. + (6-13e^2+35e^2 \cos 2\omega)\cos 3\Omega) \right. \nn \\ &\left. +4\cos 3i (6+e^2-7e^2\cos 2\omega)\cos^2 \Omega \sin\omega \sin\Omega \right. \nn \\ &\left. +\cos i \sin \omega [(26+23e^2+7e^2\cos 2 \omega )\sin \Omega \right. \nn \\ &\left. - 3(2+19e^2+35e^2\cos 2\omega)\sin 3\Omega]   \right],
\end{align}
%%%%%%%%%%%%%%%%%%%%%%%%%%%%%%%%
\begin{align} 
\label{eq:zzxM}
\langle z^2x \rangle_M & =(5/16)a^3e \sin^2 i \nn \\ & \times \left[\cos \omega (-2-5e^2+7e^2\cos 2\omega)\cos \Omega \right. \nn \\ &\left. + \cos i (6+e^2-7e^2\cos 2\omega )\sin \omega \sin \Omega \right]. 
\end{align}
%%%%%%%%%%%%%%%%%%%%%%%%%%%%%%%%
Plugging the results \eqref{eq:zzxM}-\eqref{eq:phixxxavg} in to \eqref{H2MtLK}, the resulting doubly-averaged test particle octupole Lidov-Kozai Hamiltonian is 
\begin{align} 
\overline{\langle H_\mathrm{oct} \rangle}_M &= \left( \frac{m_2-m_1}{m_2+m_1} \right) \times \frac{15}{128}GMa_\mathrm{g}^{-4}e_\mathrm{g}(1-e_\mathrm{g}^2)^{-5/2}a^3 \nn \\ &\times \Big\{ \left(e+\frac{3e^3}{4} \right)\Big[(1-11\theta-5\theta^2+15\theta^3)\cos(\omega-\Omega) \nn \\ &+ (1+11\theta -5\theta^2 -15\theta^3)\cos(\omega+\Omega)\Big] \nn \\ &-\frac{35}{4}e^3\Big[(1-\theta-\theta^2+\theta^3)\cos(3\omega-\Omega)\nn \\ &+(1+\theta-\theta^2-\theta^3)\cos(3\omega+\Omega)\Big]\Big\},  
\label{eqn:HoctLK} 
\end{align} 
where $\theta\equiv \cos i$.  Equation \eqref{eqn:HoctLK} is precisely the result found in standard LK literature (e.g. \citealt{Ford2000,Lithwick2011,Naoz2016}).

%%%%%%%%%%%%%%%%%%%%%%%%%%%%%%%%%%%%%%%%%%
%%%%%%%%%%%%%%%%%%%%%%%%%%%%%%%%%%%%%%%%%%

\section{Numerical prescription for computing time-averages} \label{Numerical_Time_Averages} 

%%%%%%%%%%%%%%%%%%%%%%%%%%%%%%%%%%%%%%%%%%

To calculate the time-averages $\overline{\Phi}_{\alpha \beta}$ numerically we use the orbit integrator in \texttt{galpy} \citep{Bovy2015}. Given the initial position $\Rg(0)$ and velocity of the binary's outer orbit around the cluster, we integrate its equation of motion (\ref{eq:R_g}) numerically in the smooth cluster potential $\Phi$. We use a constant timestep $\Delta t$ so that after $k$ timesteps the time elapsed is $t_k = k\Delta t$. Then the running time-average of a quantity $\mathcal{F}(\Rg(t))$  is \begin{align} 
\overline{\mathcal{F}}(t) = \frac{1}{t}\int_0^t \md t'\, \mathcal{F}(t') \approx \frac{\Delta t}{t}\sum_{k=0}^{t/\Delta t} \mathcal{F}(t_k).
\end{align}
In nearly all numerical examples shown in this paper we used $\Delta t \approx T_\phi/100$ where $T_\phi$ is the azimuthal period of $\Rg$, and integrated the outer orbit for approximately $100 T_\phi$.  The exception is Figure \ref{fig:App_D_Ex2}, where we used $\Delta t \approx T_Z/10$ ($T_Z$ is the vertical period of $\Rg$) and integrated the outer orbit for $5000 T_Z$.

% Don't change these lines
\bsp	% typesetting comment
\label{lastpage}
\end{document}